\documentclass{elsarticle}
\usepackage{hyperref}

\journal{Physica A}

\biboptions{compress}
 \usepackage{graphicx}
 \usepackage{amsmath}
 \usepackage{caption}
 \usepackage{mwe}
 \usepackage{subcaption}
 \usepackage{mathtools}
 \usepackage{xcolor}
\begin{document}

\begin{frontmatter}

\title{A multi-scale symmetry analysis of uninterrupted trends returns of daily financial indices}

\author[mymainaddress]{C.M. Rodr\'{\i}guez-Mart\'{\i}nez}
\author[mymainaddress,mysecondaryaddress]{H.F. Coronel-Brizio} 
\author[mymainaddress,mysecondaryaddress]{A.R. Hern\'andez-Montoya\corref{mycorrespondingauthor}}

\ead{alhernandez@uv.mx}
\ead[url]{www.uv.mx/alhernandez}
\cortext[mycorrespondingauthor]{Corresponding author}

\address[mymainaddress]{Centro de Investigaci\'on en Inteligencia Artificial. Universidad Veracruzana. Sebasti\'an Camacho 5, Xalapa Veracruz 91000, M\'exico. Tel/Fax: 52-228-8172957/8172855.}

\address[mysecondaryaddress]{Facultad de F\'{\i}sica. Universidad Veracruzana, Apdo. Postal 475. Xalapa, Veracruz. M\'{e}xico.}

\begin{abstract}

  We present a symmetry analysis of the distribution of variations of different financial indices, by means of a statistical procedure developed by the authors  based on a symmetry statistic by Einmahl and Mckeague. We applied this statistical methodology to financial  uninterrupted daily trends returns and to other derived observable. In our opinion, to study distributional symmetry, trends returns offer more advantages than the commonly used daily financial returns; the two most important being: 1) Trends returns involve sampling over different time scales and 2) By  construction, this variable time series contains practically the same number of non-negative  and negative entry values. We  also show that these time multi-scale returns display distributional bi-modality. 
Daily financial indices analyzed in this work, are the Mexican IPC, the American DJIA, DAX from Germany and the Japanese Market index Nikkei, covering a time period from 11-08-1991 to 06-30-2017. We show that, at the time scale resolution and significance considered in this paper, it is almost always feasible to find an interval of possible symmetry points containing one most plausible symmetry point denoted by $C$. Finally, we study the temporal evolution of $C$ showing that this point is seldom zero and responds with sensitivity to extreme market events.
\end{abstract}

\end{frontmatter}

\section{Introduction}

Properties of financial markets and their behavior are of great interest to academics, policy makers, traders and investors. These markets are complex systems and all of them present universal emerging properties of statistical nature, nowadays commonly called by the name of stylized facts \cite{Kaldor,Cont}.

In particular, the empirical gain/loss asymmetry, i.e.  ``draw-downs in stock prices or stock index values are larger than the upward variations'', is one of those stylized facts, and it is of great and current interest under both, practical/applied and academic points of view (i.e. symmetry based automatic trading systems, skewness induced risk, validity of Efficient Market Hypothesis, etc). Indeed, studying the symmetry of the distribution of financial returns has been researched for many years and it has always been an important and tricky business: results of symmetry measurements depend on the period of time of data records, the time scale analyzed, returns definition, leverage effect, etc. \cite{Beedles,Fogler,Bouchaud,Jensen2,Dutta,deAlmeida}.

Initially, most of the market symmetry or gain/loss asymmetry studies relied on studying the third standardized moment of the price or index variations or other similar measurement. More recently, symmetry of financial variations or related problems has been approached by very ingenious methodologies, as for example, the analysis of the returns distribution of stocks ensembles during crash and rally days \cite{Lillo}; the study of large fluctuation dynamics under time reversal (TR) symmetry (large fluctuations dynamics at daily scale are not TR symmetric, but at the scale of high frequency data they are)\cite{Jiang}; study of the investment horizons distribution \cite{Jensen,Karpio}; empirical analysis of the clustering on the asymmetry properties in financial time series \cite{Jun}; symmetry break mechanisms \cite{Savona}; symmetry in trading volume \cite{Duarte}; analysis related to time-scale effects on gain/loss asymmetry in stock indices \cite{Zoltan}; the use of the non-extensive formalism of Physics \cite{Grech}, or focusing in searching possible symmetry points of returns \cite{Coronel-Montoya} and many more interesting empirical and agents modeled studies \cite{Vitalies,Takayasu}.

Even though it has been researched for many years, the study of the symmetry of the unconditional distribution of financial returns remains, in our opinion a not fully understood subject. For instance, \cite{peiro,peiro2} have analyzed returns of a large sample of diverse financial indices without finding important symmetry deviations or fully rejecting the symmetry hypothesis. On the other hand, many studies under different conditions and points of view have reported the emergence of asymmetries in the financial returns distribution \cite{Lillo,Jiang,Jensen,Karpio,Jun,Savona,Duarte,Zoltan,Grech,Coronel-Montoya,Vitalies,Takayasu}.

By the analysis presented in this paper we expect to shed new light in what it may be grasped when we speak of symmetry of the distribution of financial variations. Usually, in finance, a symmetric distribution of financial variations is understood as a distribution with a mirrored symmetry around the $y$ vertical axis in $x=0$, i.e. the mean values of the distribution coincides with the origin. In this paper, we will see that although it is possible to find a symmetry point of the variations distribution, most of the time this point it is not situated in the origin or symmetrically around it. 

\subsection{Organization of the present paper}

This work is organized as follows: in the next section we briefly review the statistical methodology used for the symmetry analysis: we show the results of the numerical calculation of the asymptotic distribution of the $T_n$ statistic as obtained in \cite{Coronel-Montoya}. These values are important because we use them to find the interval where it is more plausible to observe symmetry points of market variations, as is explained in subsection \ref{Meth}. In section \ref{sec:Data}, and following \cite{Olivares}, we present the time multi-scale observables analyzed in this work. We explain their construction, report some interesting statistical distributional properties that they display, such as bi-modality, and we include a brief digression on the reasons these observables were chosen to study the symmetry of market variations. In section \ref{sec:analisis}, we apply our methodology for assessing symmetry to four data samples of different markets from the period of time including 11-08-1991 to 06-30-2017. Also, in this section we show how the most plausible symmetry point for a given confidence level from market variations evolves in time and behaves around extreme market movements. Finally, conclusions and a summary are discussed in section \ref{sec:Final}.

\section{The $T_{n}$ Statistic and methodology to test symmetry}
\label{sec:Tn}

Our symmetry test is based in the distribution-free test statistic $T_{n}$ for symmetry testing proposed by Einmahl and Mckeague \cite{Einmahl} and it follows an empirical likelihood approach.
Suppose a sample $X_{1},\ldots,X_{n}$ consisting of independent and identically distributed random variables with  common absolutly continuous cumulative distribution function $F$.
The null hypothesis is that there exists a point of symmetry around zero, i.e. , $H_{0}:F(0-x)=1-F(x-0)$, for all $x>0$ belonging to the sample.

We use the notation $F_n(-x) \coloneqq F_n(0-x)$ and $F_n(x-) \coloneqq F_n(x-0)$, where the empirical distribution function $F_n$ is defined as $$F_n(x): = \frac{1}{n}\sum_{i=1}^{n} I(X_{i}\leq x)$$ and the function $I$, named the indicator function, takes a value of 1 if its argument is true and 0 if its argument is false. The test statistic by Einmahl and Mckeague is:
\begin{equation}
T_{n}=-2\int_{0}^{\infty} \log H(x)dG_{n}(x)=-\frac{2}{n}\sum_{i=1}^{n} \log H \left( \left| X_{i} \right| \right). 
\label{ts1}
\end{equation}

where $G_{n}$ is the empirical distribution function of the $\left| X_{i} \right|$, and $H$ satisfies:

$ \log H(x) = nF_n \left( { - x} \right)\log \frac{{F_n \left( { - x} \right) + 1 - F_n (x - )}}{{2F_n \left( { - x} \right)}}+n\left[ {1 - F_n \left( {x - } \right)} \right]\log \frac{{F_n ( - x) + 1- F_n (x - )}}{{2\left[ {1 - F_n \left( {x - } \right)} \right]}} $

More details on the above shown, can be found in \cite{Einmahl}.

In the previous cited paper by the authors \cite{Coronel-Montoya}, the asymptotic percentage points of the statistic $T_{n}$ were calculated numerically and they are shown in table \ref{tab:Cpoints}.

\begin{table}[h!tb]
	\begin{center}
		{\renewcommand{\arraystretch}{1.2}
	\begin{tabular}{|c|c|}
		\hline
		Cumulative Probability &Percentage point\\ 
		\hline
		\hline
		0.50&0.659\\
		\hline
		0.75&1.258\\
		\hline
		0.85&1.768\\
		\hline
		0.90&2.200\\
		\hline
		0.95&2.983\\
		\hline
		0.975&3.798\\
		\hline
		0.990&4.909\\
		\hline
		0.995&5.768\\
		\hline
		0.999&7.803\\
		\hline
		\hline
	\end{tabular}
	}
	\caption{Asymptotic percentage points of $T_n$ calculated numerically from  \cite{Coronel-Montoya}.}
	\label{tab:Cpoints}
	\end{center}
\end{table}

These limiting distribution points were found by proving that $T_n$ converges to:
\[
T_n \mathop \to \limits^D \int\limits_0^1 { \frac{W(t)^2}{t} dt}
\]
where $W$ denotes a standard Wiener process.

Asymptotic points were obtained using the series representation \cite{imhof}
\[
    T_n \mathop \to \limits^D \sum\limits_{i = 1}^\infty {\lambda _i \nu _i },
\] 
where $\nu_1,\nu_2,\ldots$ are independent chi-squared random variables with one degree of freedom, and $\lambda_1,\lambda_2,\ldots$ are the eigenvalues of the integral equation:
\begin{equation}
\int\limits_0^1 {\sigma (s,t)f_i ds = \lambda _i f_i (t)} 
\label{inteq}
\end{equation}

where $\sigma(s,t)$ denotes the covariance function of the process $\frac{W(t)}{\sqrt{t}}$. Finally, the asymptotic percentage points of the distribution of $T_n$ were found solving equation~\ref{inteq} numerically.

It is worth mentioning that in table A1 of \cite{Einmahl2}, Einmahl and Mckeague had only provided critical values of $T_{n}$ based on 100,000 simulations for selected sample sizes up to n=150. Their simulation results showed a fast convergence to the asymptotic percentage points.

Values of table \ref{tab:Cpoints} will be used in our statistical procedure to assessing the symmetry of the financial indices variations as summarized in section \ref{sec:Data}. For further details on this methodology see again \cite{Coronel-Montoya}.

\subsection{Statistical Methodology}
\label{Meth}
Let us now summarize the statistical methodology based on $T_n$ to assess symmetry of a given  set of observations from an unknown probability distribution. For more details, consult \cite{Coronel-Montoya}.

As usual, given a daily financial time series of prices or indice values $P_1,\ldots P_{N+1}$ we define its corresponding time series of daily logarithmic returns, called in this paper only Returns, as $R_i := \log(P_{i+1})-\log(P_i$), $i=1,\ldots N$.

Let us define for Returns time series their ``{\it shifted returns around} $c$'' as $S_{R_{t}}(c)=R_{t}-c$, with $t=1,\ldots,N$, where $c$ is a real number, not necessarily a symmetry point. In particular, for the present explanation, $R_{t}$ may represent either daily logarithmic returns or the multi-scale returns, defined in next section.

We will use the time series of shifted returns to obtain the corresponding values of their test-statistic $T_{n}$ calculated from $S_{R_{1}}(c),\ldots,S_{R_{N}}(c)$, and denoted $T_{n}(c)$ for a particular value of $c$.

Now, let us suppose the data to test have at least a symmetry point. Then a {\em plausible} value of the symmetry point (for a given significance level $\alpha$), will be any real number $c$, such that
\begin{equation}
T_{n}(c)<T(\alpha),
\label{eq:SL}
\end{equation}
where $T(\alpha)$ denotes the $\alpha-$level upper point of the distribution of $T_{n}$ selected from table \ref{tab:Cpoints}.

The interval of symmetry of our returns distribution, with a significance level $\alpha$, will be the interval $(C_{min},C_{max})$, where $C_{min}$ and $C_{max}$ are, respectively the infimum and supremum of the set of all plausible symmetry points i.e. those points satisfying equation \ref{eq:SL}.

Finally, the most plausible symmetry point for the chosen significance level $\alpha$, is the one that minimizes $T_{n}(c)$ in the interval  $(C_{min},C_{max})$ and it will denoted $C$.

To obtain $(C_{min},C_{max})$ it is necessary to determine the intersection of the curve of $T_{n}(c)$ versus $c$ and the horizontal line of significance level calculated from  table \ref{tab:Cpoints} in an interval that contains the set of all possible values of $c$ which would not lead to the rejection of the null hypothesis of symmetry for the probability distribution of the random variable $R_{t}$ at the given significance level $\alpha$. Of course, $C$ is the point in this interval that minimizes $T_{n}(c)$.

The previously  described methodology will be illustrated for all our data samples in section \ref{sec:Data}.

\subsection{Skewness vs $T_n$}

We close this section with a few words explaining our motivation to choose a methodology based on the $T_n$ statistic over  more traditional tests based on skewness:
\begin{itemize}
\item It is well known that a small skewness of a distribution is a necessary condition for symmetry, but it is not a sufficient condition, i.e., a small value of skewness does not guarantee the symmetry of the corresponding distribution.
\item A statistical test of symmetry or a statistic based on skewness has to overcome the above mentioned problem and for that reason should be more complex than our test making necessary to set up additional conditions to the tested distribution.
\item Our assessment of symmetry is independent of the distribution of data. No additional constrains are required for its application and $T_n$ is based in the empirical distribution function only.
\item Finally, and as previously mentioned at the beginning of this work, since many years, most of the statistical studies of symmetry of financial data were elaborated by analyzing skewness of data distribution. In our opinion, new statistic perspectives to face this important and interesting problem are needed.
\end{itemize}

\section{Data Sample and construction of our observables}
\label{sec:Data}

We construct the observables of market variations from the following financial indices daily time series:
\begin{enumerate}
	\item American DJIA (Dow Jones Industrial Average).
	\item American Nasdaq (National Association of Securities Dealers Automated Quotation).
	\item Mexican IPC (Indice de Precios y Cotizaciones in Spanish of Index of Prices and Quotations in English).
	\item Japanese Nikkei 225.
\end{enumerate}
All data samples were downloaded from \url{https://finance.yahoo.com/} website, the period of time considered in our analysis is from 11-08-1991 to  06-30-2017.

\subsection{Observables construction}
\begin{figure}[h!tb!]
    \includegraphics[scale=0.4]{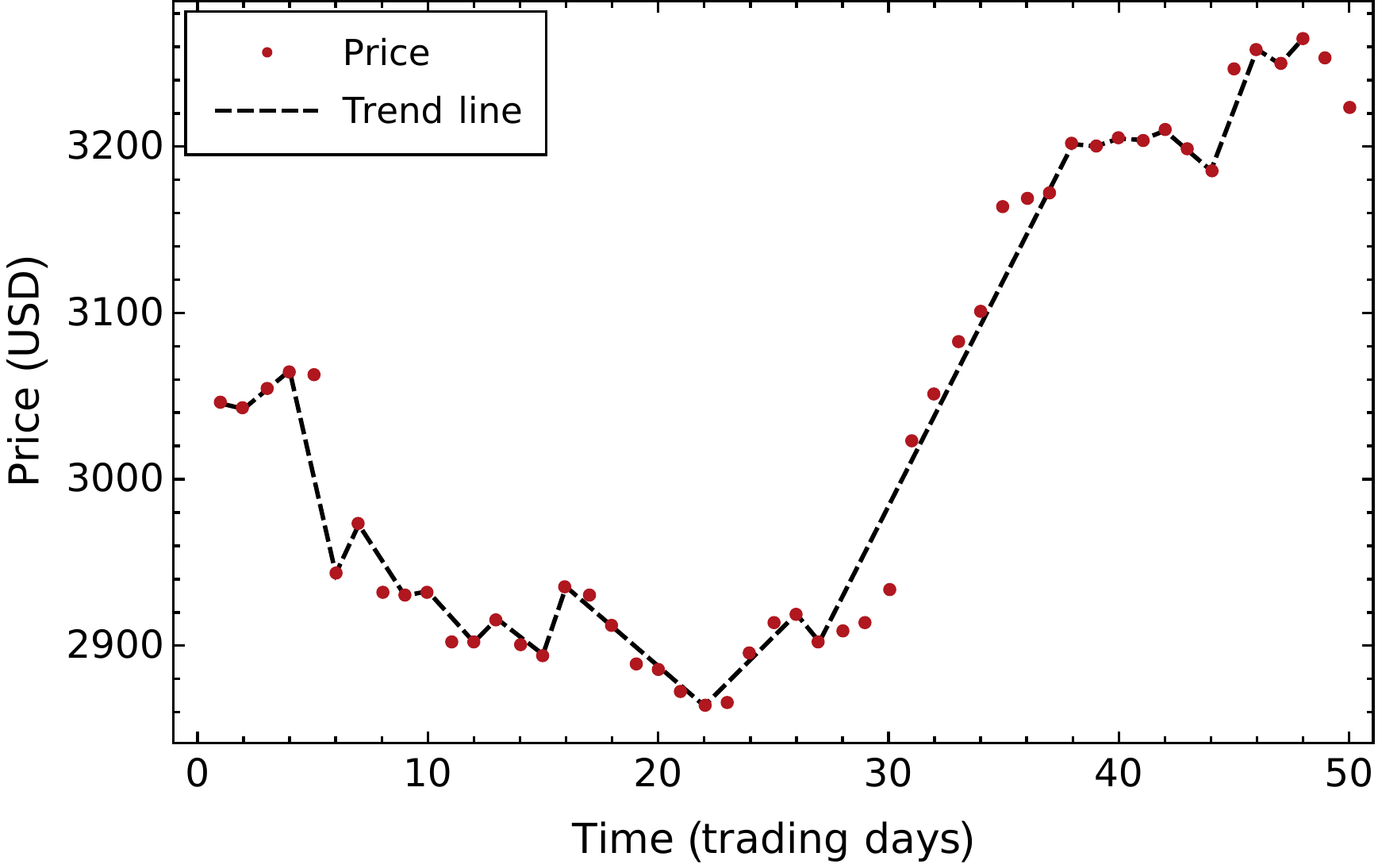}
    \centering
    \caption{\protect\raggedright Uninterrupted daily trends for DJIA. Time period from Nov/08/1991 to Jan/21/1992.}
    \label{fig:TReturns}
\end{figure}

In order to study the symmetry of financial markets variations, we construct two time multi-scale observables of financial returns calculated by analogy with daily returns and using the logarithmic differences of the extreme values of daily uninterrupted trends \footnote{In probability theory, this kind of consecutive events are called ``runs''.}\cite{Wilks}. This has the advantage of not having to define a fixed and arbitrary scale for returns, since that time scale is naturally defined by the different uninterrupted trends durations. For a graphic illustration using DJIA data, showing what we mean by uninterrupted trends, see figure \ref{fig:TReturns}. 

More formally, given a price time series $P_{1},\ldots,P_{n}$, we define an ``uninterrupted trend'' of duration $k$, as a succession of $k+1$ consecutive values of the given time series where each value is greater than the preceding one. In this case we say that these consecutive $k+1$ increasing values form an uninterrupted uptrend. On the other hand, if each one of the $k+1$ consecutive values is smaller or equal than the preceding one we say we have an uninterrupted downtrend of length $k$.

Now, let us suppose that for a fixed integer $m>0$, an uninterrupted trend (uptrend or downtrend), starts at $P_m$ with a duration of $k$ days: $P_{m},\ldots,P_{m+k}$. In this case, we define their corresponding trend return as:
	\begin{equation}
		S_m^k := \log(P_{m+k}) - \log(P_{m})
	\end{equation}
Where $m$ indexes the different trends and $k$ indicates the duration in days of the $m$-th trend.

An other useful and closely related observable we analyze in this paper is defined as:
	\begin{equation}
		TS_m^k :=\frac{ \log(P_{m+k}) - \log(P_{m})}{k}
	\end{equation}
        i.e. the terms of the time series of TReturns divided by their corresponding  duration. It gives us an estimation of how fast trends\footnote{From now on, for simplicity, we will refer to uninterrupted trends only as trends} are. Let us name $TS_m^k$ as TVReturns, meaning Trends Velocity \footnote{Independently of knowing that TVReturns is a discrete observable, we use this term because every TVReturn is obtained dividing its corresponding TReturn by time duration in number of days} returns.

\begin{figure}[htb!]
          \begin{subfigure}[b]{0.48\textwidth}
            \centering 
            \includegraphics[width=\textwidth]{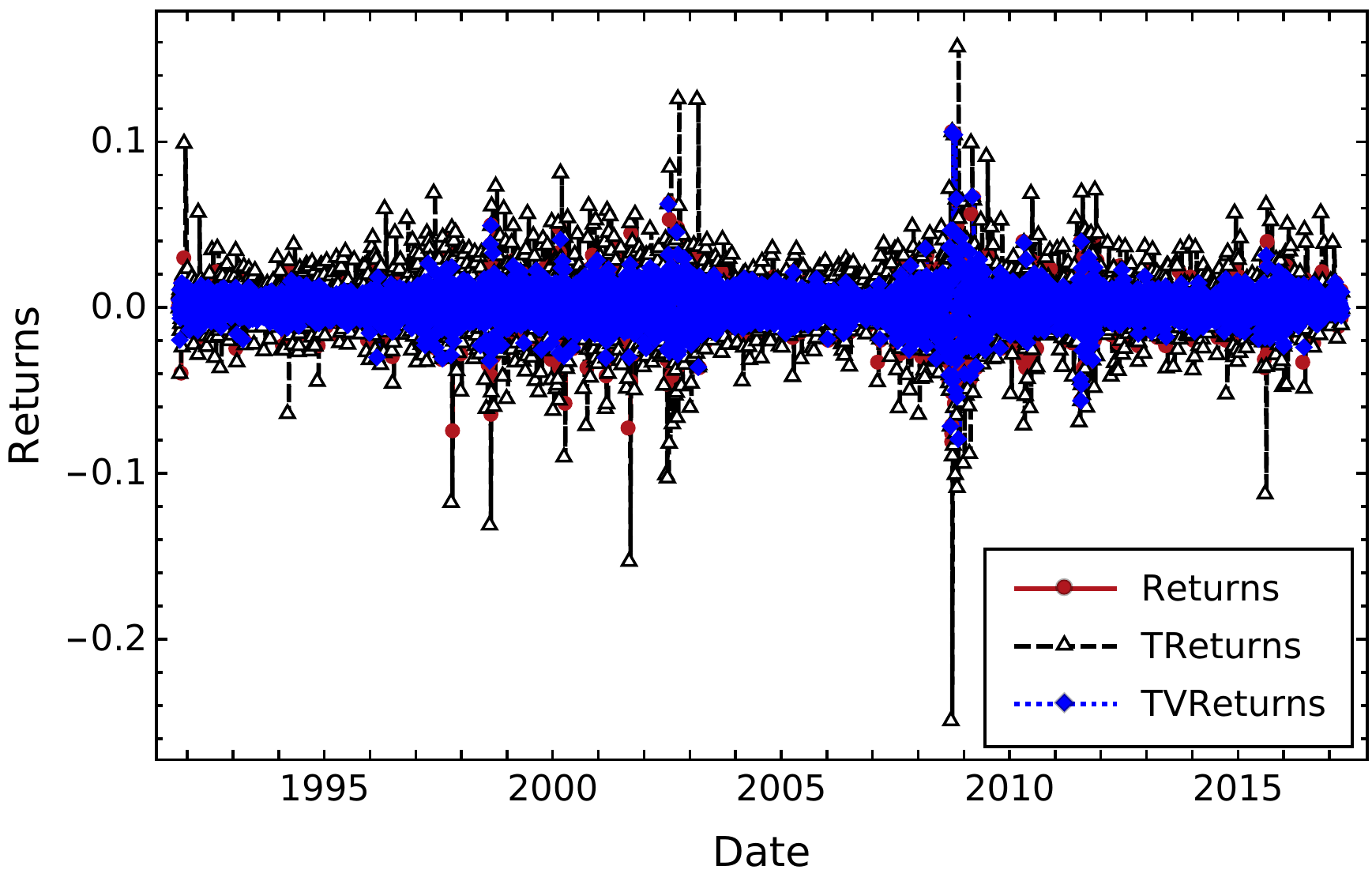}
            \caption[ ]{{Time evolution of TReturns, TVReturns and Returns  for DJIA data sample.}}
            \label{fig:TRandTVRdista}
          \end{subfigure}
                  \hfill
          \begin{subfigure}[b]{0.475\textwidth}
            \centering 
            \includegraphics[width=\textwidth]{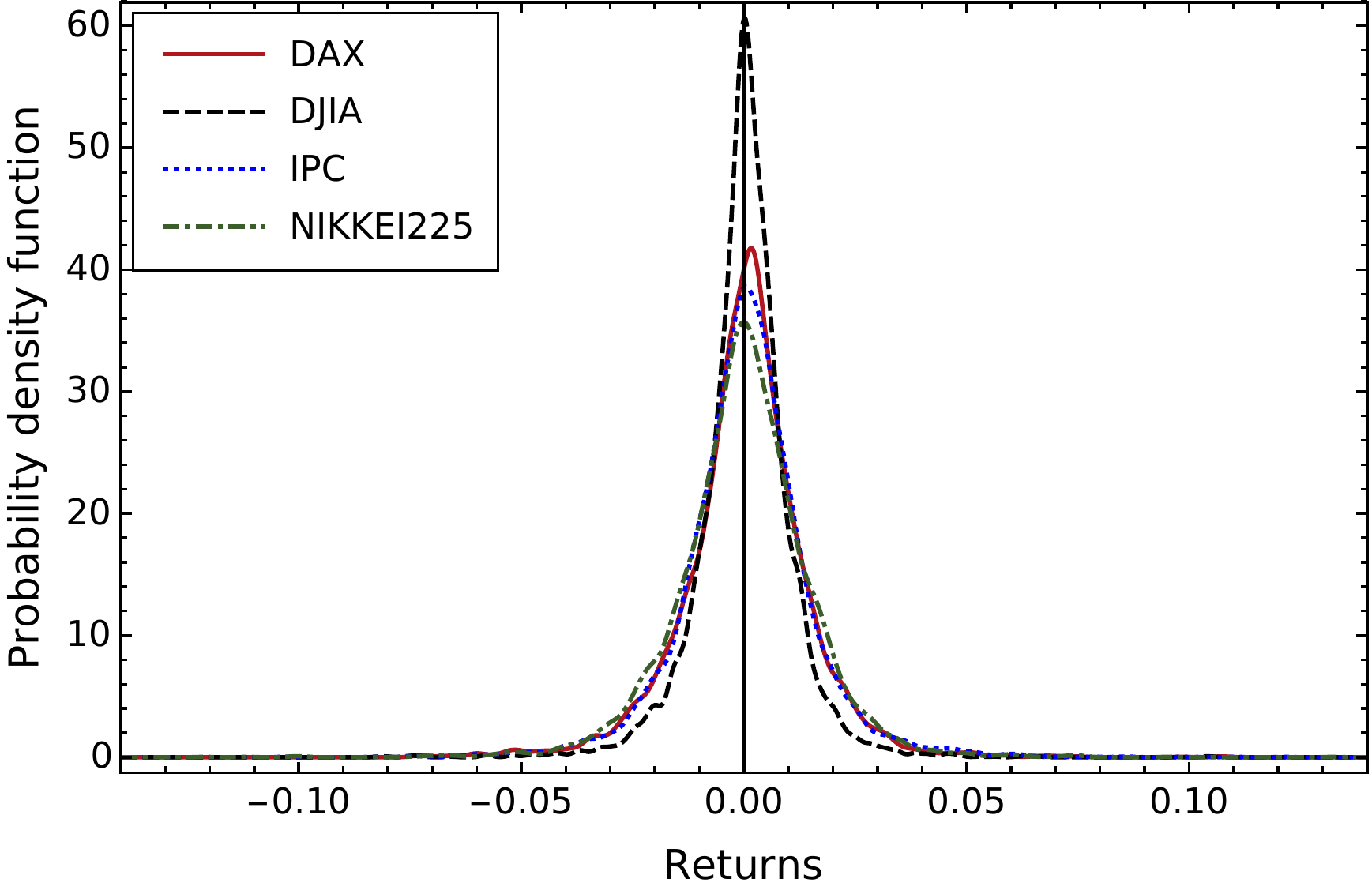}
            \caption[ ]{{Returns probability density function for the four analysed data samples.}}
            \label{fig:TRandTVRdistb}
          \end{subfigure}
                  \vskip\baselineskip
        \centering
        \begin{subfigure}[b]{0.475\textwidth}
            \centering
            \includegraphics[width=\textwidth]{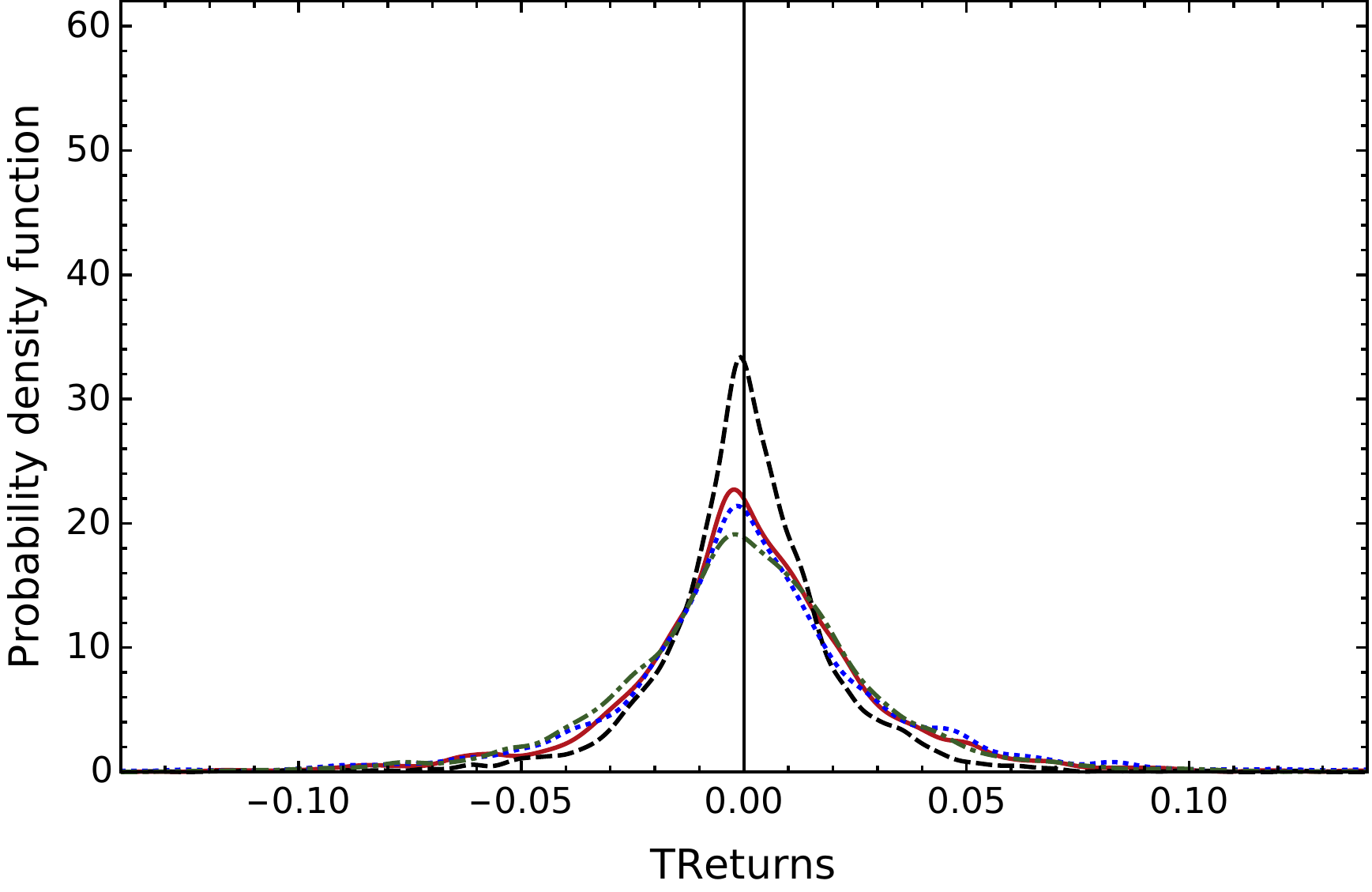}
            \caption[ ]{{TReturns probability density function for all samples.}}
            \label{fig:TRandTVRdistc}
        \end{subfigure}
        \hfill
        \begin{subfigure}[b]{0.475\textwidth}
            \centering 
            \includegraphics[width=\textwidth]{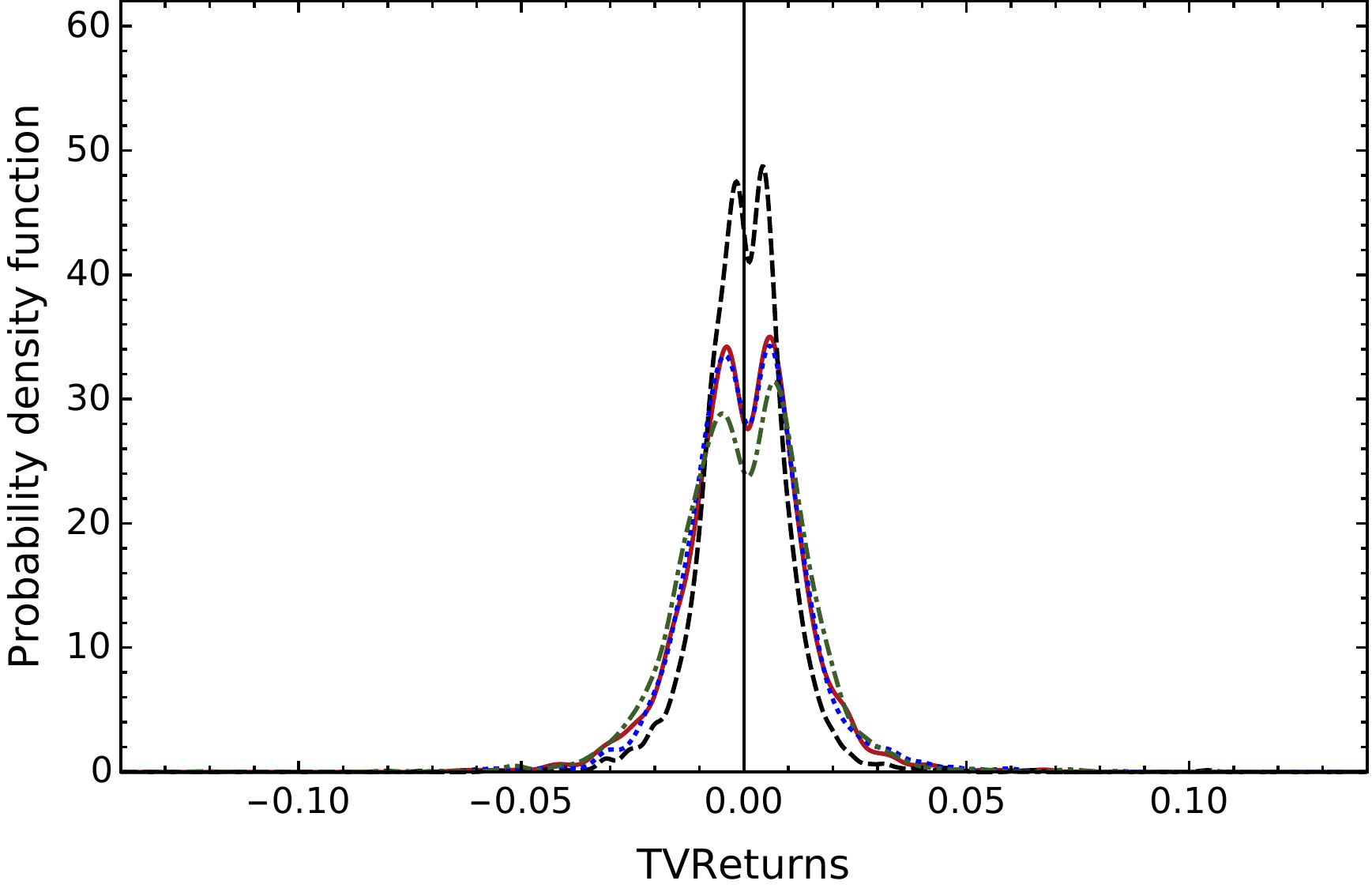}
            \caption[ ]{{TVReturns probability density function, again for all data samples}}
            \label{fig:TRandTVRdistd}
        \end{subfigure}

\caption{a) Returns, TReturns and TVReturns evolution  for DJIA sample. Figures (b),(c) and (d)  Smoothed probability density function of TReturns and the constructed  Multi-scale returns for all data samples.Descriptive statistics of these distributions is show in table \ref{Tab:TabStats} }
\label{fig:TRandTVRdist}
\end{figure}

      Figure \ref{fig:TRandTVRdist} is divided in four panels and they will be referred as figures \ref{fig:TRandTVRdista} to \ref{fig:TRandTVRdistd} respectively. 

For illustration purposes figure \ref{fig:TRandTVRdista}  displays only for DJIA data, the evolution of Returns, TReturns and TVReturns for the period of time analysed in the present work. The another figures \ref{fig:TRandTVRdistb} to  \ref{fig:TRandTVRdistd} show Returns, TReturns and TVReturns  smoothed probability density functions respectively.  A summary of their descriptive statistics can be found in table \ref{Tab:TabStats}, showing number of entries, mean, standard deviation, skewness and kurtosis values.

As expected from the idea of aggregating same signed variations, TReturns reach the  bigger amplitude fluctuations, then Returns and finally TVReturns. Compare their RMS values in  fifth column of table \ref{Tab:TabStats}.

\begin{table}[htb]
	\begin{center}
		\setlength\tabcolsep{3.2pt} 
		{\renewcommand{\arraystretch}{1.1}
		\begin{tabular}{|c|c|c|c|c|c|c|c|c|c|}
			\hline
			Market&Observable&Entries&$\mu$&$\sigma$&Skewness&Kurtosis\\
			\hline
			\hline
			DJIA&Returns& 6447& $(2.9 \pm 1.3) \!\times\!10^{-4}$&0.011&-0.176&11.56\\
			\hline
			DJIA&TReturns& 3340& $(5.6 \pm 3.5) \!\times\!10^{-4}$&0.021&-0.693&14.95\\
			\hline
			DJIA&TVReturns& 3340& $(1.6 \pm 1.7) \!\times\!10^{-4}$&0.010&0.378&13.96\\
			\hline
      IPC&Returns&6409& $(5.5 \pm 1.8) \!\times\!10^{-4}$&0.015&0.022&9.355\\
			\hline
			IPC&TReturns&2942& $(1.2 \pm 0.6) \!\times\!10^{-3}$&0.034&-0.241&9.678\\
			\hline
			IPC&TVReturns&2942& $(4.6 \pm 2.4) \!\times\!10^{-4}$&0.013&0.0152&5.845\\
			\hline
			DAX&Returns&6465& $(3.1 \pm 1.7) \!\times\!10^{-4}$&0.014&-0.106&7.300\\
			\hline
			DAX&TReturns&3274& $(6.3 \pm 5.1) \!\times\!10^{-4}$&0.029&-0.729&10.83\\
			\hline
			DAX&TVReturns&3274& $(1.2 \pm 2.2) \!\times\!10^{-4}$&0.013&-0.027&5.431\\
			\hline
			Nikkei&Returns&6282& $(-0.3 \pm 1.9) \!\times\!10^{-4}$&0.015&-0.203&8.094\\
			\hline
			Nikkei&TReturns&3272& $(-0.5 \pm 5.2) \!\times\!10^{-4}$&0.030&-0.569&11.95\\
			\hline
			Nikkei&TVReturns&3272& $(-0.6 \pm 2.4) \!\times\!10^{-4}$&0.014&-0.257&6.813\\
			\hline
			\hline
		\end{tabular}
		}
	\caption[]{\small Descriptive statistics of Returns, TReturns and TVReturns for all data samples.}
	\label{Tab:TabStats}
	\end{center}
\end{table}

\begin{figure}[h!tb]
    \centering
    \begin{subfigure}[b]{0.45\textwidth}
        \centering
        \includegraphics[scale=0.30]{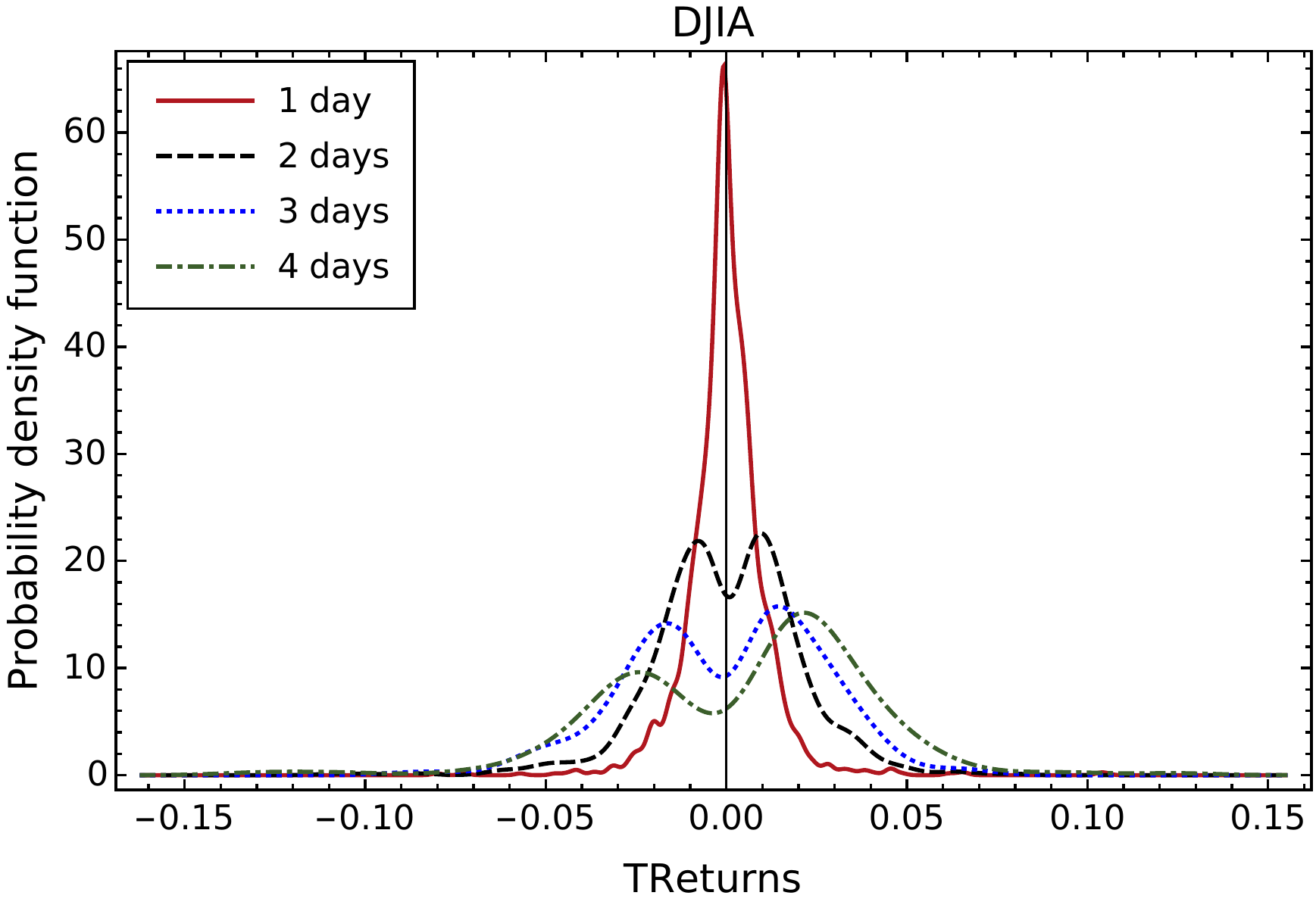}
    \end{subfigure}
    \quad
    \begin{subfigure}[b]{0.45\textwidth}
        \centering 
        \includegraphics[scale=0.30]{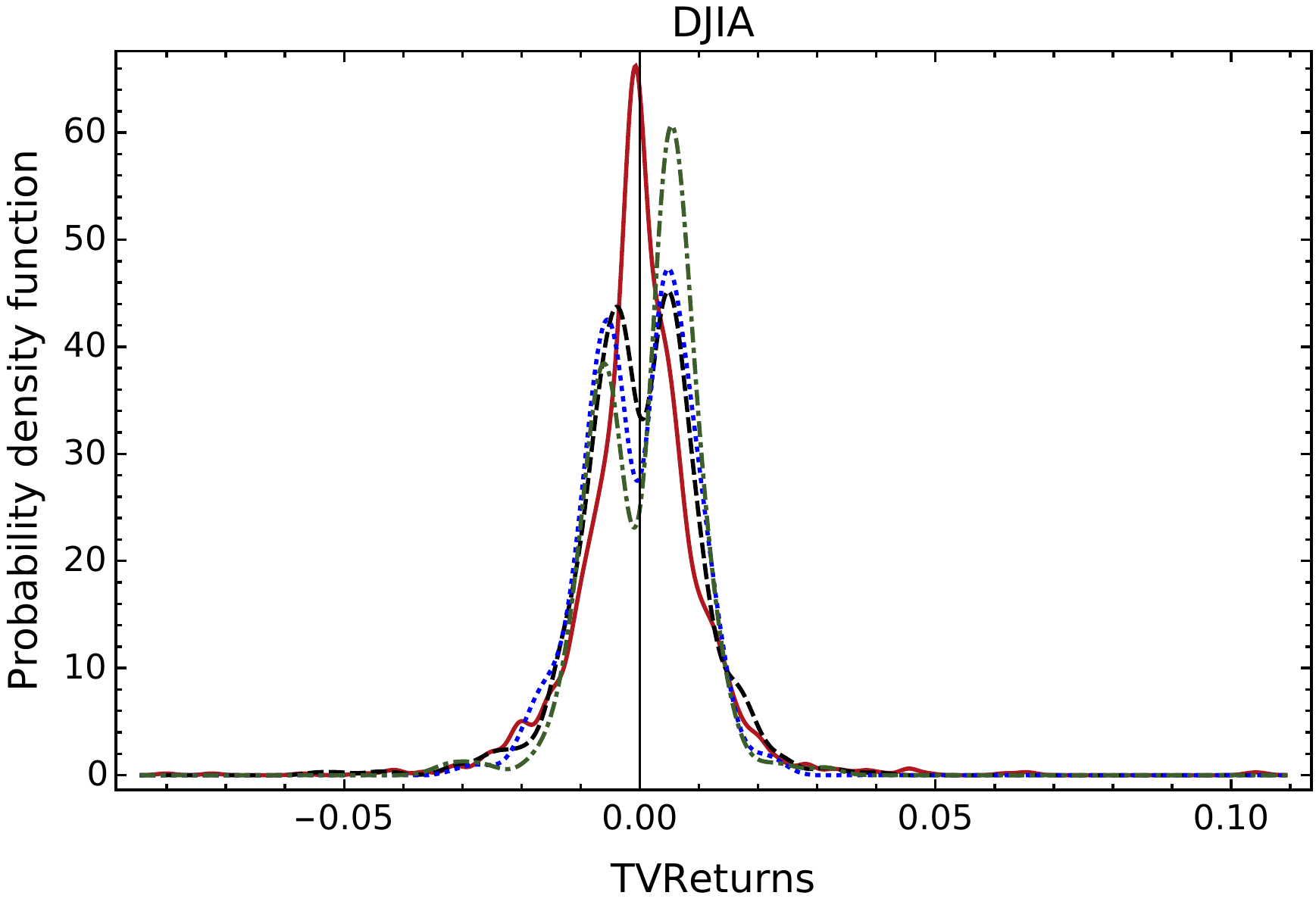}
    \end{subfigure}
    \vskip\baselineskip
    \begin{subfigure}[b]{0.45\textwidth}
        \centering
        \includegraphics[scale=0.3]{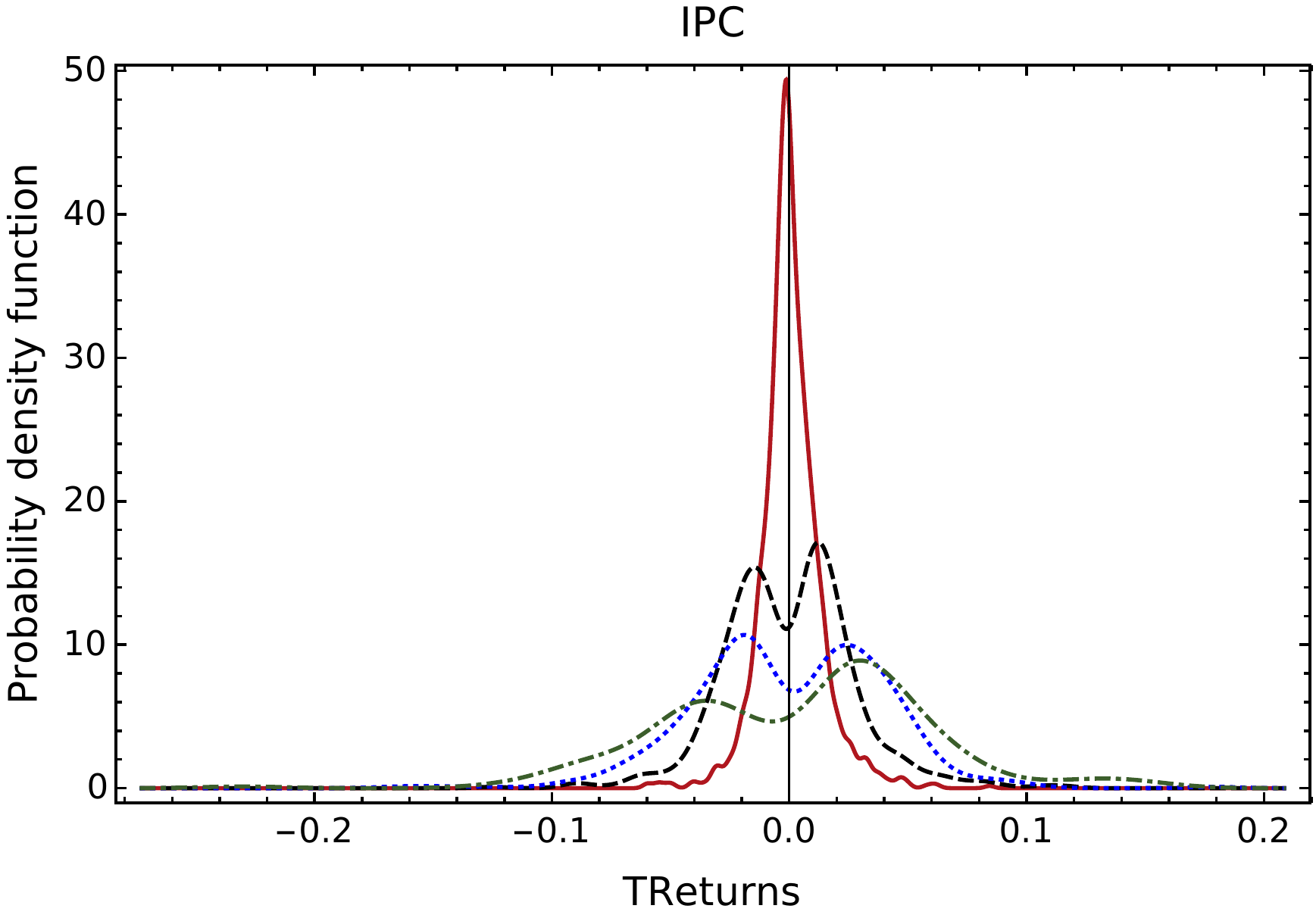}
    \end{subfigure}
    \quad
    \begin{subfigure}[b]{0.45\textwidth}
        \centering 
        \includegraphics[scale=0.3]{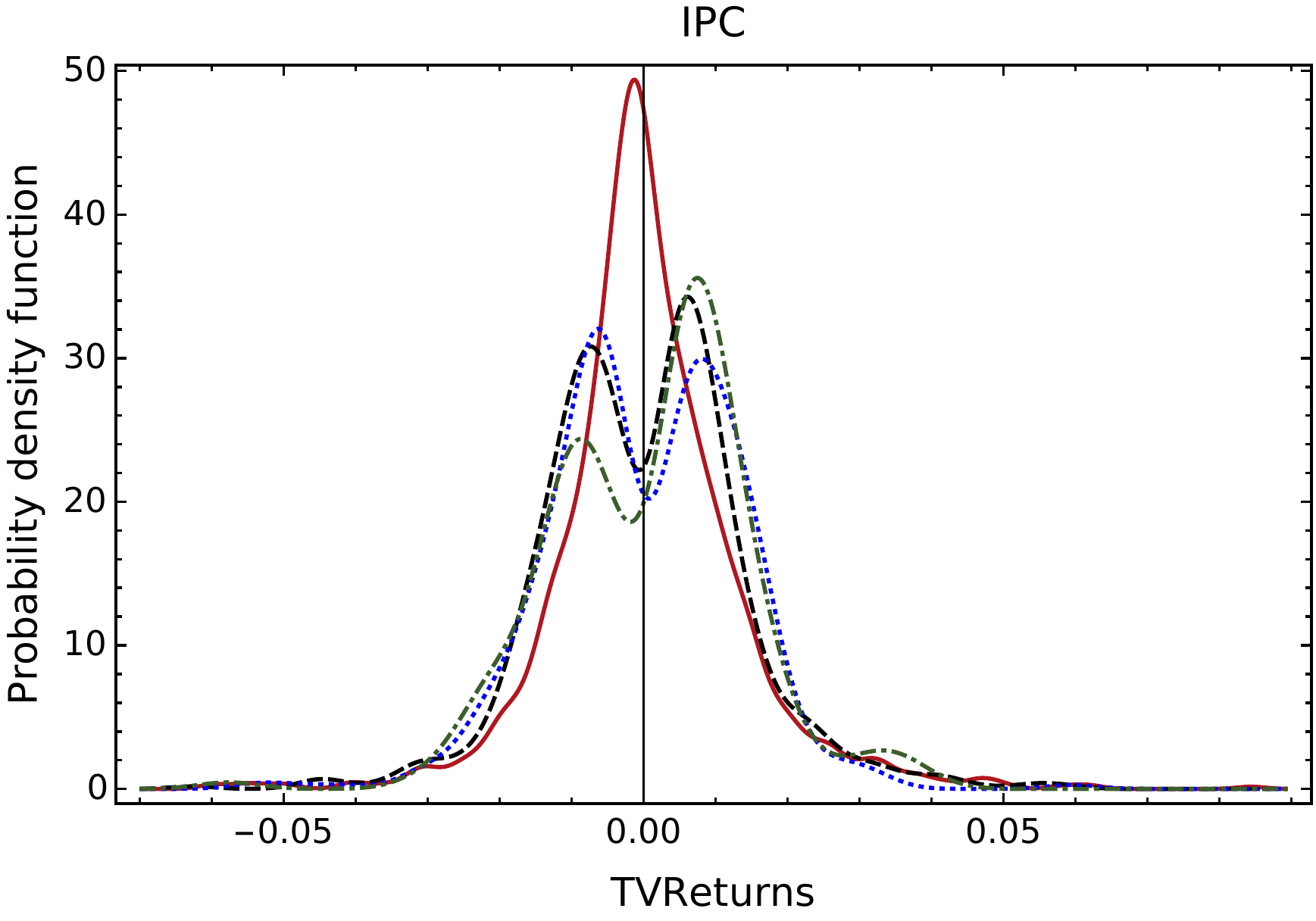}
    \end{subfigure}
    \vskip\baselineskip
    \begin{subfigure}[b]{0.45\textwidth}
        \centering
        \includegraphics[scale=0.3]{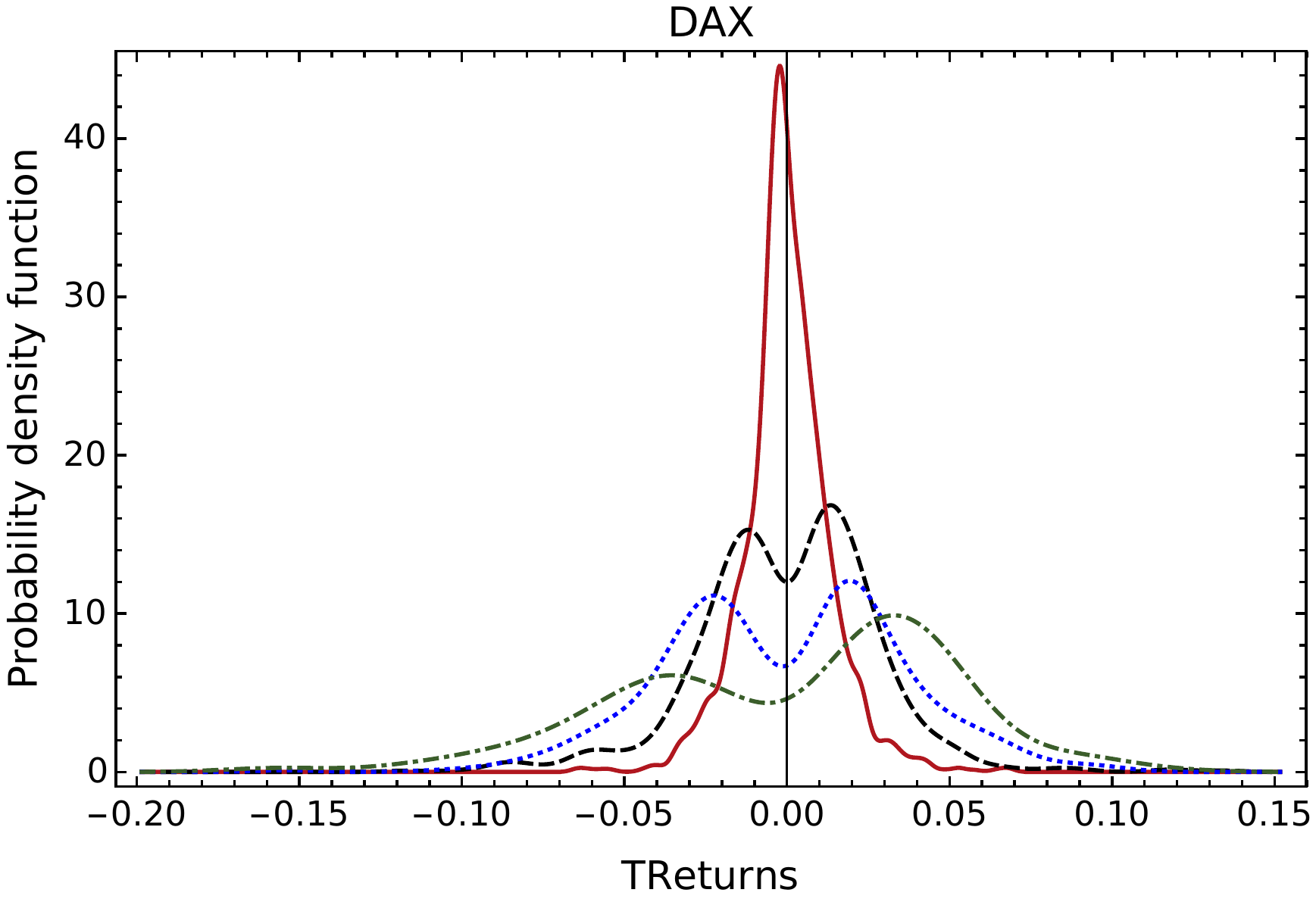}
    \end{subfigure}
    \quad
    \begin{subfigure}[b]{0.45\textwidth}
        \centering 
        \includegraphics[scale=0.3]{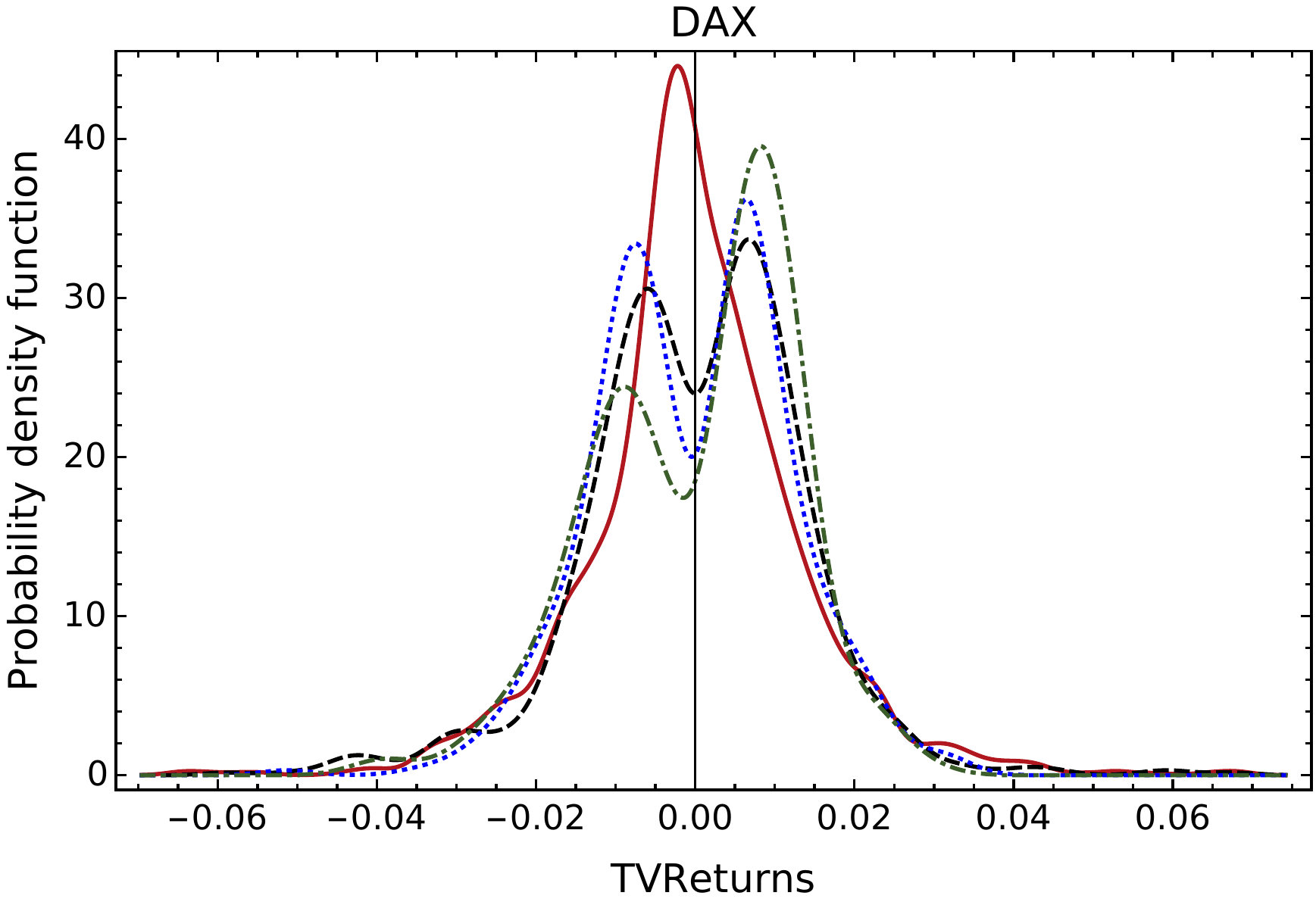}
    \end{subfigure}
    \vskip\baselineskip
    \begin{subfigure}[b]{0.45\textwidth}
        \centering
        \includegraphics[scale=0.3]{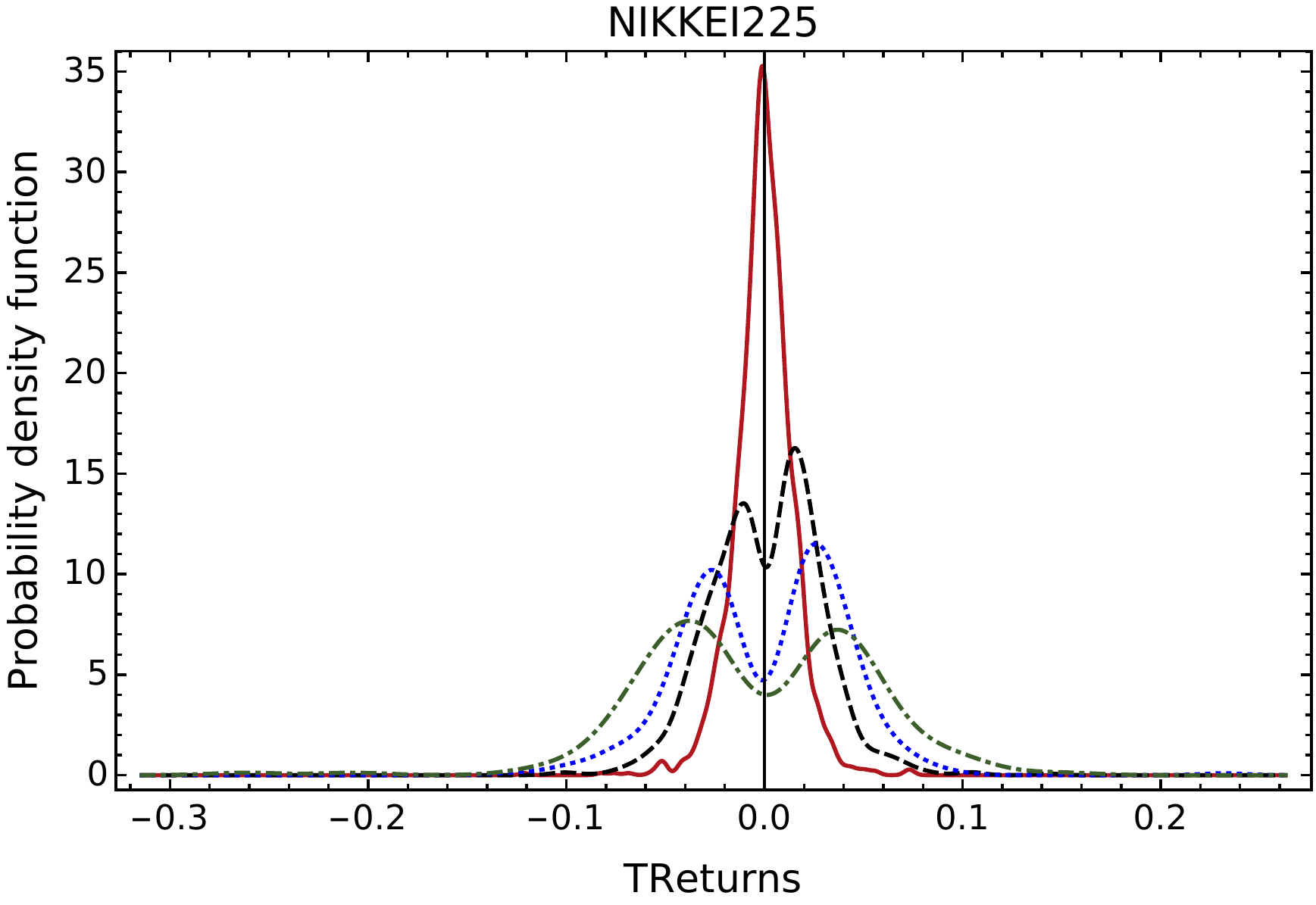}
    \end{subfigure}
    \quad
    \begin{subfigure}[b]{0.45\textwidth}
        \centering 
        \includegraphics[scale=0.3]{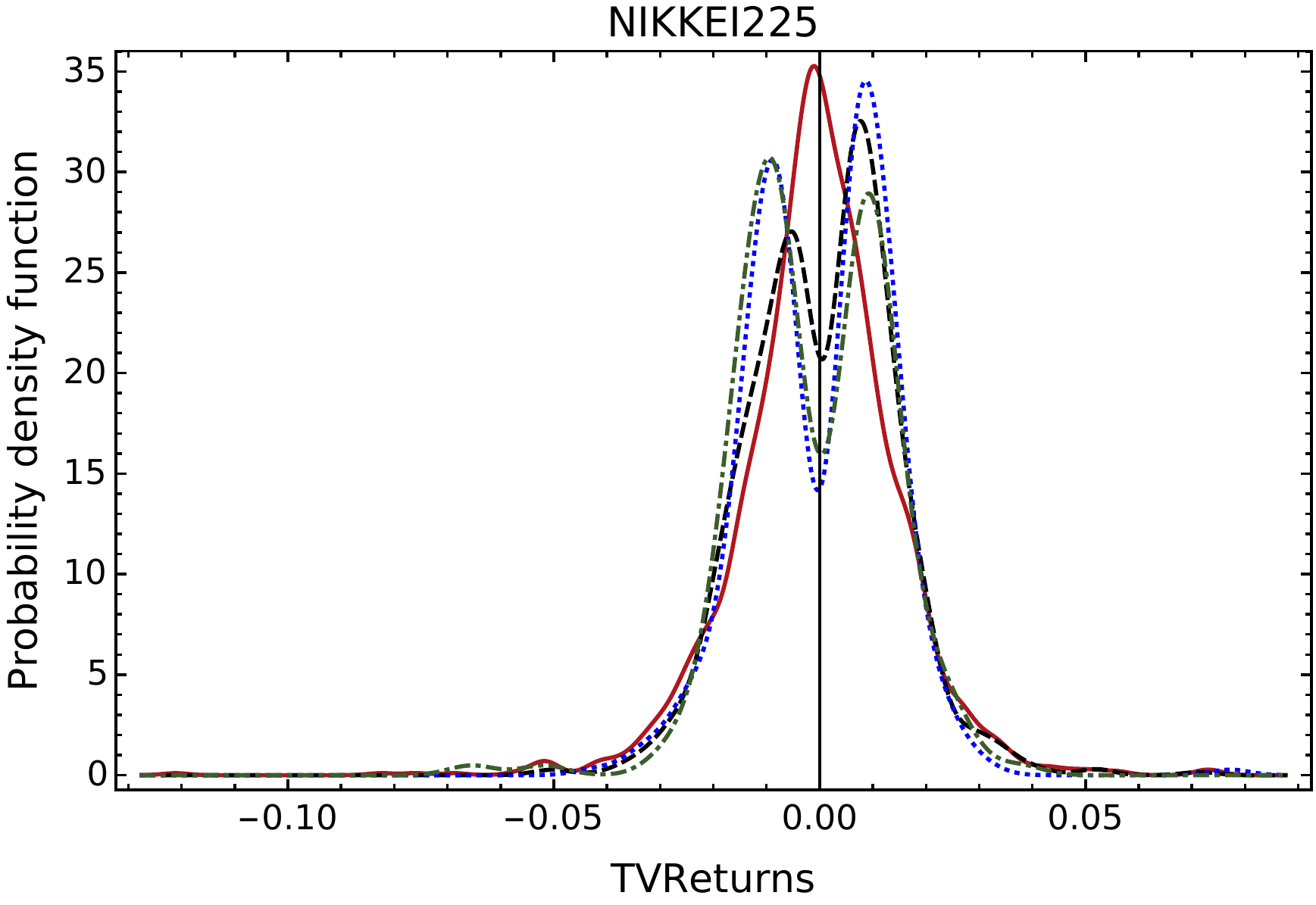}
    \end{subfigure}
    
    \caption{Bi-Modal behavior of TReturns (left column) and TVReturns (right column). We show only variations from trends not longer than 4 days.}
    \label{fig:TRetsMulTiTrends}
\end{figure}

It is interesting to observe in the figure \ref{fig:TRandTVRdistd} that TVReturns for the four samples have a bi-modal distribution. Even more, bi-modality is also observed in the distributions of TReturns obtained from trends with duration longer than one day as is shown in all left sub-figures of figure \ref{fig:TRetsMulTiTrends}, where for clarity, we have plotted TReturns and TVReturns distributions only with duration from one to four days.

The bi-modality of TReturns from trends longer than one day and TVReturns for all trends durations is explained as follows:

TReturns for a trend duration equal to one day, are a sub-sample of the uni-modal, daily returns distribution of any of our data samples; the TReturns distribution becomes bi-modal for trends with duration bigger than one day as  shown in all left sub-figures of figure \ref{fig:TRetsMulTiTrends}, because as trend duration increases, the magnitude of TReturns tends to increase and the probability of observing small positive and negative fluctuations around zero becomes smaller, giving place to the bi-modal shape of TReturns distributions for trends duration bigger than one day. See left sub-figures of figure \ref{fig:TRetsMulTiTrends}.

For the case of TVReturns, as shown in figure \ref{fig:TRandTVRdistd},the bi-modal effect is clearly observed at the overall of TVReturns distributions. The reason of this is because the definition of TVReturns implies the division or TReturns by their corresponding duration in days, this operation narrows the obtained TVReturns distribution with trend duration bigger than one day, with respect to the distribution of TReturns, giving more entries closer to zero, without canceling the bi-modal behavior of TReturns and increasing the amplitude of peaks for trends durations bigger than one day. Finally, all right sub-figures on figure \ref{fig:TRetsMulTiTrends} show that the subsample of TVReturns for trends of length equal to one day, i.e. uni-modal daily returns, unlike the TReturns case, has not enough statistics to mask the bi-modal distribution of TVReturns for trends with duration bigger than one day, as is evidenced again in figure \ref{fig:TRandTVRdistd}.

It is important to point out that in the case of the bi-modal distributions of figure \ref{fig:TRandTVRdistd}, our symmetry test results in the non rejection of the assumption of symmetry, due to the fact that, even though the TVReturns series shows two modes with slightly different heights, the difference was not large enough for rejecting the null hypothesis, according to the distribution of the test statistic $T_n$. Additional studies have to be performed with much larger data samples to examine in detail this behavior. To have an idea of the difference in peaks amplitude of TVReturns samples, table \ref{tab:heights} indicates their higher to lower peaks ratio.

\begin{table}[h!tb]
	\begin{center}
		\setlength\tabcolsep{3pt}
		{\renewcommand{\arraystretch}{1.2}
		\begin{tabular}{|c|c|c|}
			\hline
			Name& Height ratio\\
			\hline
			\hline
			DJIA & 1.0246\\
			\hline
			IPC &1.0233\\
			\hline
			DAX &1.0235\\
			\hline
			Nikkei& 1.0878\\
			\hline
			\hline
		\end{tabular}
		}
	\caption[]{\small Ratio between higher to lower peaks for TVReturns samples. See figure \ref{fig:TRandTVRdistd}.}
	\label{tab:heights}
	\end{center}
\end{table}

\subsubsection{Justification of our selection of observables}
\label{Observables}
Since in this paper we are interested in the symmetry of market indices variations, in our opinion, our selection of observables TReturns and TVReturns is more interesting and realistic for this goal than using only standard daily returns for the following reasons:

\begin{enumerate}

\item Sign symmetry: By construction the time series of TReturns and TVReturns
alternate in sign, having the same number of negative and non-negative
terms, or differing in the number of common signed entries only by one unit
if the time series have an odd number of terms. Entries of the TReturns series are
obtained by aggregating consecutive same signed variations, spreading its
distribution in relation to the Returns distribution, as may be appreciated in
figure 2 and confirmed from RMS values from Table 3. We expect this
aggregation process to make the observable more sensitive to symmetry
fluctuations by unmasking and emphasizing the effect on symmetry
of large fluctuations composed by smaller ones. In fact small consecutive
same signed variations in real life can cause large financial gains or
losses.
The TVReturns, obtained by the same process of aggregating consecutive common signed
variations, become smaller and spread less than the TReturns and Returns  due to the effect of dividing
the latter by their corresponding durations. See again figure \ref{fig:TRandTVRdist} and the corresponding RMS  values from table \ref{tab:heights}.

\item Time multi-scale observables: The time series of TReturns and TVReturns
involve in their construction different and non arbitrary time scales determined by trends duration. This makes our analysis more general, allowing us to explore a more realistic situation and giving us the opportunity to construct new indicators and observables based on this, for example market volatility from TReturns or TVReturns.
\end{enumerate}

\section{Data analyses}
\label{sec:analisis}
\begin{figure}[h!tb]
	\centering
        \begin{subfigure}[b]{0.45\textwidth}
            \centering
            \includegraphics[width=\textwidth]{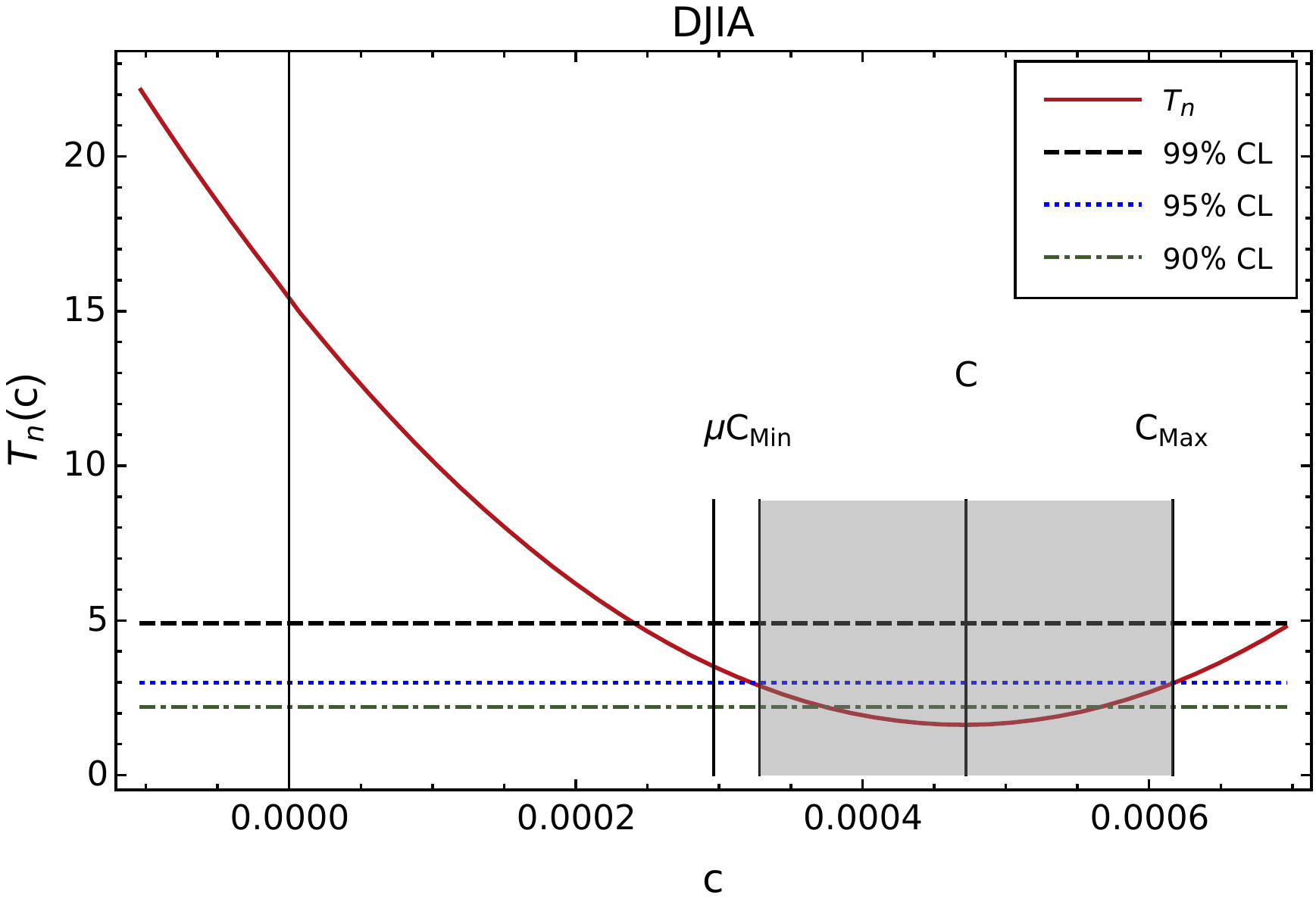}
            \caption[ ]%
            {{\small DJIA (Asymmetric around zero)}}
            \label{fig:RetsFigsDJIA}
        \end{subfigure}
        \quad
        \begin{subfigure}[b]{0.45\textwidth}
            \centering 
            \includegraphics[width=\textwidth]{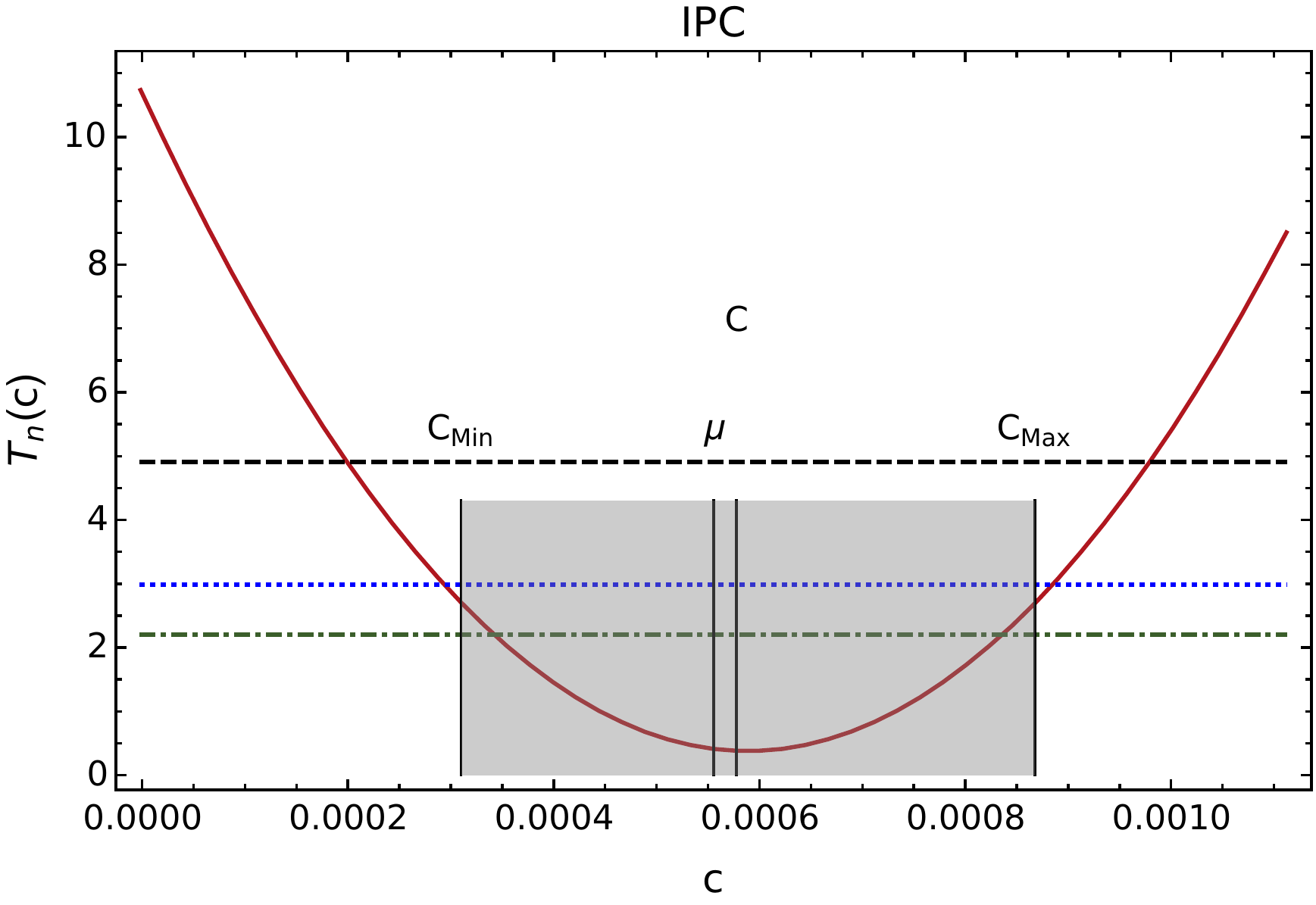}
            \caption[ ]%
            {{\small IPC (Asymmetric around zero)}}
            \label{fig:RetsFigsIPC}
        \end{subfigure}
        \begin{subfigure}[b]{0.45\textwidth}
            \centering 
            \includegraphics[width=\textwidth]{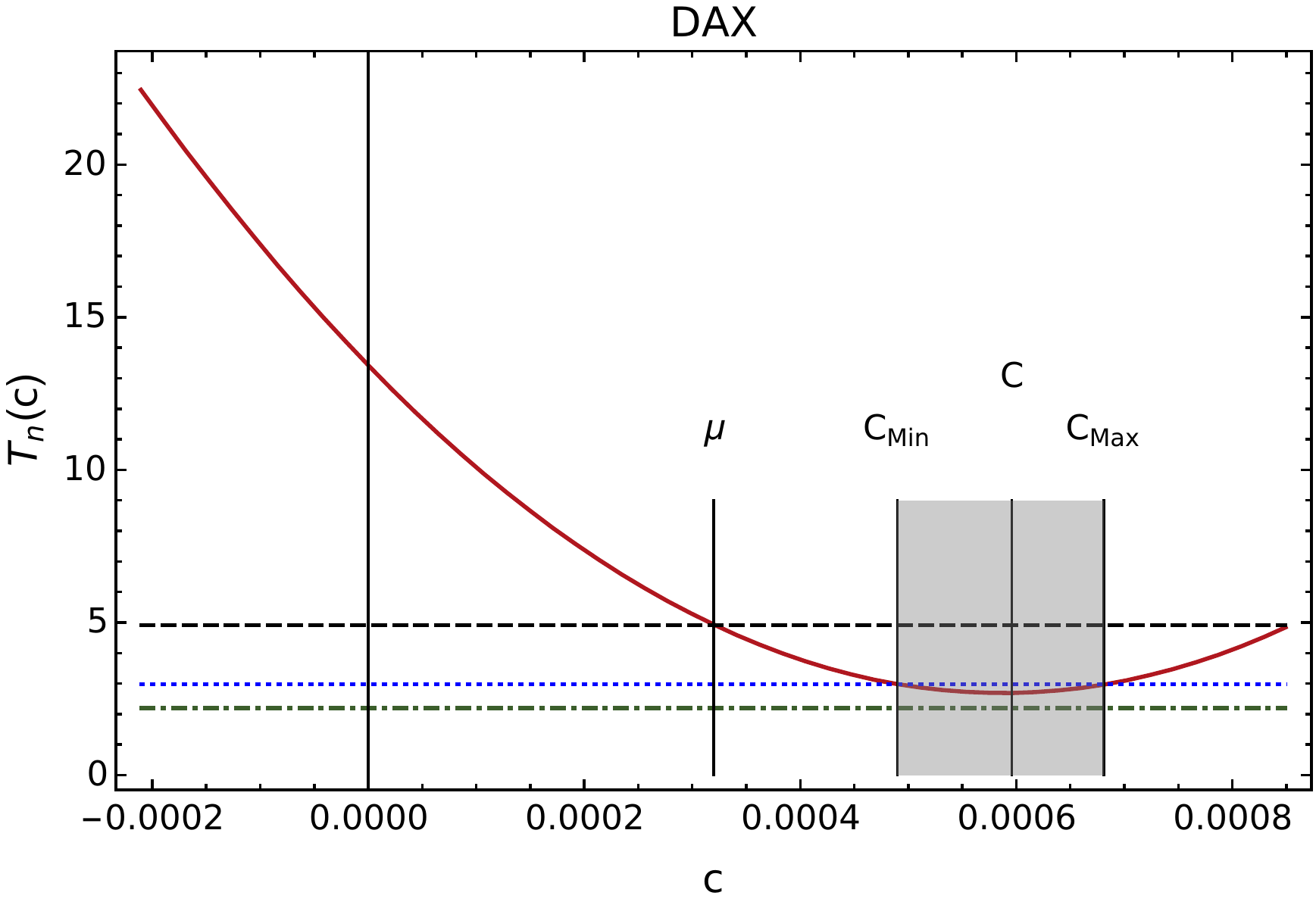}
            \caption[]%
            {{\small DAX (Asymmetric around zero)}}
            \label{fig:RetsFigsDAX}
        \end{subfigure}
        \quad
        \begin{subfigure}[b]{0.45\textwidth}
            \centering 
            \includegraphics[width=\textwidth]{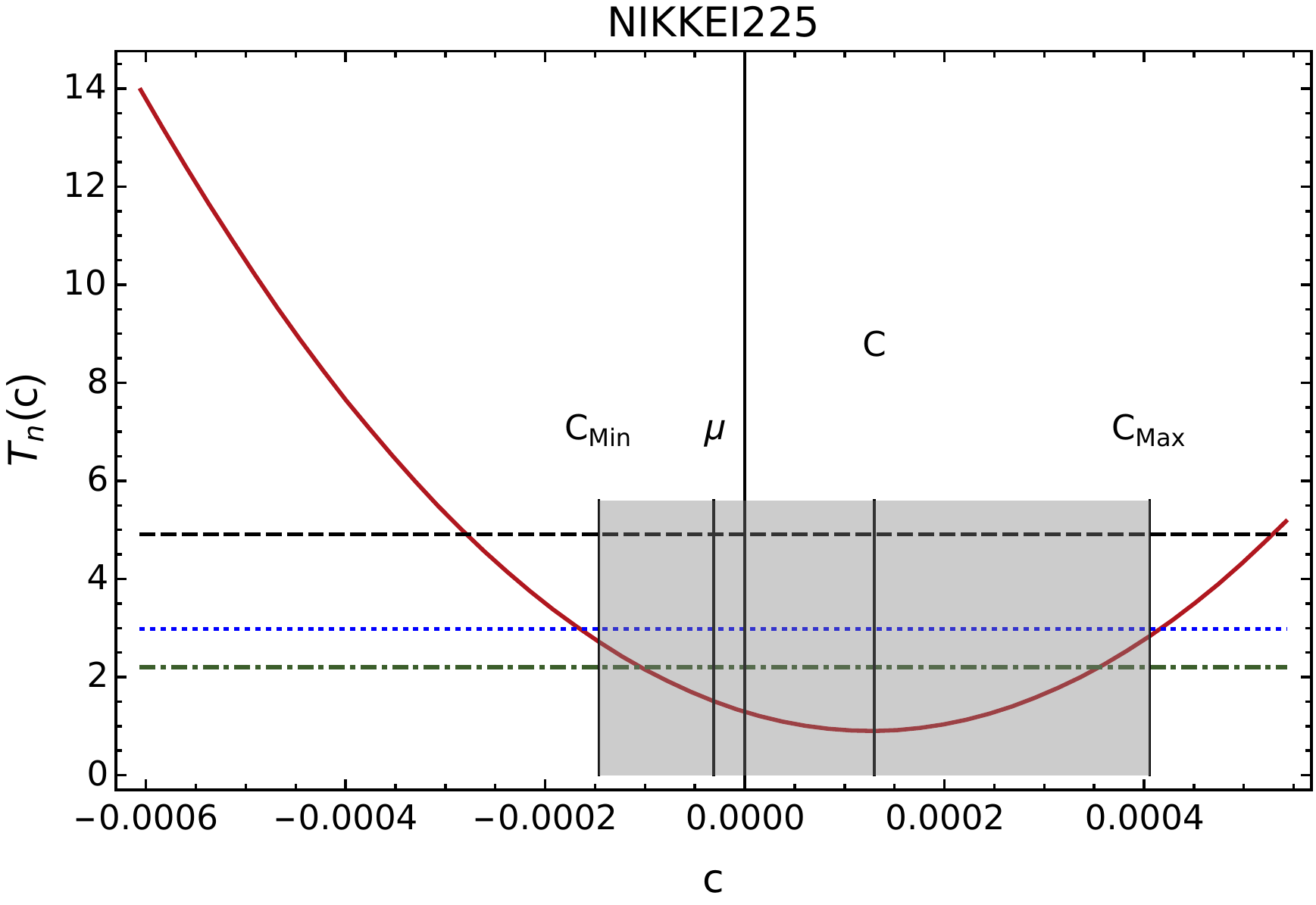}
            \caption[]%
            {{\small Nikkei (Symmetric around zero)}}
            \label{fig:RetsFigsNikkei}
        \end{subfigure}
        \caption[Plots of statistic $T_{n}(c)$ versus selected values of the symmetry point $c$ for daily simple logarithmic returns of our four analyzed markets]
        {\small Returns statistic $T_{n}(c)$ versus selected values of the symmetry point $c$. Horizontal straight lines correspond to the 99, 95 and 90 upper percentage points, as indicated. Returns mean value $\mu$, the origin, and the most plausible symmetry point $C$ are indicated by vertical lines. Gray rectangles determine the interval $(C_{min},C_{max})$ for $\alpha=0.05$.} 
	\label{fig:RetsFigs}
\end{figure}

\begin{figure}[h!tb]
        \centering
        \begin{subfigure}[b]{0.45\textwidth}
            \centering
            \includegraphics[width=\textwidth]{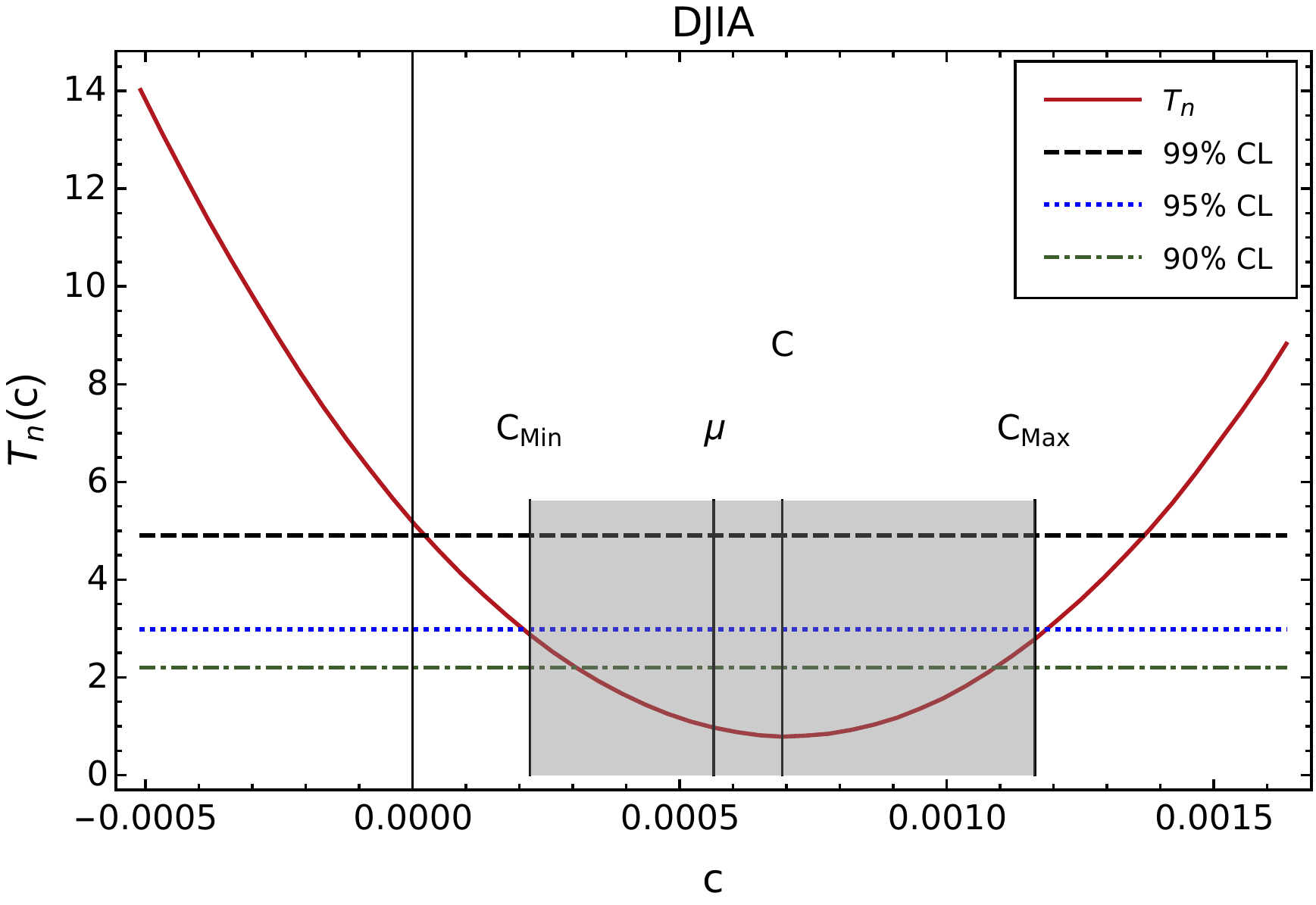}
            \caption[ ]%
            {{\small DJIA (Asymmetric around zero)}}
            \label{fig:TRetsFigsDJIA}
        \end{subfigure}
        \quad
        \begin{subfigure}[b]{0.45\textwidth}
            \centering 
            \includegraphics[width=\textwidth]{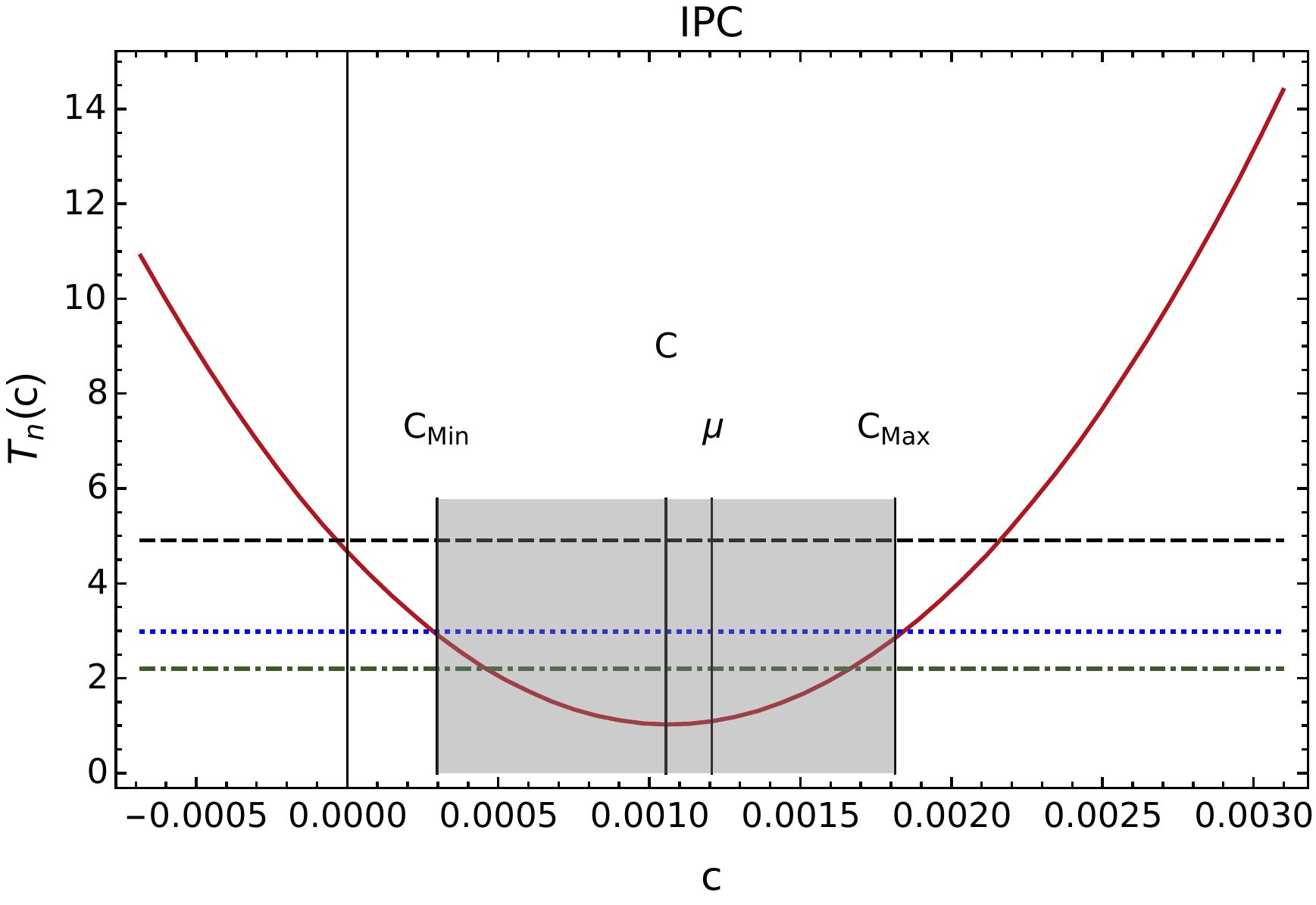}
            \caption[ ]%
            {{\small IPC (Asymmetric around zero)}}
            \label{fig:TRetsFigsIPC}
        \end{subfigure}
        \begin{subfigure}[b]{0.45\textwidth}
            \centering 
            \includegraphics[width=\textwidth]{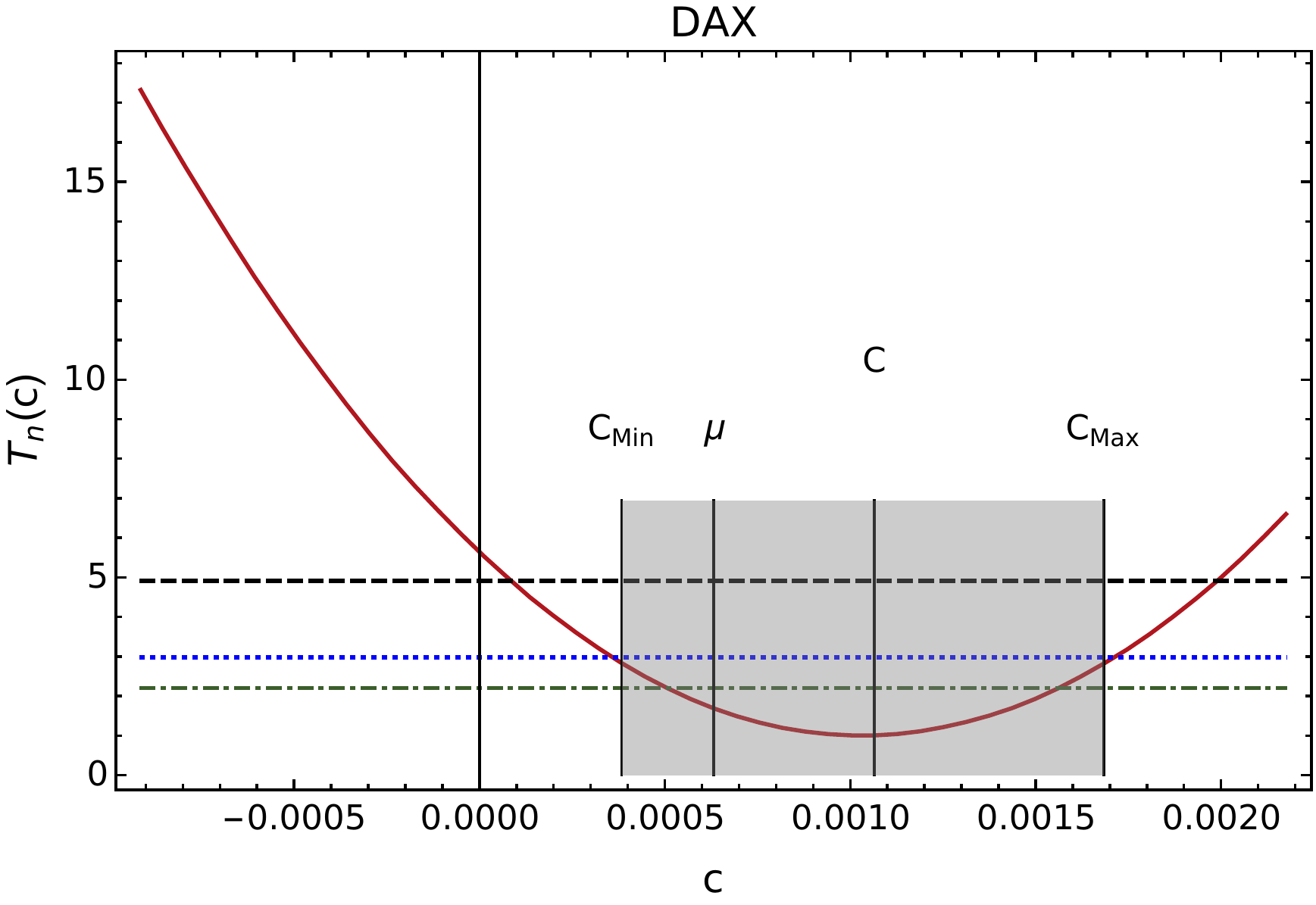}
            \caption[]%
            {{\small DAX (Asymmetric around zero)}}
            \label{fig:TRetsFigsDAX}
        \end{subfigure}
        \quad
        \begin{subfigure}[b]{0.45\textwidth}
            \centering 
            \includegraphics[width=\textwidth]{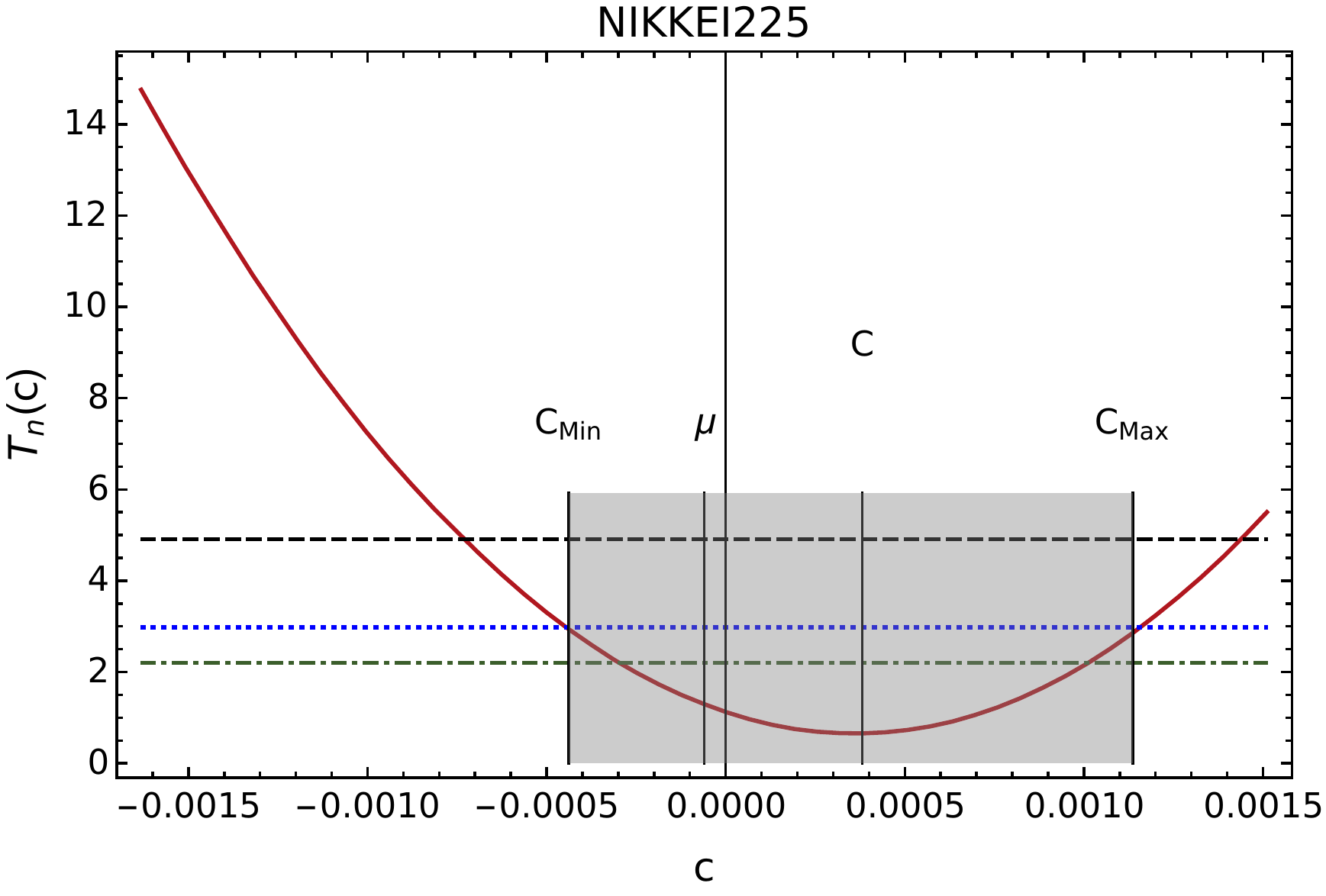}
            \caption[]%
            {{\small Nikkei (Symmetric around zero)}}
            \label{fig:TRetsFigsNikkei}
        \end{subfigure}
        \caption[Plots of statistic $T_{n}(c)$ versus selected values of the symmetry point $c$ for our four different markets TReturns series data]
        {\small TReturns Statistic $T_{n}(c)$ versus selected values of the symmetry point $c$. Again, horizontal straight lines indicate the 99, 95 and 90 upper percentage points, TReturns mean $\mu$, the origin and $C$ are signaled by vertical lines. Again the symmetry interval is marked in gray for $\alpha=0.05$.} 
        \label{fig:TRetsFigs}
\end{figure}

\begin{figure}[h!tb]
        \centering
        \begin{subfigure}[b]{0.45\textwidth}
            \centering
            \includegraphics[width=\textwidth]{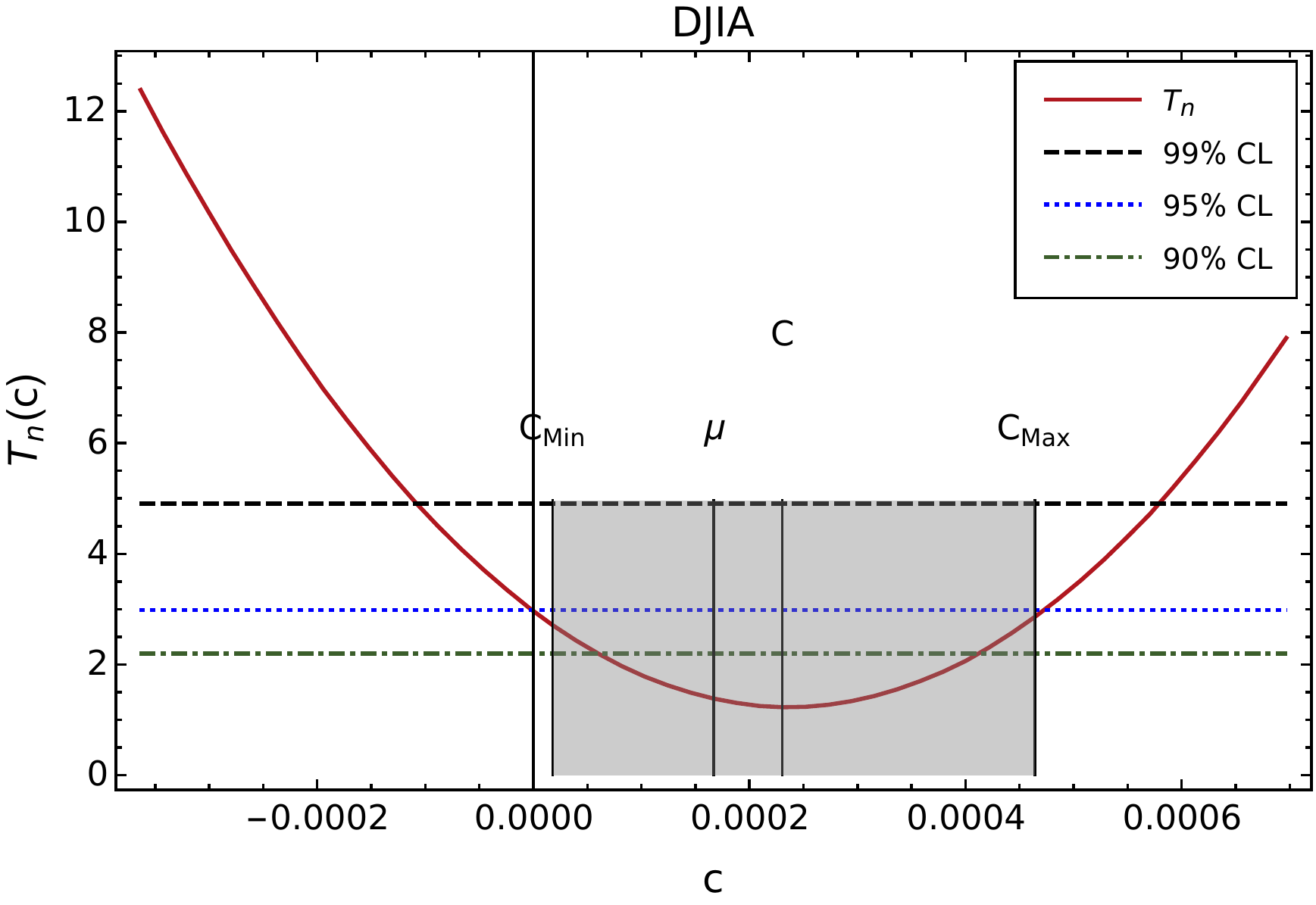}
            \caption[ ]%
            {{\small DJIA (Asymmetric around zero)}}
            \label{fig:TVRetsFigsDJIA}
        \end{subfigure}
        \quad
        \begin{subfigure}[b]{0.45\textwidth}
            \centering 
            \includegraphics[width=\textwidth]{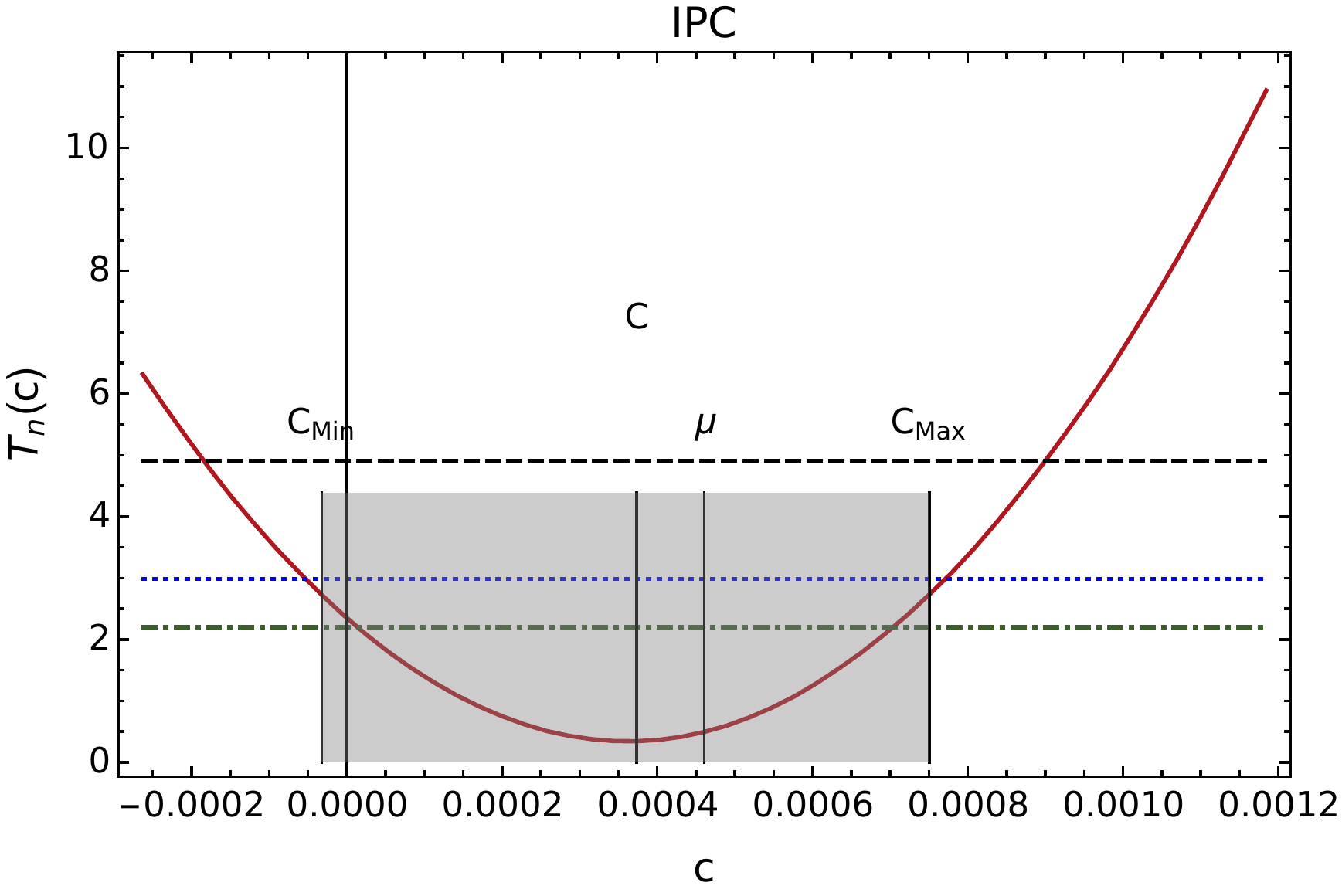}
            \caption[ ]%
            {{\small IPC (Symmetric around zero)}}
            \label{fig:TVRetsFigsIPC}
        \end{subfigure}
        \begin{subfigure}[b]{0.45\textwidth}
            \centering 
            \includegraphics[width=\textwidth]{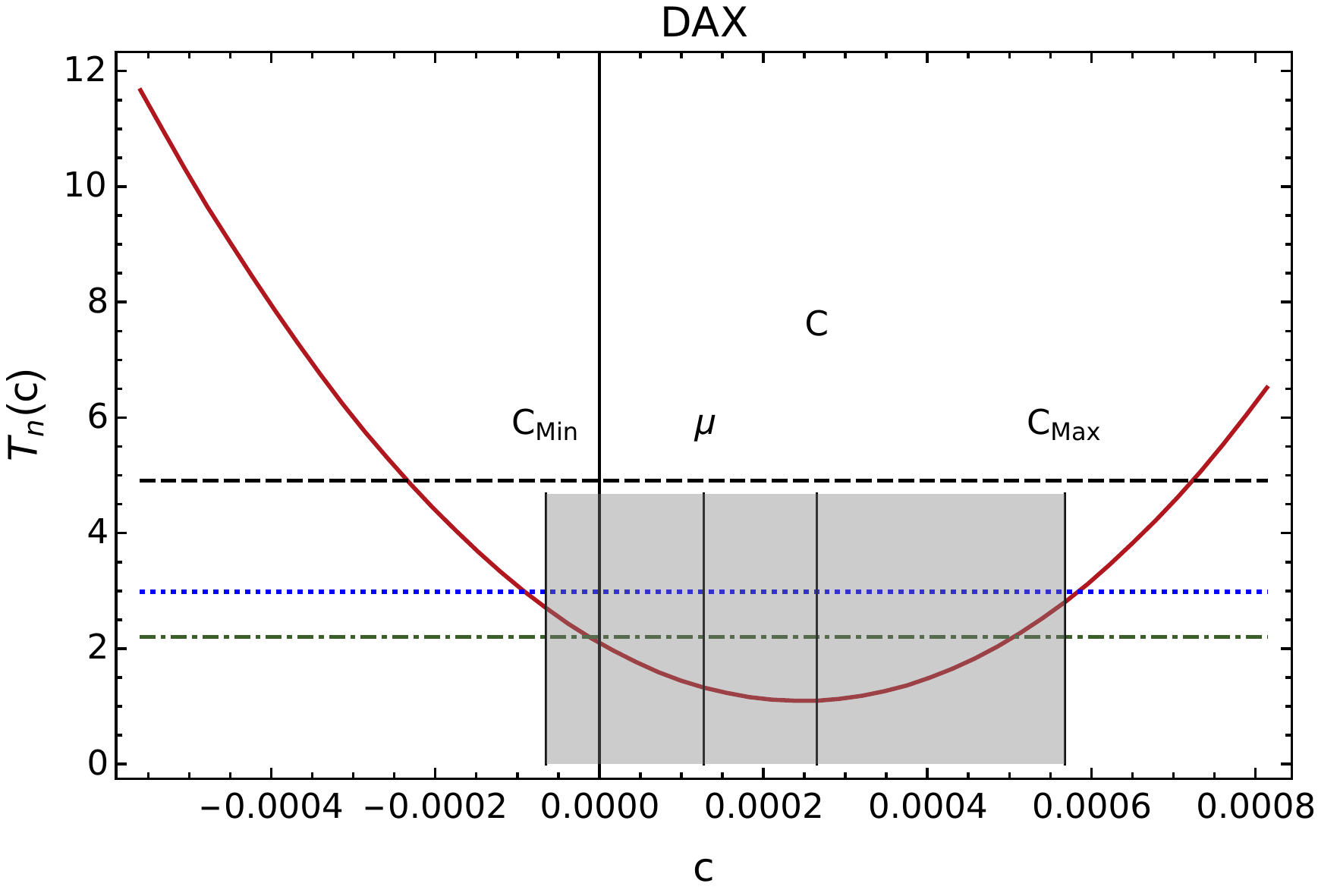}
            \caption[]%
            {{\small DAX (Asymmetric around zero)}}
            \label{fig:TVRetsFigsDAX}
        \end{subfigure}
        \quad
        \begin{subfigure}[b]{0.45\textwidth}
            \centering 
            \includegraphics[width=\textwidth]{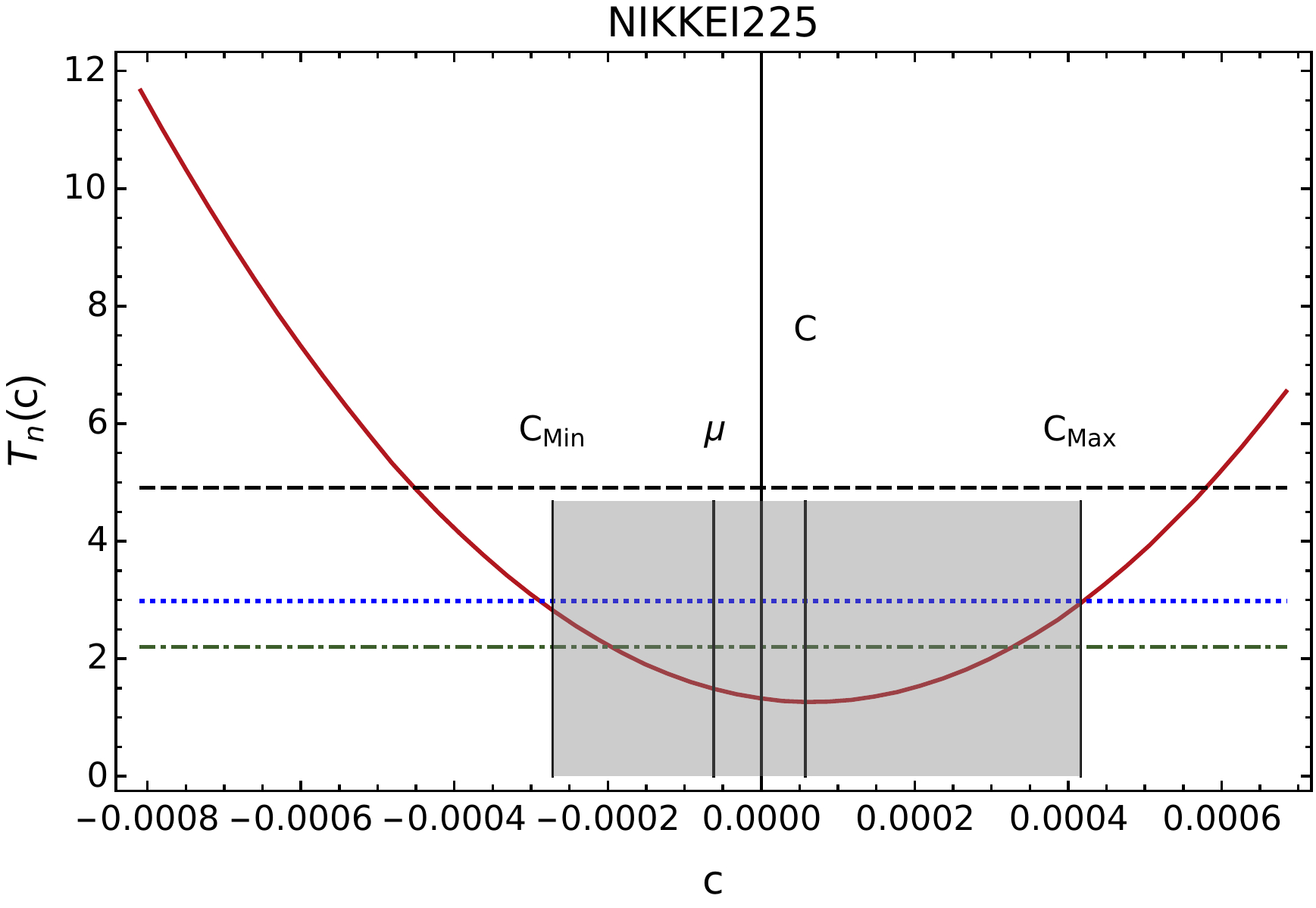}
            \caption[]%
            {{\small Nikkei (Symmetric around zero)}}
            \label{fig:TVRetsFigsNikkei}
        \end{subfigure}
        \caption[TVReturns statistic $T_{n}(c)$ versus selected values of the symmetry point $c$ ]
        {\small TVReturns statistic $T_{n}(c)$ versus selected values of the symmetry point $c$. As before, horizontal straight lines correspond to the 99, 95 and 90 upper percentage points. Mean value $\mu$, $C$ and the origin are again shown by vertical lines. As two previous plots $(C_{min},C_{max})$ for $\alpha=0.05$ is emphasized in gray. } 
        \label{fig:TVRetsFigs}
\end{figure}

Applying the statistical methodology summarized in subsection \ref{Meth}, the interval of symmetry $(C_{min},C_{max})$, i.e. the {\em plausible} set of values of the unknown symmetry points $c$, for different values of significance level $\alpha$ can be seen in the plots of $T_{n}(c)$ versus $c$ for our multi-scale returns samples in figures \ref{fig:RetsFigs} to \ref{fig:TVRetsFigs}.

As mentioned in subsection \ref{Meth}, the symmetry interval $(C_{min},C_{max})$ is determined by the intersection of the curve of $T_{n}(c)$ and the horizontal line of significance level calculated in table \ref{tab:Cpoints} in an interval that contains the set of all possible values of $c$ which would not lead to the rejection of the null hypothesis of symmetry for the probability distribution of the analysed random variable. This interval is shown in gray for our different multi-scale returns data, in figures \ref{fig:RetsFigs}, \ref{fig:TRetsFigs} and \ref{fig:TVRetsFigs}, where the horizontal lines $T_n=4.909$, $T_n=2.983$ and $T_n=2.200$, correspond to the asymptotic 0.99, 0.95 and 0.90 percentiles of the distribution of the $T_{n}$ statistic corresponding to $\alpha=0.01$, $0.05$ and $0.1$ respectively, displayed in table \ref{tab:Cpoints}. For a significance level $\alpha=0.10$ (or lower), in each one of our data samples, it was always possible to find an interval of plausible values for the unknown point of symmetry which would not lead us to the rejection of the assumption of symmetry. The table \ref{tab:confidence} shows all this symmetry intervals for $\alpha=0.05$, where each point belonging to them can be statistically considered as a plausible point around which the distribution of the market variations (Returns, TReturns and TVReturns) is symmetric.

Table \ref{tab:confidence} shows a summary of the information presented visually in figures \ref{fig:RetsFigs}, \ref{fig:TRetsFigs} and \ref{fig:TVRetsFigs}, including observables mean values, symmetry intervals, confidence level and whether or not the daily and multi-scale returns distributions past the symmetry test.

\begin{table}[h!tb]
	\begin{center}
		\setlength\tabcolsep{1.6pt} 
		{\renewcommand{\arraystretch}{1} 
		\begin{tabular}{|c|c|c|c|c|c|c|c|c|}
			\hline
			Market&Observable&$\mu$&Symmetry&$C$&Zero\\ 
			& & &interval& & symmetric\\
			\hline
			\hline
			DJIA&Returns&$2.9\!\times\!10^{-4}$&($3.2\!\times\!10^{-4}$,$6.1\!\times\!10^{-4}$)&$4.7\!\times\!10^{-4}$&No\\
			\hline
			DJIA&TReturns&$5.6\!\times\!10^{-4}$&($2.0\!\times\!10^{-4}$,$1.1\!\times\!10^{-3}$)&$6.9\!\times\!10^{-4}$&No\\
			\hline
			DJIA&TVReturns&$1.6\!\times\!10^{-4}$&($-1.5\!\times\!10^{-7}$,$4.7\!\times\!10^{-4}$)&$2.3\!\times\!10^{-4}$&Yes\\
			\hline
			IPC&Returns&$5.5\!\times\!10^{-4}$&($2.9\!\times\!10^{-4}$,$8.8\!\times\!1.0^{-4}$)&$5.9\!\times\!10^{-4}$&No\\
			\hline
			IPC&TReturns&$1.2\!\times\!10^{-3}$&($2.8\!\times\!10^{-4}$,$1.8\!\times\!10^{-3}$)&$1.0\!\times\!10^{-3}$&No\\
			\hline
			IPC&TVReturns&$4.6\!\times\!10^{-4}$&($-5.3\!\times\!10^{-5}$,$7.6\!\times\!10^{-4}$)&$3.6\!\times\!10^{-4}$&Yes\\
			\hline
			DAX&Returns&$3.1\!\times\!10^{-4}$&($4.8\!\times\!10^{-4}$,$6.8\!\times\!10^{-4}$)&$5.8\!\times\!10^{-4}$&No\\
			\hline
			DAX&TReturns&$6.3\!\times\!10^{-4}$&($3.6\!\times\!10^{-4}$,$1.7\!\times\!10^{-3}$)&$1.0\!\times\!10^{-3}$&No\\
			\hline
			DAX&TVReturns&$1.2\!\times\!10^{-4}$&($-8.8\!\times\!10^{-5}$,$5.8\!\times\!10^{-4}$)&$2.4\!\times\!10^{-4}$&Yes\\
			\hline
			Nikkei&Returns&$-3.1\!\times\!10^{-5}$&($-1.6\!\times\!10^{-4}$,$4.1\!\times\!10^{-4}$)&$1.2\!\times\!10^{-4}$&Yes\\
			\hline
			Nikkei&TReturns&$-5.9\!\times\!10^{-5}$&($-4.3\!\times\!10^{-4}$,$1.1\!\times\!10^{-3}$)&$3.6\!\times\!10^{-4}$&Yes\\
			\hline
			Nikkei&TVReturns&$-6.2\!\times\!10^{-5}$&($-2.8\!\times\!10^{-4}$,$4.1\!\times\!10^{-4}$)&$6.1 \times 10^{-5}$&Yes\\
			\hline
			\hline
		\end{tabular}
		}
	\caption[]{\small Mean value $\mu$, symmetry interval, most probable symmetry point and result of our symmetry test around zero for all data samples analyzed and our three observables. Measurements were performed with a significance level $\alpha = 0.05$.}
	\label{tab:confidence}
	\end{center}
\end{table}

Again, as mentioned in \cite{Coronel-Montoya} it is important to have in mind that our statistical approach is different from that of maximizing a test-statistic, as it is done generally, see for example in \cite{Karsten,Coronel_tesis,Coronel}. Instead, our methodology statistically sustains the idea that whenever there exist one or many plausible values for the points of symmetry, these plausible symmetry points may be found, and if $T_n(c)$ has a minimum value, denoted in this paper by  $C$, this point will be the most plausible symmetry point for a given significance level $\alpha$.

\subsection{Time evolution of $T_n$ and the most plausible symmetry point $C$}
\begin{figure}[htb!]
        \centering
        \begin{subfigure}[b]{0.475\textwidth}
            \centering
            \includegraphics[width=\textwidth]{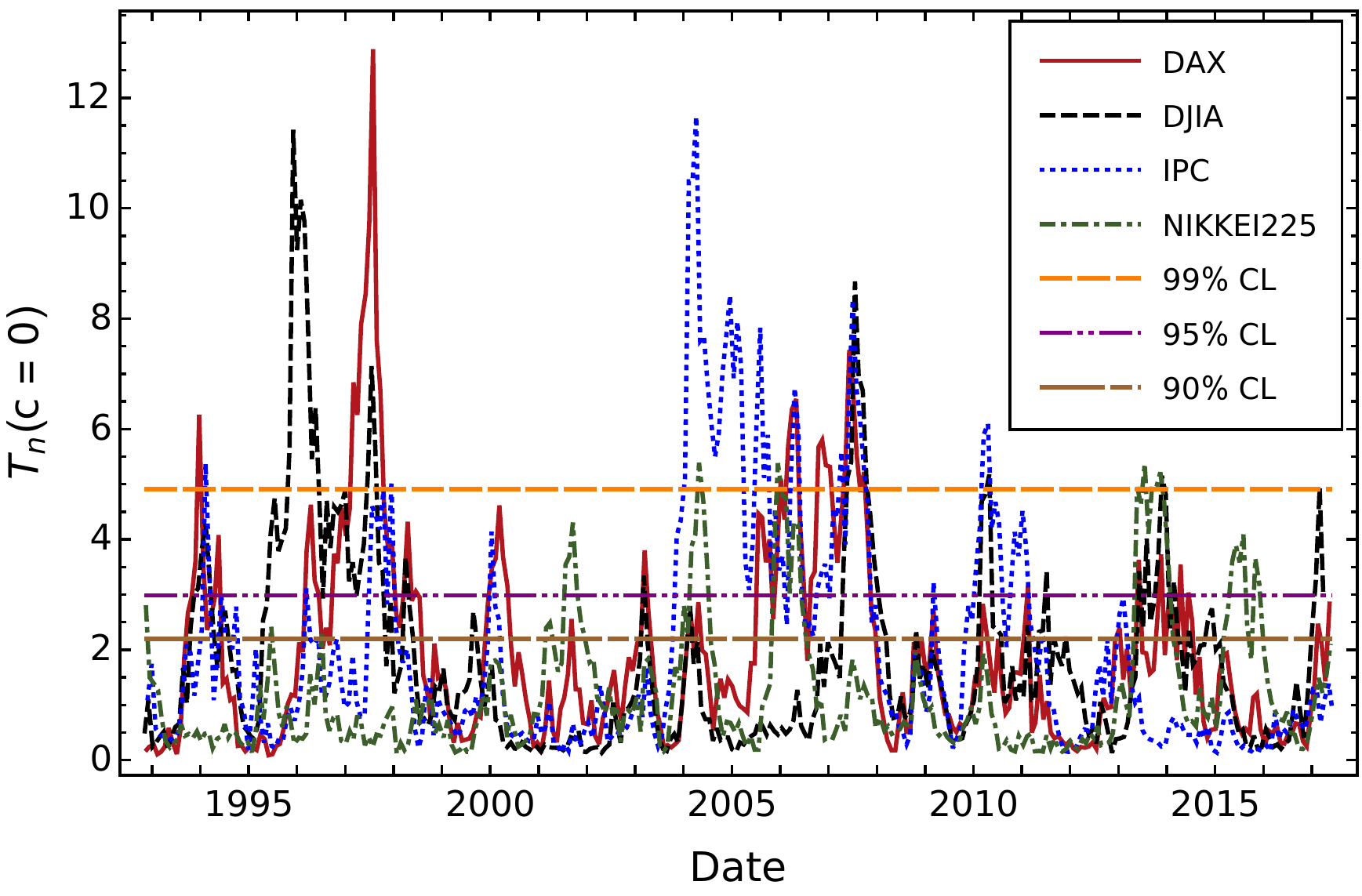}
            \caption[ ]{{\small Returns}}
            \label{fig:ZeroSymmReturns}
        \end{subfigure}
        \hfill
        \begin{subfigure}[b]{0.475\textwidth}
            \centering 
            \includegraphics[width=\textwidth]{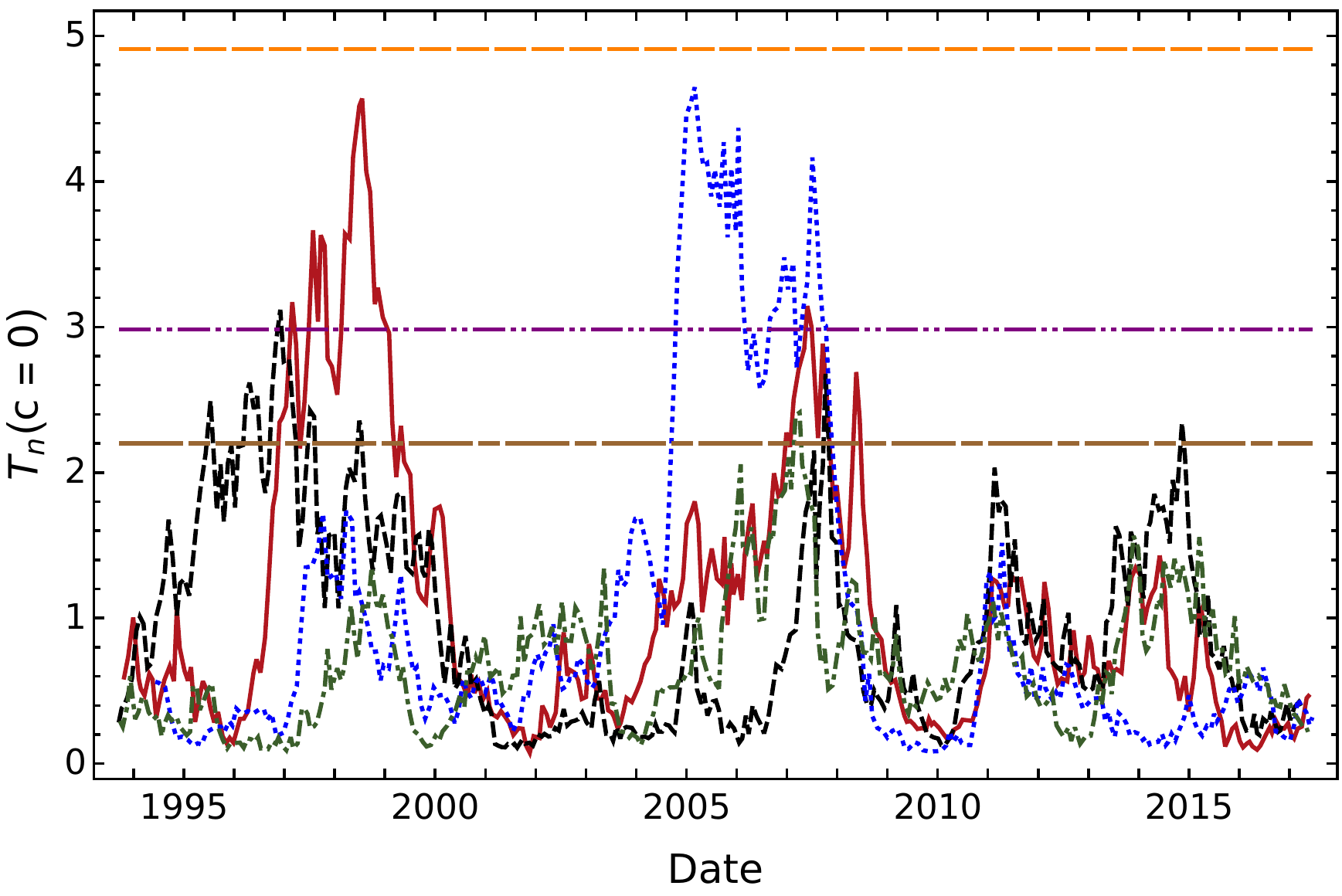}
            \caption[ ]{{\small TReturns }}
            \label{fig:ZeroSymmTReturns}
        \end{subfigure}
        \vskip\baselineskip
        \begin{subfigure}[b]{0.47\textwidth}
            \centering 
            \includegraphics[width=\textwidth]{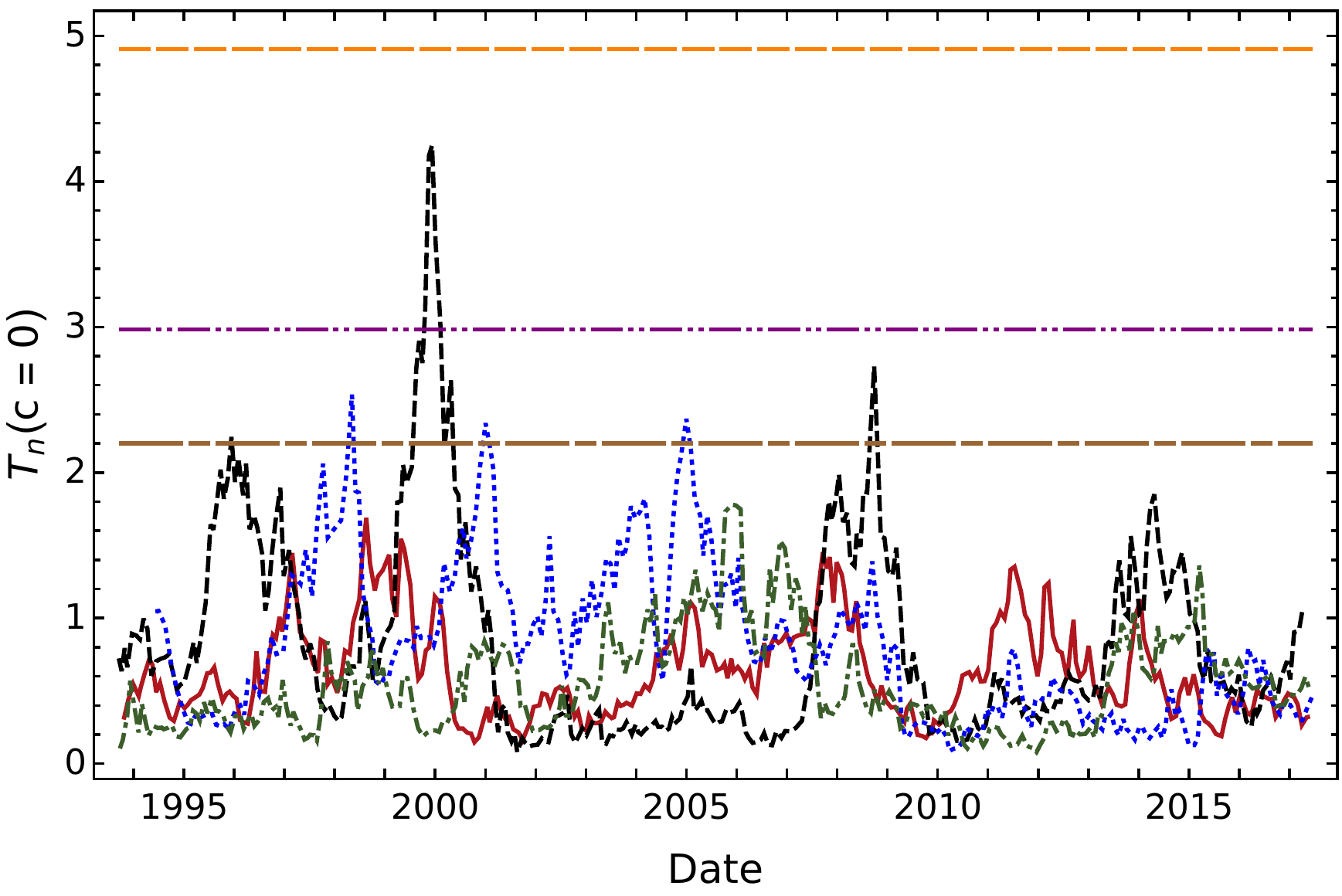}
            \caption[]{{\small TVReturns}}
            \label{fig:ZeroSymmTVReturns}
        \end{subfigure}
        \caption[Plots of statistic $T_{n}(c = 0)$ versus date for our four different markets series data]
        {\small Plots of statistic $T_{n}(c = 0)$ versus date for our four different markets series data. Also in this plot, horizontal straight lines that correspond to the 99, 95 and 90 upper percentage points, are indicated.} 
        \label{fig:ZeroSymm}
\end{figure}

 Until now we have only assessed the overall  symmetry  of our four data samples for three observables. We would like to have a more complete image of symmetry changes on time, i.e. we would like  to see how $T_n$ and the  most plausible symmetry point $C$ found with our method,  evolve over time, at least at a coarse grained level as determined by the chosen observables. To achieve this we use a movable time window for all our samples and observables and perform the statistical test successively. We chose a time window of $252$ trading days since it is the average number of trading days in a year for the selected markets and its resolution is not enough noisy to allow us  appreciate some market days with extreme movements as it will be shown in subsection \ref{extreme}.  Results on the dynamics of of $T_n$ are shown in figure \ref{fig:ZeroSymm}, where these plots show that at the chosen time window  resolution level,  the plausibility of symmetry around $c=0$ is variable, being rejected at different periods of time for the different significance levels. We will show explicitly the dynamics of $C$ and its confidence interval next.

The corresponding coarse grained images of the time evolution of $C$ for all  data samples and observables were obtained. Figure \ref{fig:CEvolution} displays the desired results. In \ref{appendixa} we show these plots separately including its confidence interval evolution. 

\begin{figure}[htb!]
        \centering
        \begin{subfigure}[b]{0.475\textwidth}
            \centering
            \includegraphics[width=\textwidth]{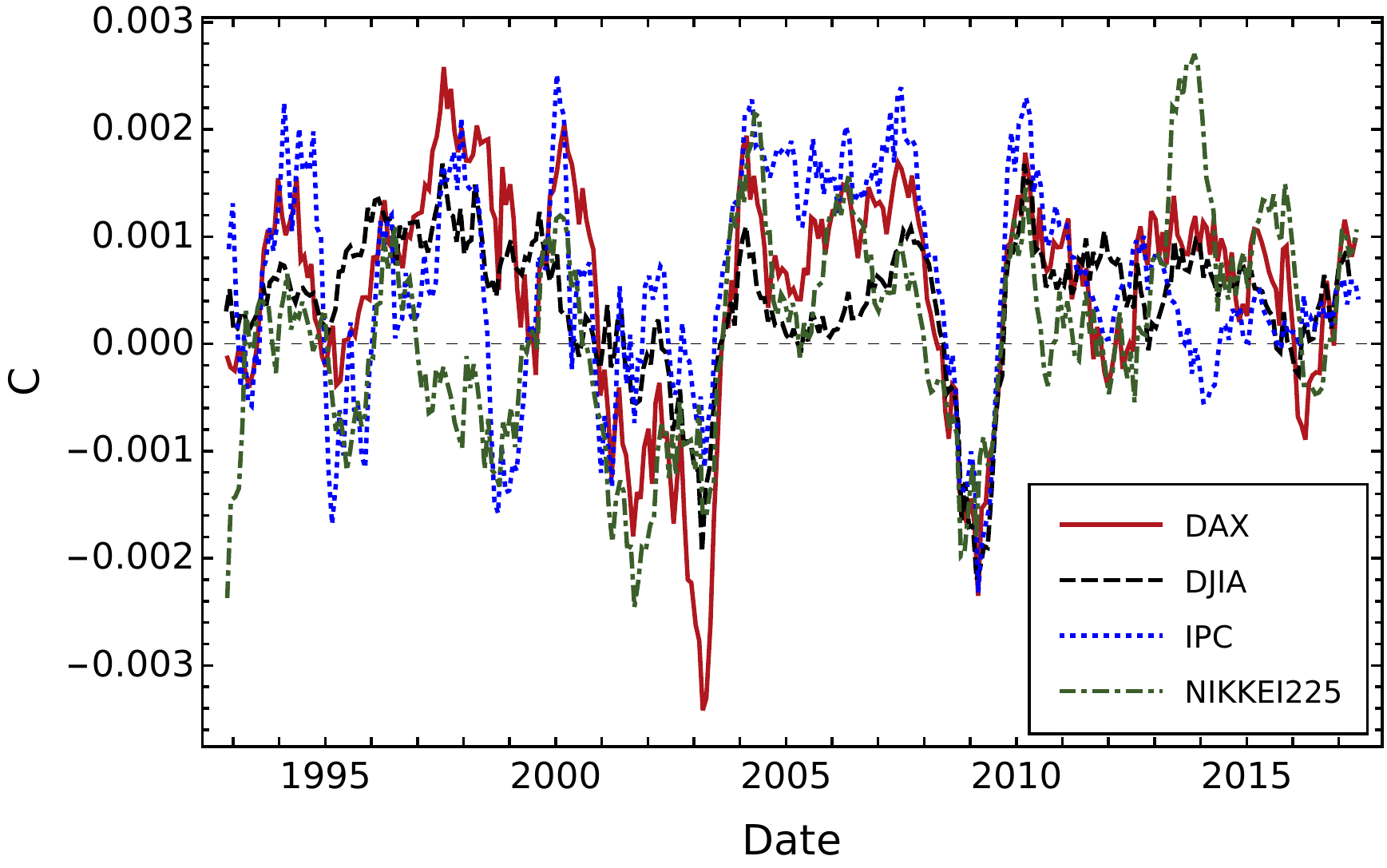}
            \caption[ ]{{Returns}}
            \label{fig:CEvolutionSimpleret}
        \end{subfigure}
        \hfill
        \begin{subfigure}[b]{0.475\textwidth}
            \centering 
            \includegraphics[width=\textwidth]{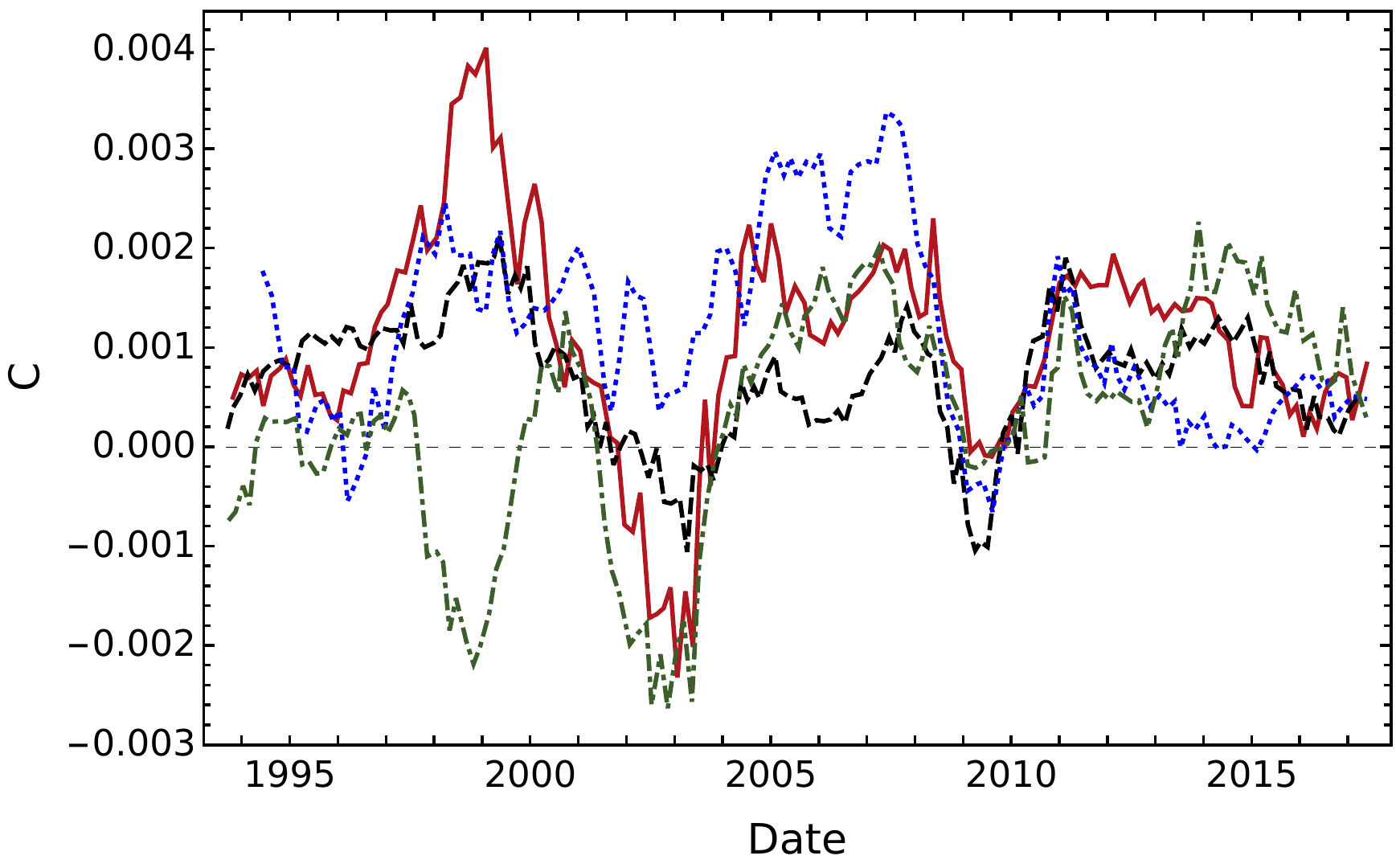}
            \caption[ ]{{TReturns}}
            \label{fig:CEvolutionTrendret}
        \end{subfigure}
        \vskip\baselineskip
        \begin{subfigure}[b]{0.475\textwidth}
            \centering 
            \includegraphics[width=\textwidth]{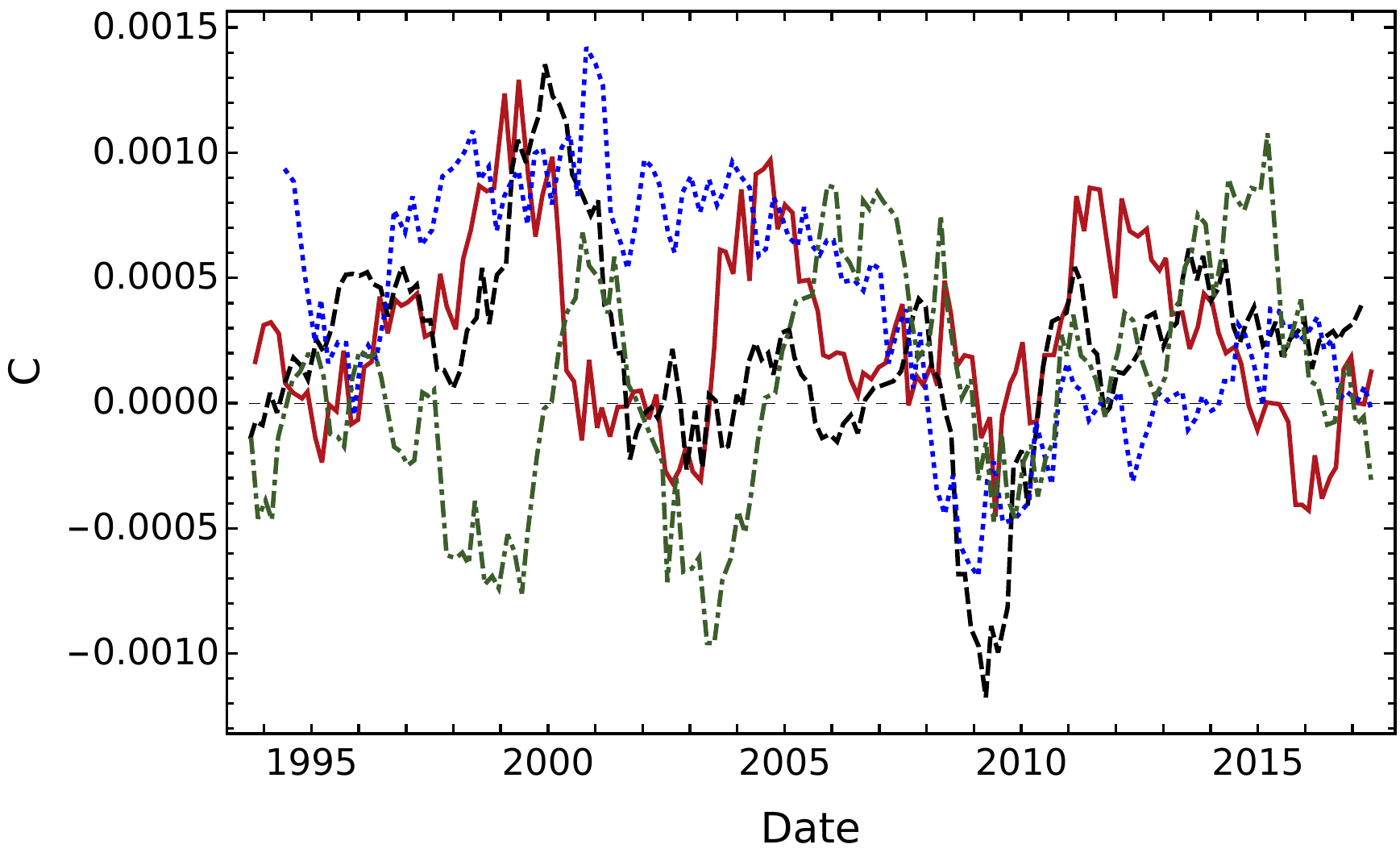}
            \caption[ ]{{TVReturns}}
            \label{fig:CEvolutionTrendvel}
        \end{subfigure}
        \caption[Plots of the most plausible point of symmetry $C$ versus time for our four different markets series data]
        {\small Plots of the most plausible point of symmetry $C$ versus time for the three different observables and four data samples. A 252 trading days time window was used. To appreciate independently each plot, see \ref{appendixa}.} 
        \label{fig:CEvolution}
\end{figure}

\subsection{Behavior of the Symmetry point $C$ around extreme market movements dates}
\label{extreme}
By their construction, since the  observables TReturns and TVReturns are coarse grained, in order to clarify the sensitivity of $C$ on market big movements, we show only plots for daily returns from figure \ref{fig:CEvolutionSimpleret}, this time separately in figure \ref{fig:CvsTRets}, where in the graphs of evolution of $C$ on time, we appreciate clearly the days listed in table \ref{tab:BigMovs} when market experienced extreme events, as well as the dates and duration of important crisis, such as the dotcom bubble and the subprime bear market. Because it is not the goal of this research, we do not compare the reliability of this methodology with other financial indicators to study extreme events or market cycles, but it is in our opinion that it can be applied to this kind of financial analyses. In fact traders use many indicators simultaneously to follow market behavior and construct trading systems. Implications for trading applications of our methodology are clear.

\begin{table}[h!tb]
	\begin{center}
		\setlength\tabcolsep{3pt} 
		{\renewcommand{\arraystretch}{1.2}
		\begin{tabular}{|c|c|c|}
			\hline
			Name&date\\
			\hline
			\hline
			a) Japanese asset price bubble& 1-1-1990\\
			\hline
			b) Tequila Effect& 12-20-1994\\
			\hline
			c) Dotcom bubble&03-10-2000\\
			\hline
			d) Subprime crisis&08-09-2007\\
			\hline
			e) Brexit&06-23-2018\\
			\hline
			\hline
		\end{tabular}
		}
	\caption[]{\small Recent crisis and critical market days chronologically ordered and pointed out in figures \ref{fig:CvsTRets}. }
	\label{tab:BigMovs}
	\end{center}
\end{table}

\subsubsection{Note on the variation ranges of $C$ and $\mu$}
\begin{figure}[h!tb]
        \centering
        \begin{subfigure}[b]{0.45\textwidth}
            \centering
            \includegraphics[width=\textwidth]{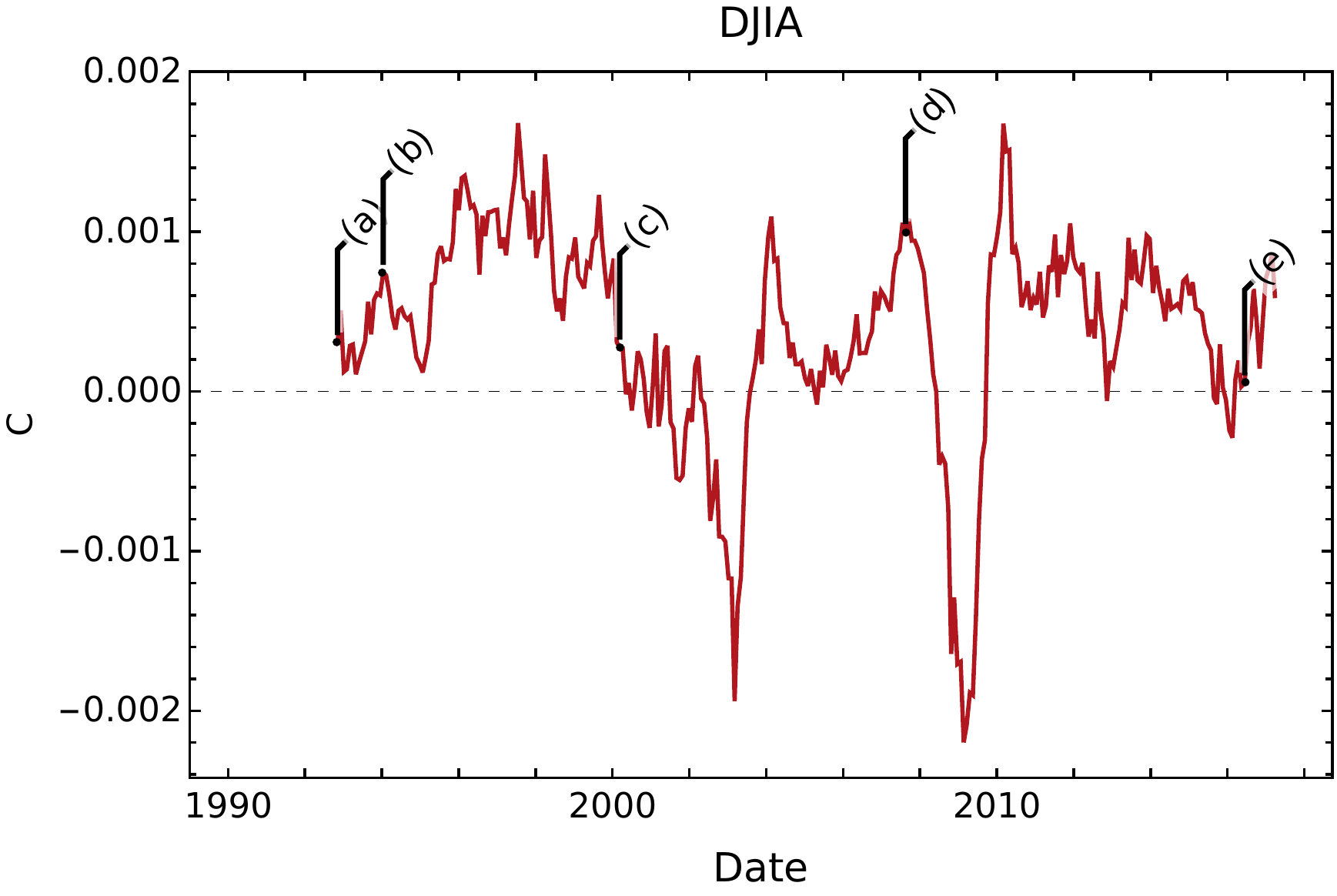}
            \caption[ ]%
            {{\small DJIA: $C$ vs time}}
            \label{fig:TVRetsTimeFigsDJIA}
        \end{subfigure}
        \quad
        \begin{subfigure}[b]{0.45\textwidth}
            \centering 
            \includegraphics[width=\textwidth]{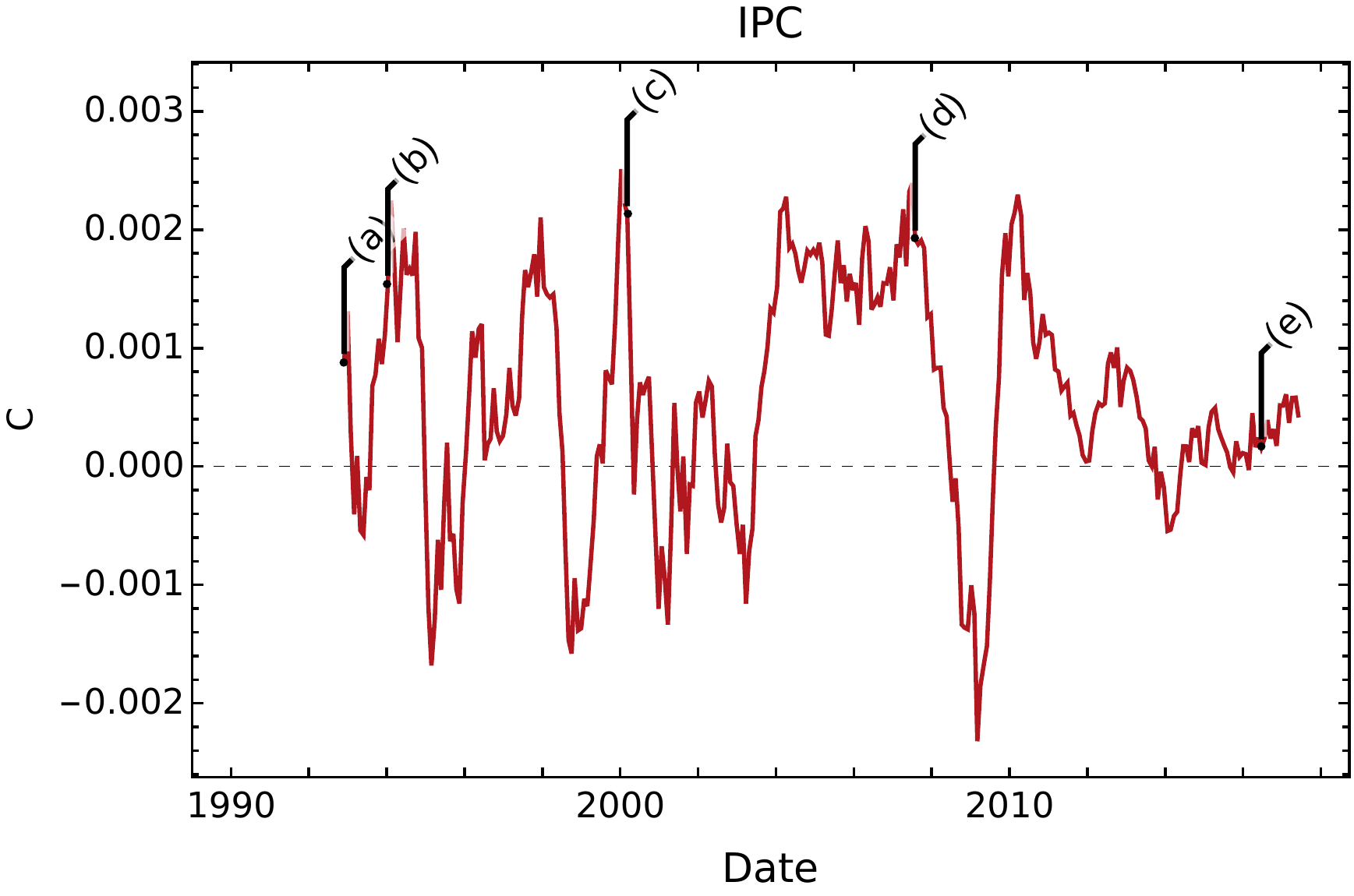}
            \caption[ ]%
            {{\small IPC: $C$ vs time}}
            \label{fig:TVRetsTimeFigsIPC}
        \end{subfigure}
        \vskip\baselineskip
        \begin{subfigure}[b]{0.45\textwidth}
            \centering 
            \includegraphics[width=\textwidth]{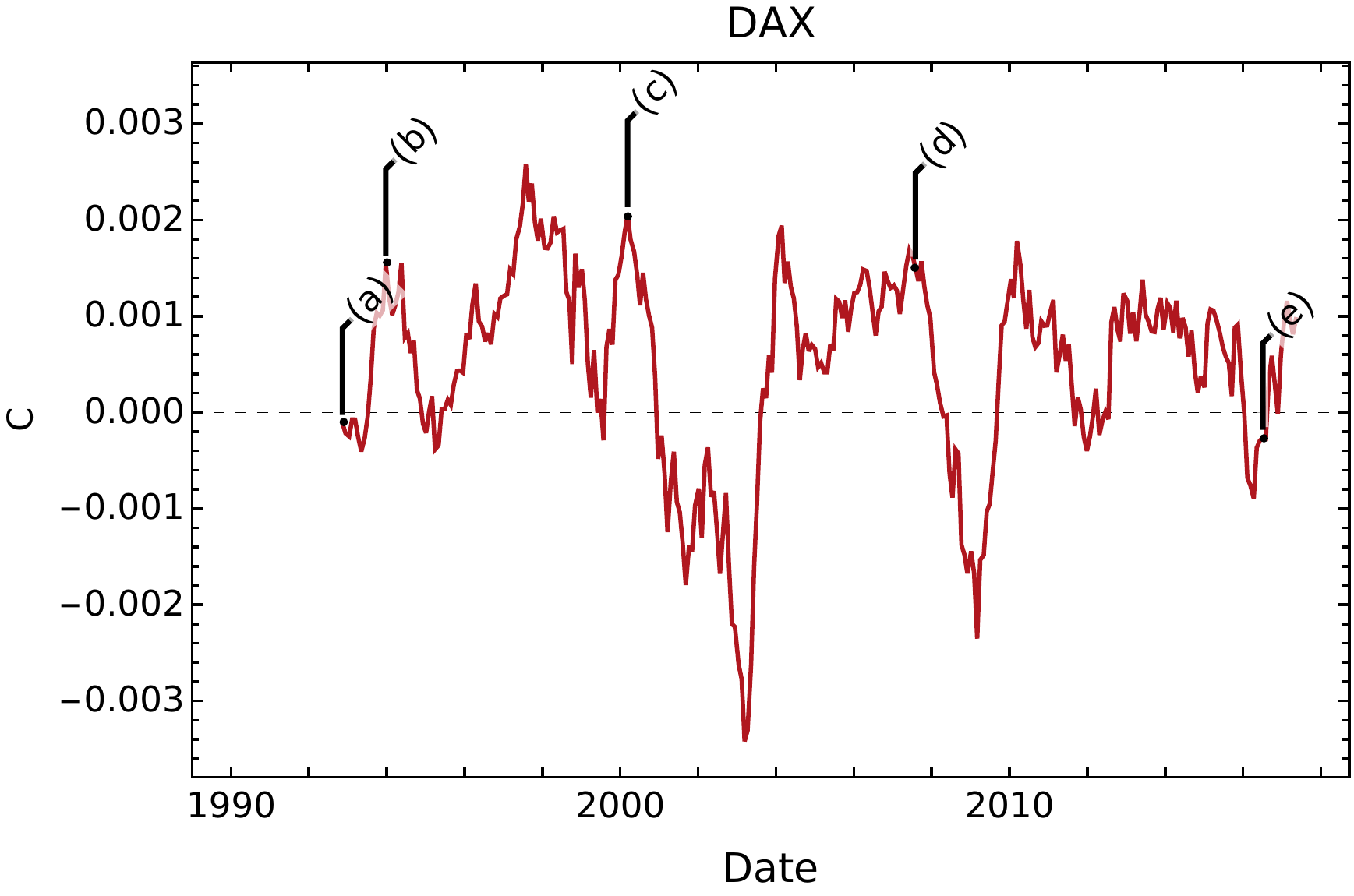}
            \caption[]%
            {{\small DAX: $C$ vs time}}
            \label{fig:TVRetsTimeFigsDAX}
        \end{subfigure}
        \quad
        \begin{subfigure}[b]{0.45\textwidth}
            \centering 
            \includegraphics[width=\textwidth]{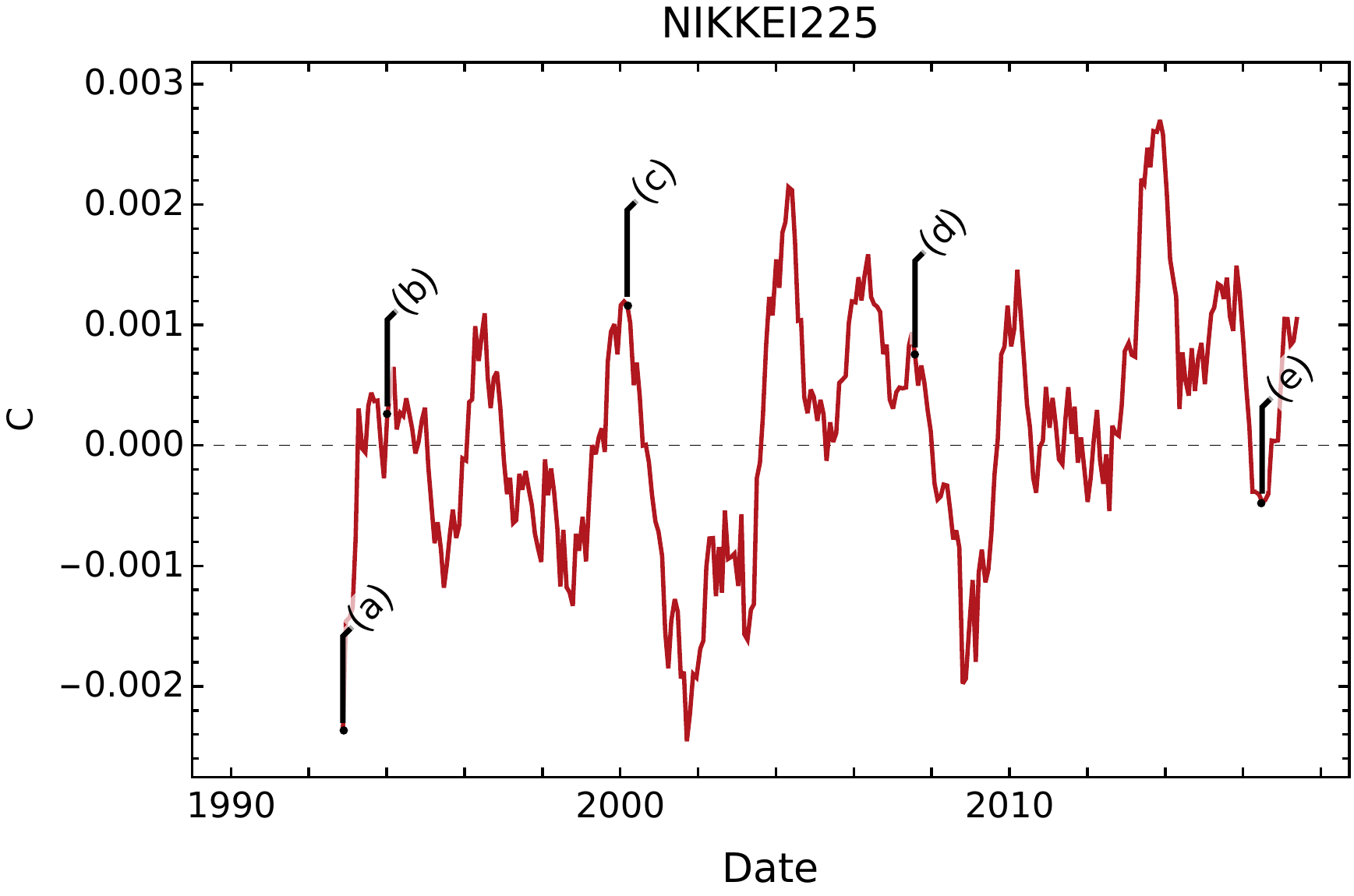}
            \caption[]%
            {{\small Nikkei: $C$ vs time}}
            \label{fig:TVRetsTimeFigsNikkei}
        \end{subfigure}
        \caption[Plots of most plausible symmetry point $C$ time evolution for simple returns..]
        {\small Most plausible symmetry point $C$ time evolution only for daily returns. Same time window of 252 trading days was used and graphs are displayed separately. Some important market extreme events are pointed out.} 
        \label{fig:CvsTRets}
\end{figure}

It is pertinent to include here a brief comment on the small range of variation of our variable $C$ displayed in figures \ref{fig:RetsFigs} to \ref{fig:TVRetsFigs},  \ref{fig:CEvolution} and \ref{fig:CvsTRets}. Table \ref{tab:confidence} shows the $C$ interval range of all our observables for all our data samples.

Figures \ref{fig:SymmReturns005} to \ref{fig:SymmTVReturns005} included in \ref{appendixa} show the evolution of $C$ and it confidence interval for all our observables and all analyzed data samples. Small values of $C$ must be expected, since that bigger the $C$ value,  the further the market would be from efficiency. Since the market seems to be efficient most of the time, $C$ usually must take  very small values. A quantitative study of the above mentioned could be addressed in another paper. Same comment applies also to the variation range of the  mean value $\mu$ of the observables analyzed here.

\section{Conclusions}
\label{sec:Final}

 In this article  we study the symmetry of distributions of two coarse-grained,  multi-scale observables  computed from daily uninterrupted trends, named TReturns and TVReturns of four financial 
daily data samples: DJIA, DAX, Nikkei and IPC indices. Per se, these distributions display
interesting properties such as bi-modality, smoother variations,
a time multi-scale nature, etc., as discussed in section 3, deserving a deeper
and additional analysis to be presented in a following paper. For comparison
and completeness purposes, the same analyses are performed on the usual daily
returns, named here only Returns.

Studying the symmetry of the distribution of variations of stocks prices, financial
index or other assets is a topic of great interest in finance. It gives us information of market fairness and how easily wealth can be created or destroyed. Also,
it is strongly linked to the Efficient Market Hypothesis (EMH), which implies
that for efficient markets, variations of assets prices or indices values must be
symmetric. It is precisely due to this reason that people from the financial
community almost automatically consider that these kind of distributions must
be symmetrical around the origin. In this work we show that this is not the general case: the most plausible symmetry point $C$ of financial variations fluctuates
around and close to zero. Furthermore, there exist intervals containing possible
symmetry points, varying over time and, although the mean value $\mu$ of financial
variations may belong to the interval of symmetry, $\mu \subseteq (C_{min},C_{max})$, it is not always necessarily the most plausible symmetry point. See figures \ref{fig:RetsFigs} to \ref{fig:TVRetsFigs}.

Results of our test for symmetry around zero, with a  significance level $\alpha=0.05$ 
and a time window of 252 trading days can be seen in table \ref{tab:confidence}, showing that
for all the analyzed data samples, with the exception of Nikkei index, Returns
and TReturns are not zero symmetric and that TVReturns are zero symmetric
according to the symmetry test applied to all data samples. In the same table, 
the intervals where these observables, Returns and the constructed multi-scale returns,
have symmetry points, together with their mean values and their corresponding most 
plausible symmetry points found with our method are displayed. We find that, in the same way as log-returns have greater plausibility of being symmetric than plain returns (defined as just price differences) \cite{Beedles}, TVReturns have greater plausibility of being symmetric than TReturns.

We have also examined the behavior of the most plausible symmetry point $C$ of our samples of financial variations around dates with extreme market movements, see figure \ref{fig:CvsTRets}, showing that $C$ displays a good response and sensitivity to market fluctuations. In our opinion, the sensitivity of $C$ to these fluctuations make our methodology a good candidate to construct automatic trading systems and may
be attractive to people interested in this aspect of financial markets.

According to our statistical test, it is not always possible to find symmetry intervals around zero for certain  significance levels, as can be seen in figure \ref{fig:ZeroSymm}. The same can be said more generally for other points that are different but close to zero.

In fact, at the confidence limits  and  time window resolution used in our analyses, most of the time markets seem to be close to efficiency, and then the most plausible symmetry point $C$ as well as $\mu$ must adopt very small values and increase when the markets moves away from efficiency. Of course, using a small enough time windows would result in periods of time around big market movements  where would  not be  possible to find an interval of symmetry points and neither a most plausible symmetry point. This issue  deserves a proper and additional study and as we mentioned before, we consider that this observation could be applied to study quantitatively the efficiency of financial markets.

On the other hand, and under a different research perspective, such as studies performed using
agents simulation models, our  results are of interest because all acceptable agent models or of
any other kind must reproduce the empirical facts of our findings.

\subsection*{\bf Acknowledgments}

We thank  Ms. Selene Jim\'enez for her \LaTeX \, revision and writing, Alejandro Aguilar for his comments and Dr. Horacio Tapia-McClung by his detailed revision to this manuscript.
This work has been endorsed by Conacyt-Mexico under Grants 283815 and 427582 and project number 5150 supported by FOINS.

\appendix

\section{Time evolution of $C$ and its confidence interval}
\label{appendixa}

Below, we display the plots of evolution on time of the most plausible symmetry point $C$ and its confidence interval for all our samples and observables. We include these plots in this appendix to make our article more readable. A time window of 252 trading days and an significance level $\alpha=0.05$ were set up in all figures.

\begin{figure}[h!tb]
        \centering
        \begin{subfigure}[b]{0.44\textwidth}
            \centering
            \includegraphics[width=\textwidth]{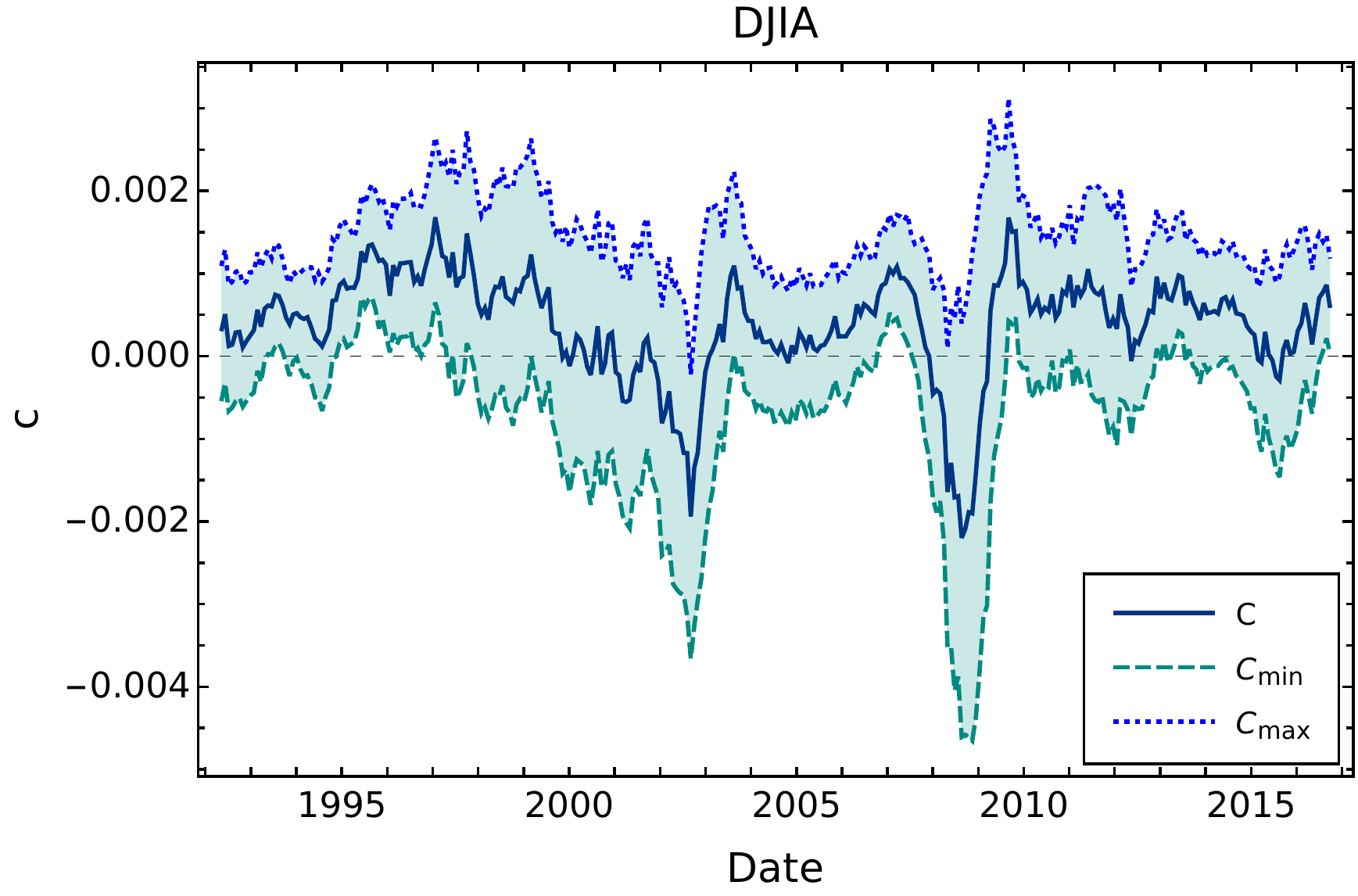}
            \label{fig:SymmReturns005DJIA}
        \end{subfigure}
        \quad
        \begin{subfigure}[b]{0.44\textwidth}
            \centering 
            \includegraphics[width=\textwidth]{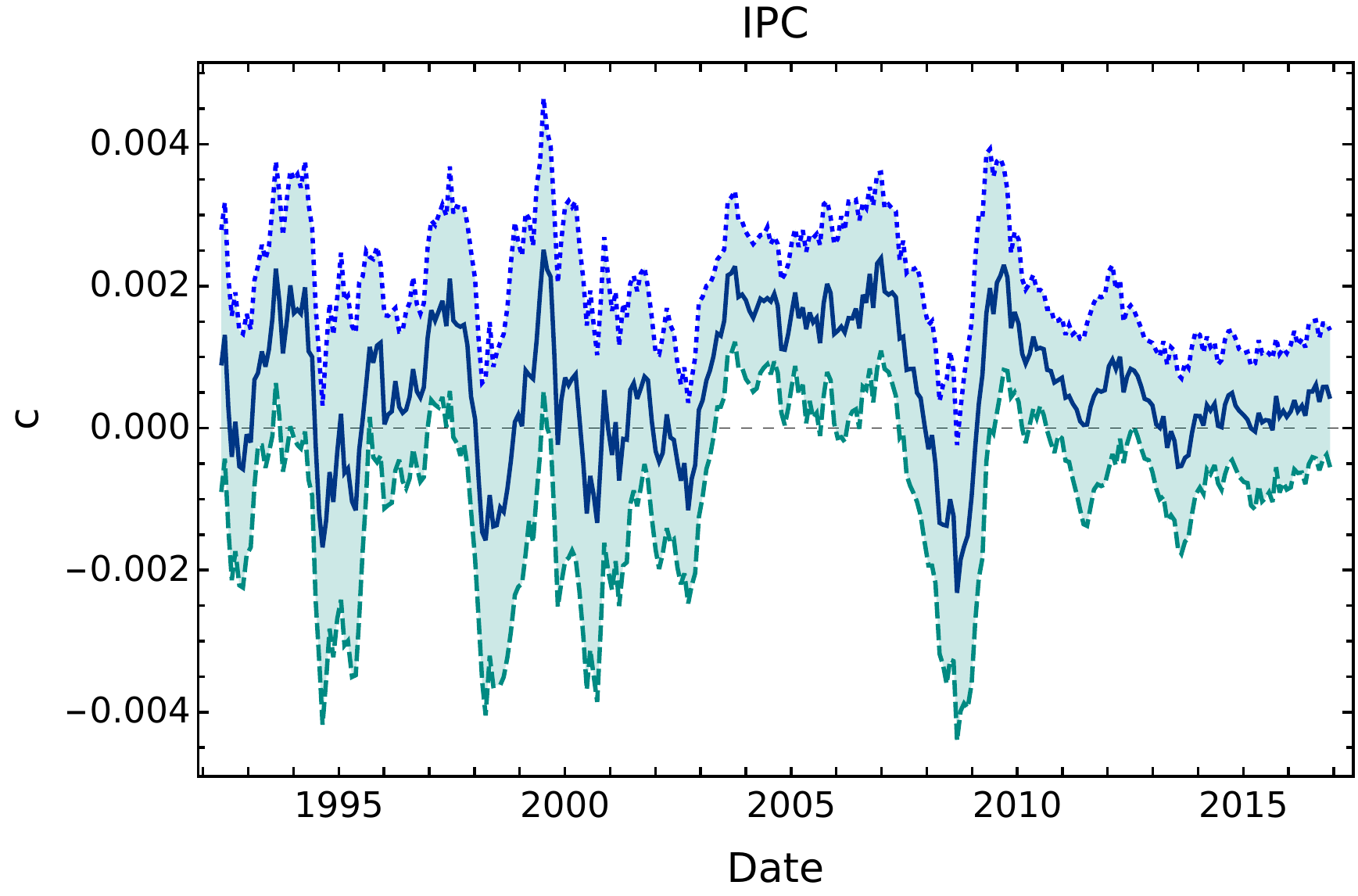}
            \label{fig:SymmReturns005IPC}
        \end{subfigure}
        \begin{subfigure}[b]{0.44\textwidth}
            \centering 
            \includegraphics[width=\textwidth]{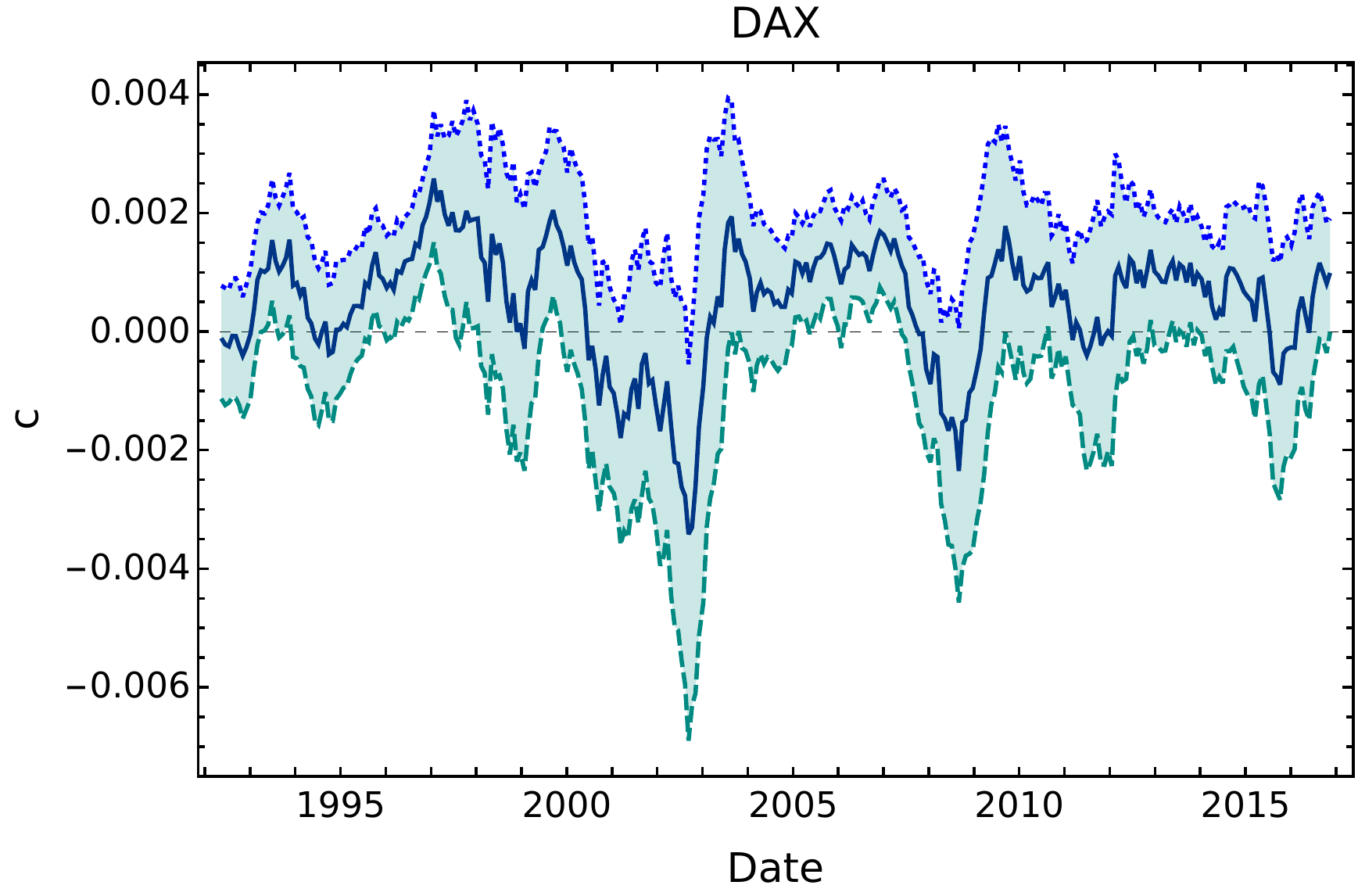}
            \label{fig:SymmReturns005DAX}
            \end{subfigure}
        \quad
        \begin{subfigure}[b]{0.44\textwidth}
            \centering 
            \includegraphics[width=\textwidth]{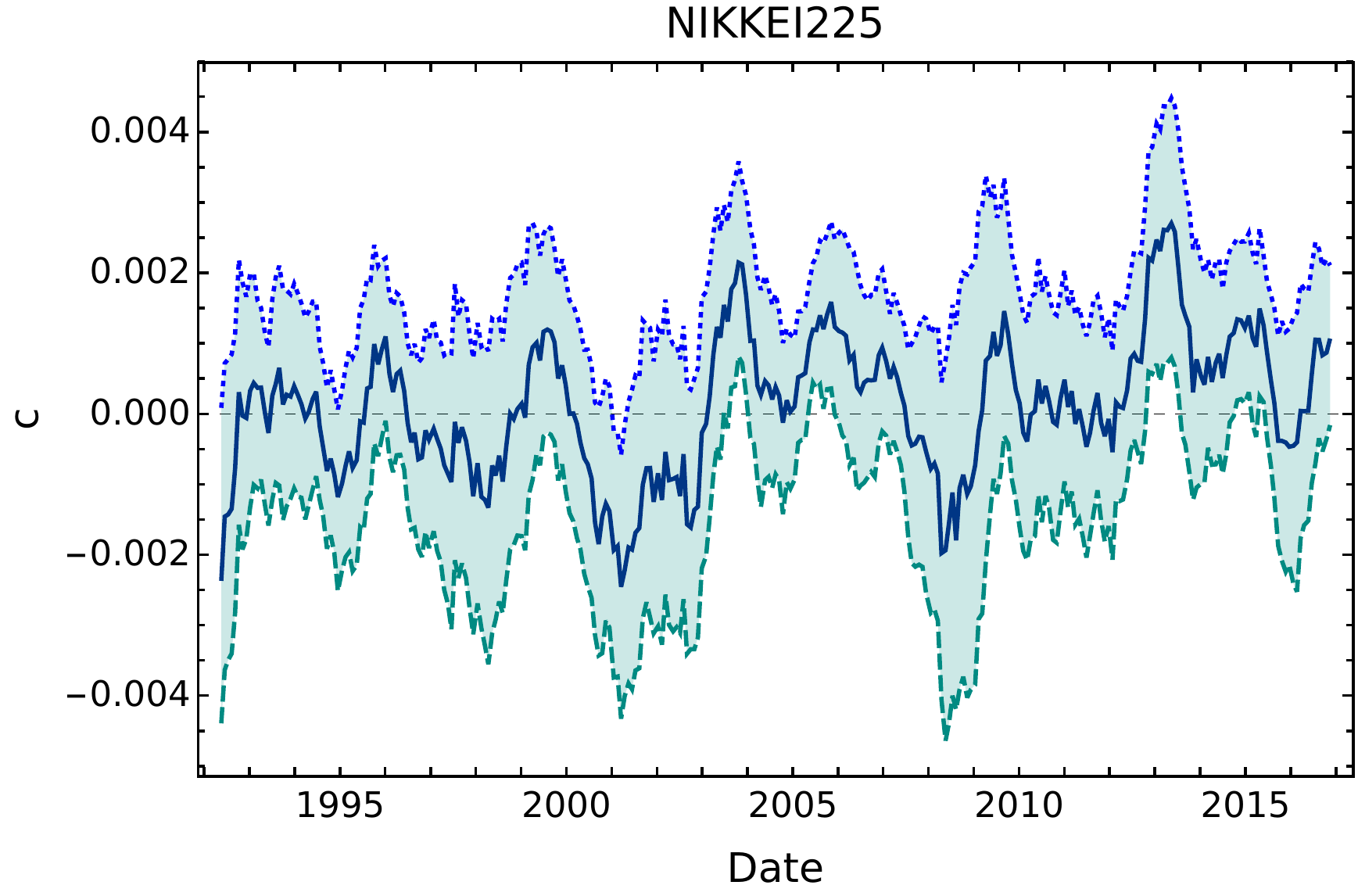}
            \label{fig:SymmReturns005Nikkei}
        \end{subfigure}
        \caption[Plots of the most plausible symmetry point $C$ and the confidence interval ]
        {\small Plots of the most plausible symmetry point $C$ and the confidence interval from the lower bound $C_{min}$ to the upper bound $C_{max}$ for the time series of the Returns. } 
        \label{fig:SymmReturns005}
\end{figure}
    
\begin{figure}[h!tb]
        \centering
        \begin{subfigure}[b]{0.44\textwidth}
            \centering
            \includegraphics[width=\textwidth]{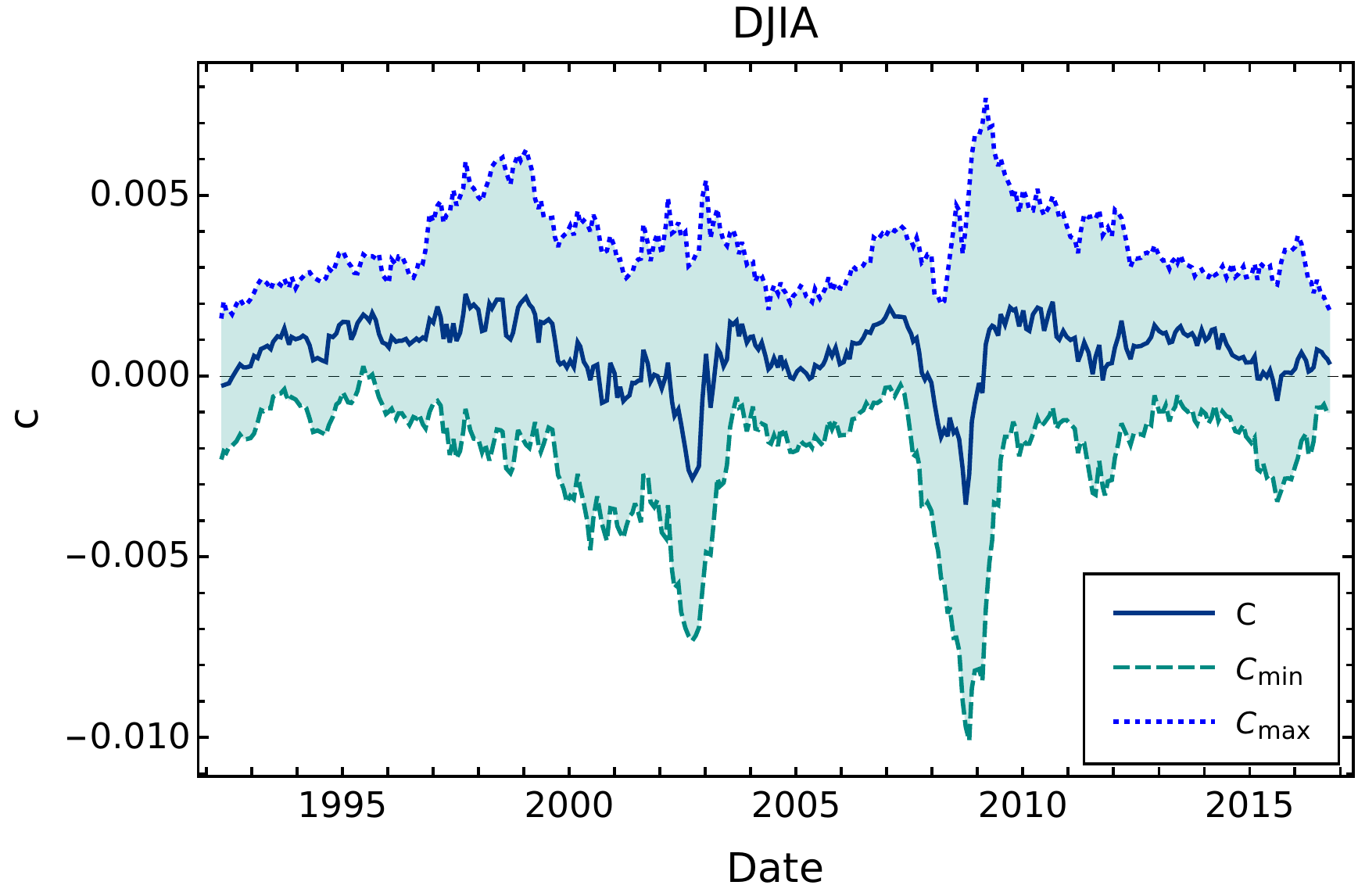}
            \label{fig:SymmTReturns005DJIA}
        \end{subfigure}
        \quad
        \begin{subfigure}[b]{0.44\textwidth}
            \centering 
            \includegraphics[width=\textwidth]{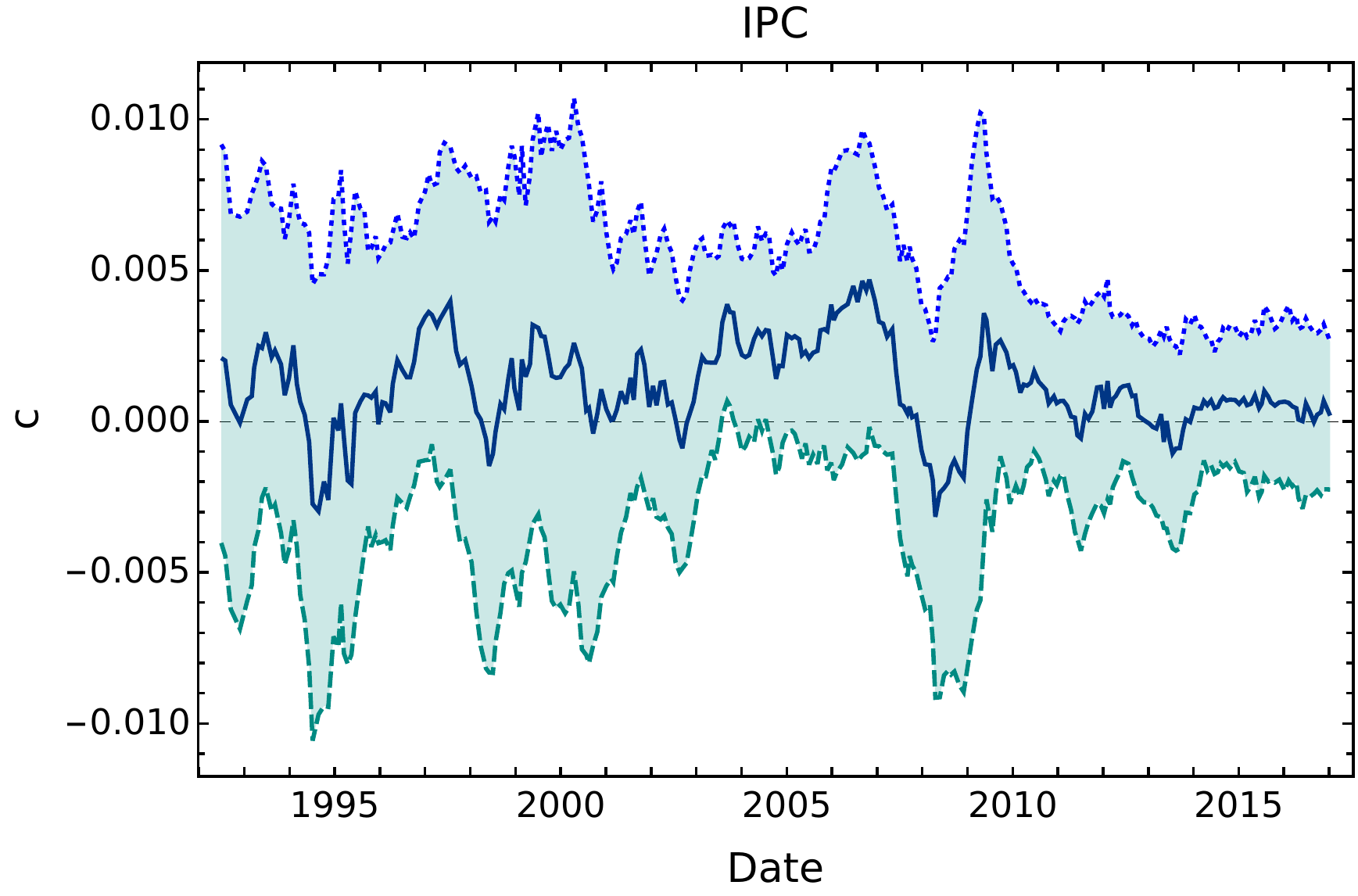}
            \label{fig:SymmTReturns005IPC}
        \end{subfigure}
        \begin{subfigure}[b]{0.44\textwidth}
            \centering 
            \includegraphics[width=\textwidth]{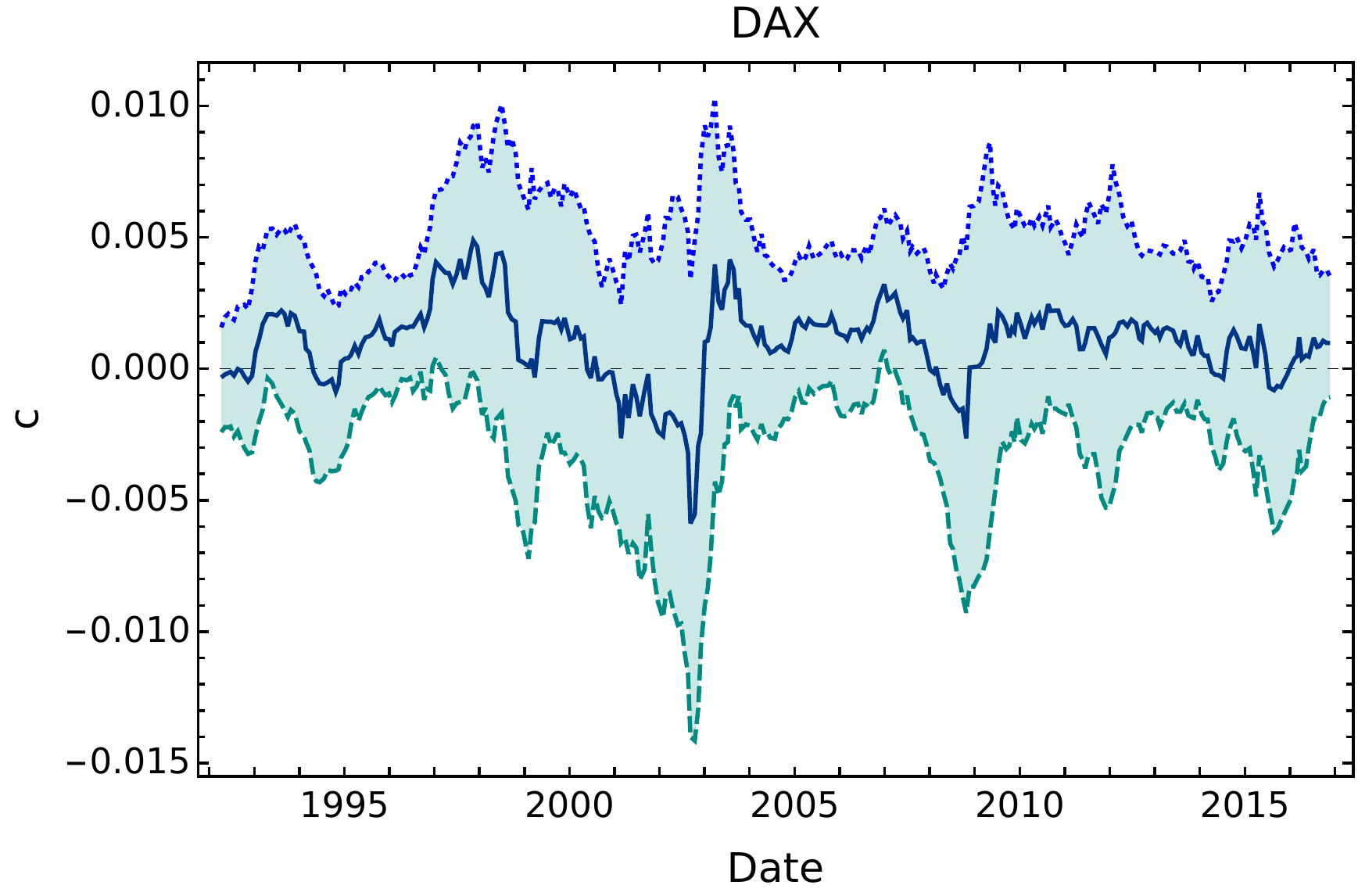}
            \label{fig:SymmTReturns005DAX}
            \end{subfigure}
        \quad
        \begin{subfigure}[b]{0.44\textwidth}
            \centering 
            \includegraphics[width=\textwidth]{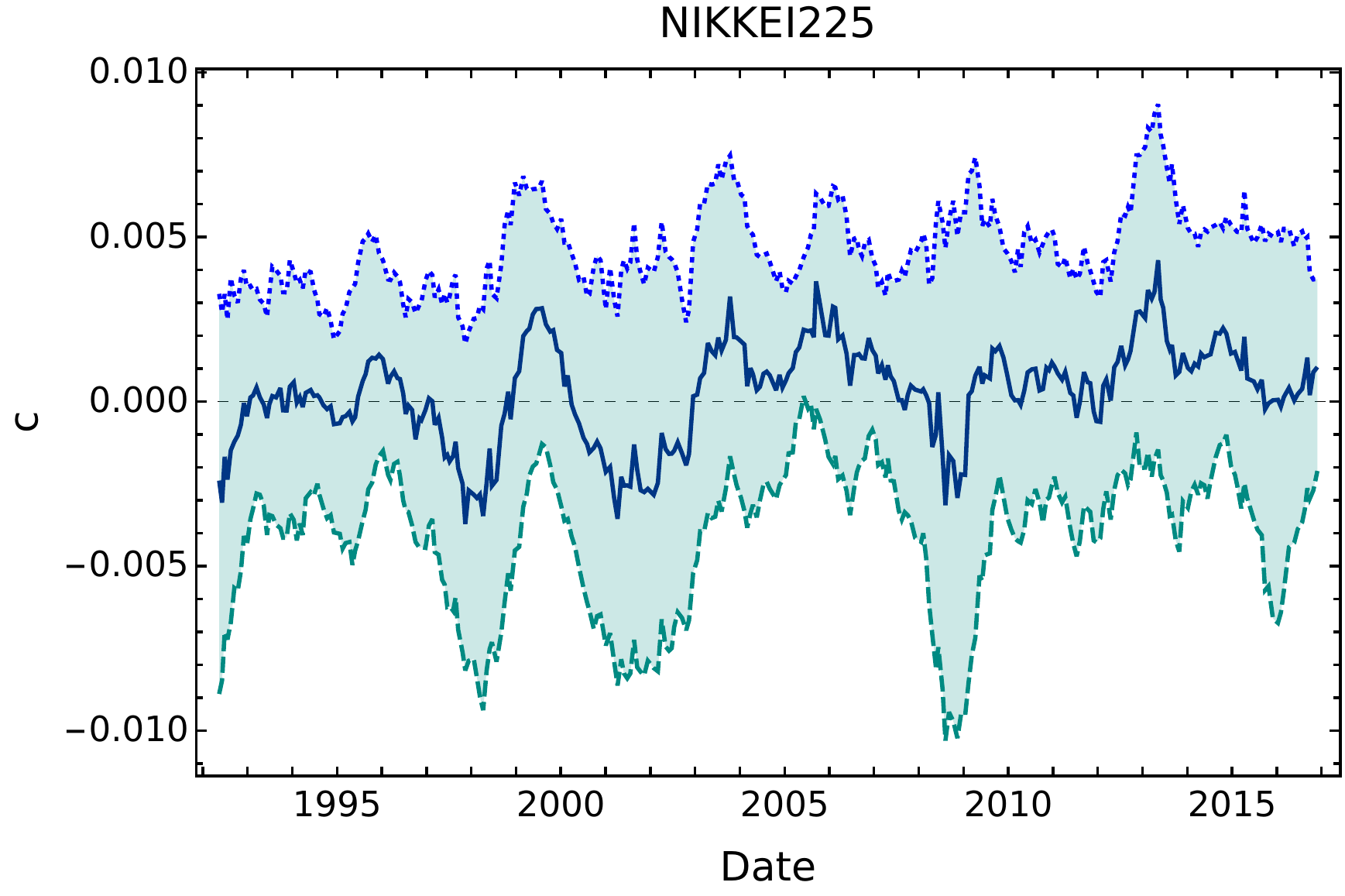}
            \label{fig:SymmTReturns005Nikkei}
        \end{subfigure}
        \caption[Plots of the most plausible symmetry point $C$ and the confidence interval TReturns]
        {\small Plots of the most plausible symmetry point $C$ and the confidence interval from the lower bound $C_{min}$ to the upper bound $C_{max}$ for the time series of the TReturns.} 
        \label{fig:SymmTReturns005}
\end{figure}

\begin{figure}[h!tb]
        \centering
        \begin{subfigure}[b]{0.44\textwidth}
            \centering
            \includegraphics[width=\textwidth]{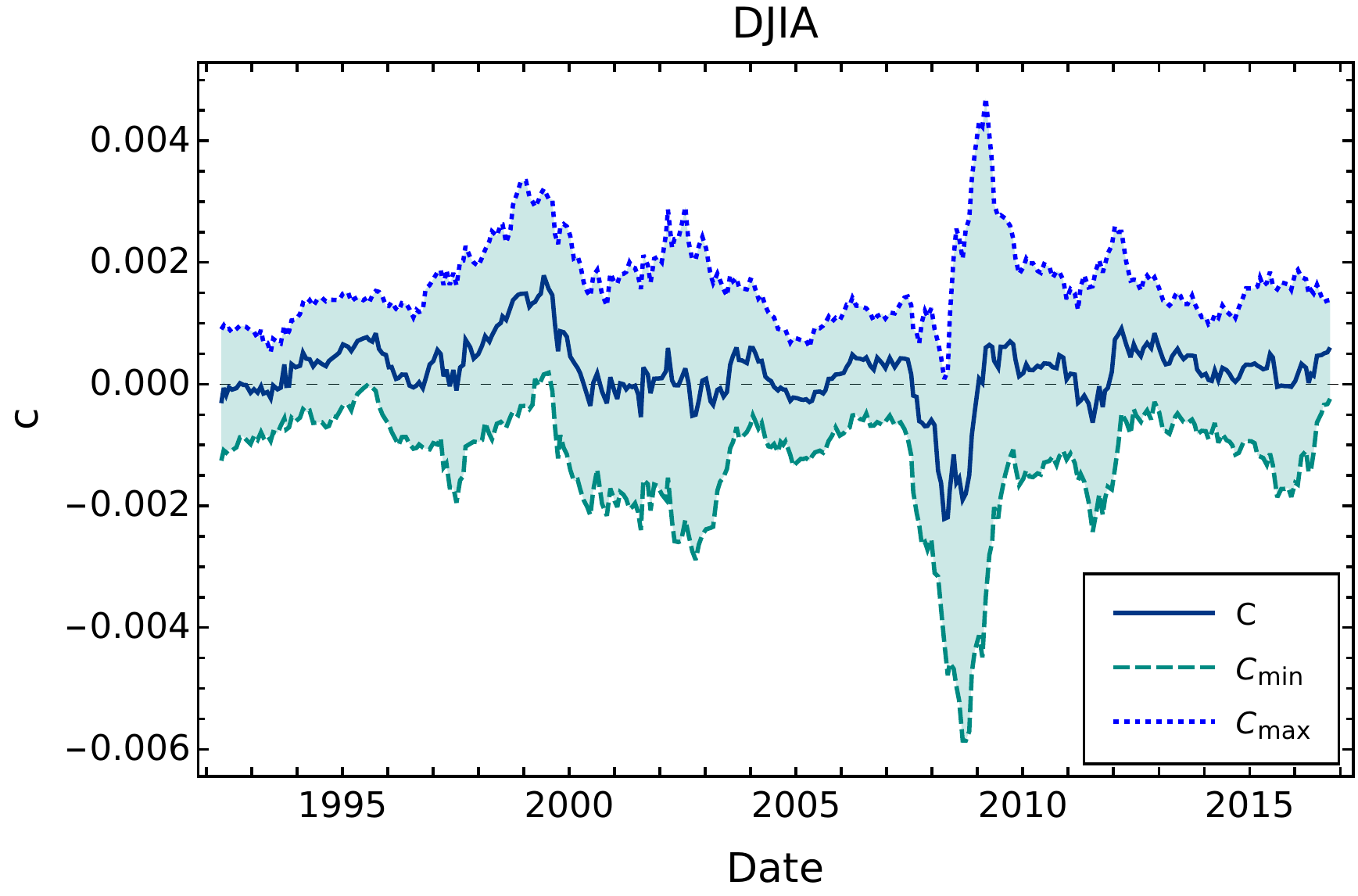}
            \label{fig:SymmTVReturns005DJIA}
        \end{subfigure}
        \quad
        \begin{subfigure}[b]{0.44\textwidth}
            \centering 
            \includegraphics[width=\textwidth]{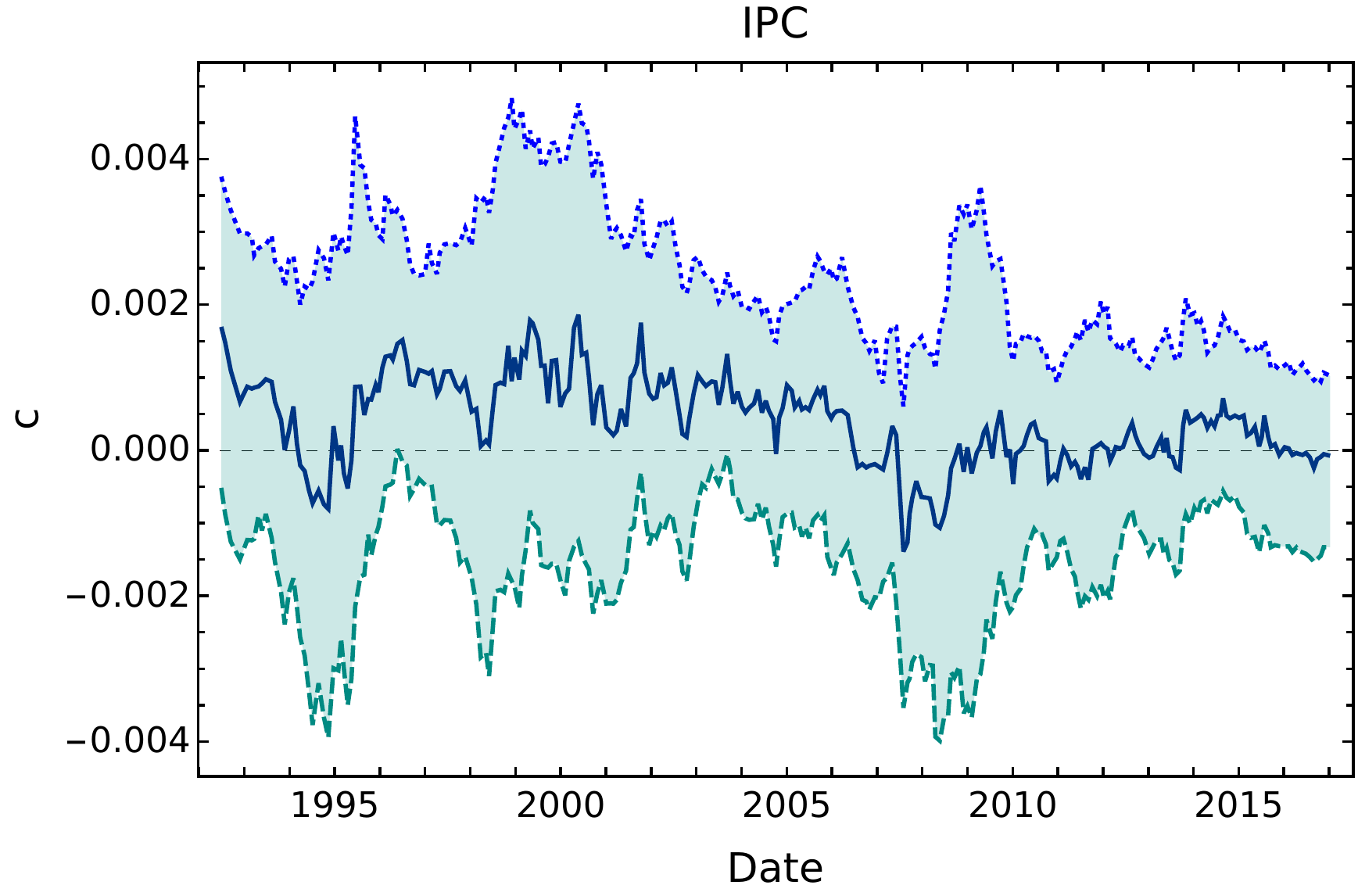}
            \label{fig:SymmTVReturns005IPC}
        \end{subfigure}
        \begin{subfigure}[b]{0.44\textwidth}
            \centering 
            \includegraphics[width=\textwidth]{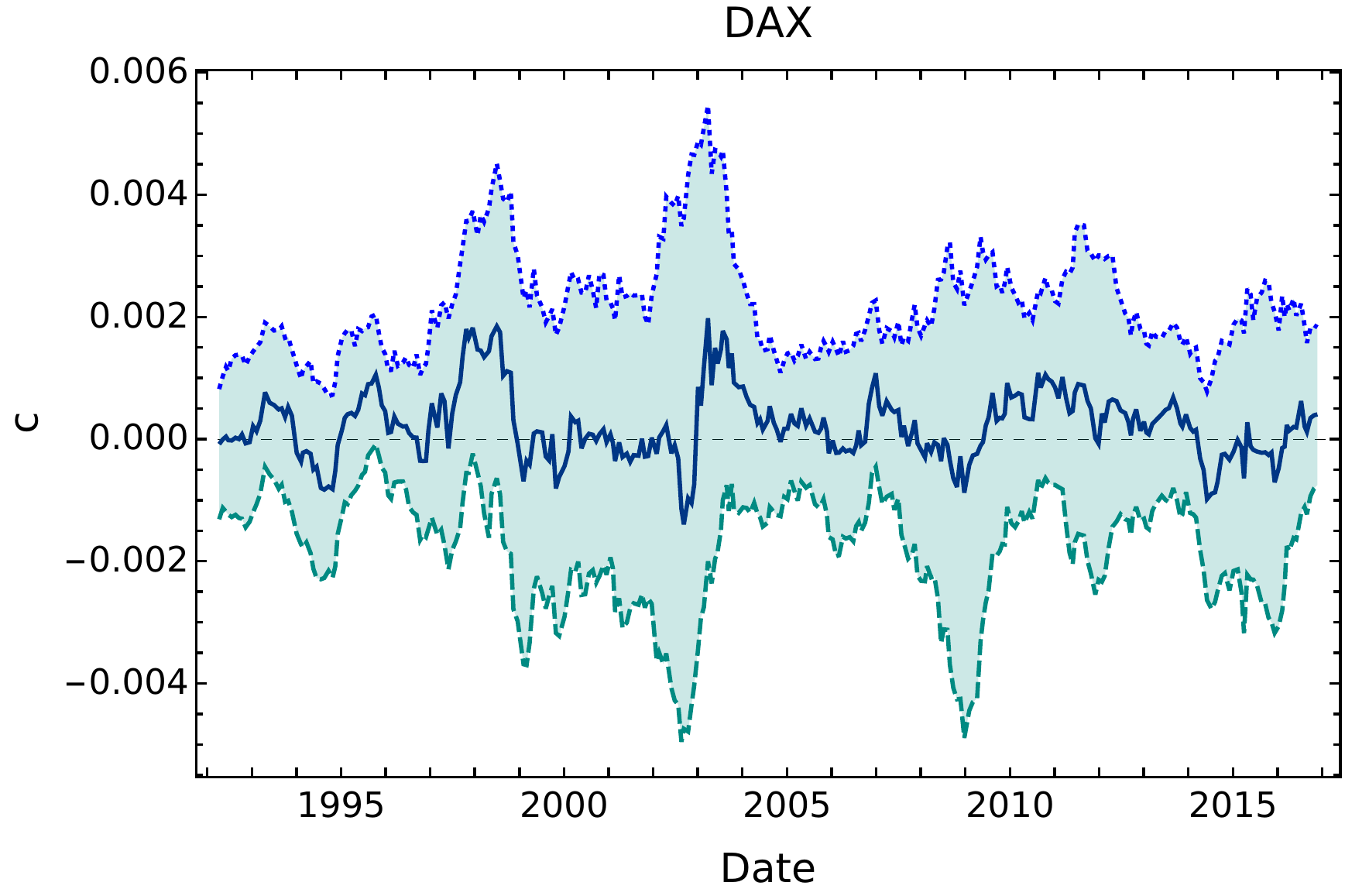}
            \label{fig:SymmTVReturns005DAX}
            \end{subfigure}
        \quad
        \begin{subfigure}[b]{0.44\textwidth}
            \centering 
            \includegraphics[width=\textwidth]{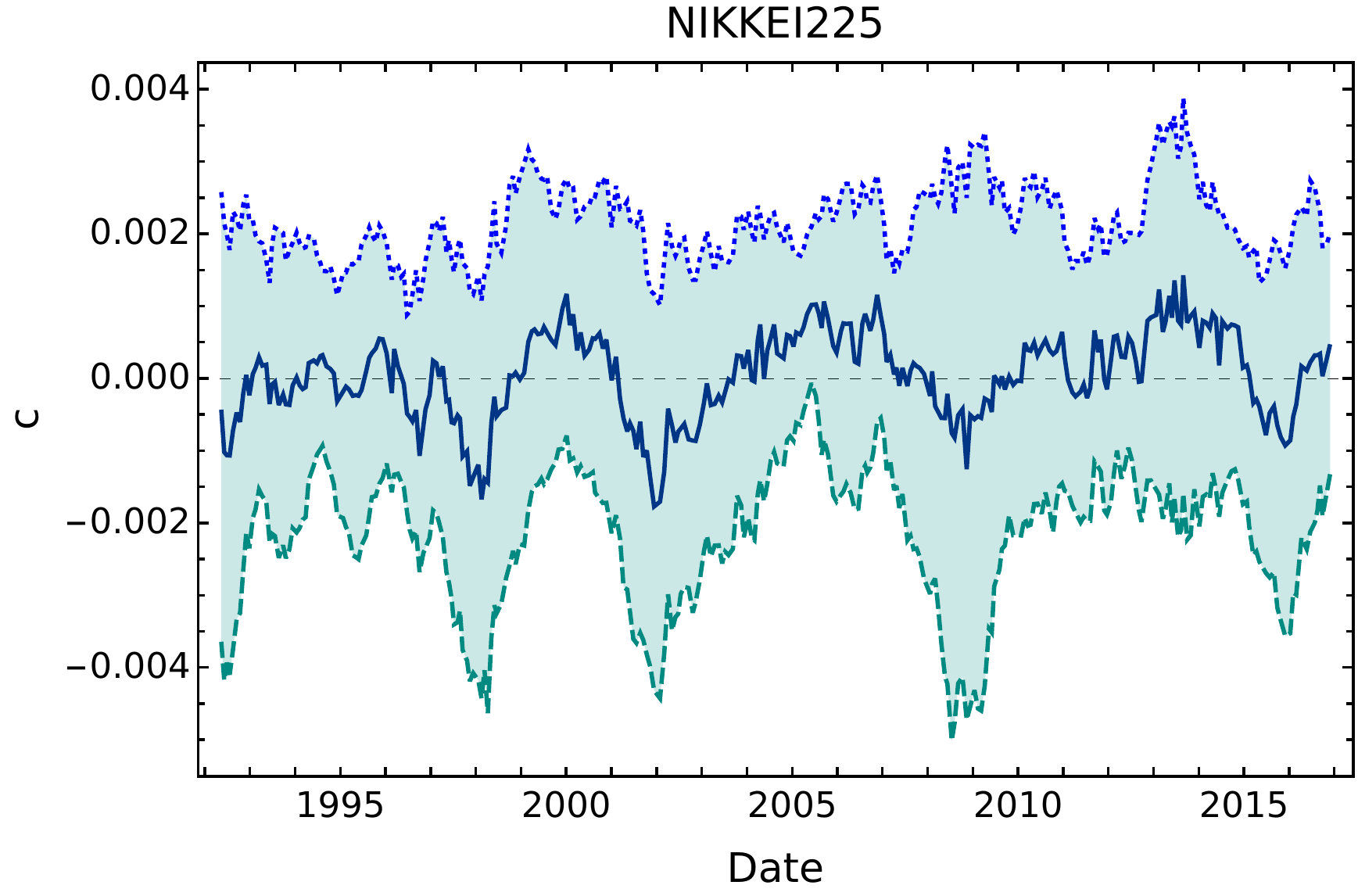}
            \label{fig:SymmTVReturns005Nikkei}
        \end{subfigure}
        \caption[Plots of the most plausible symmetry point $C$ and the confidence interval TReturns]
        {\small Plots of the most plausible symmetry point $C$ and the confidence interval from the lower bound $C_{min}$ to the upper bound $C_{max}$ for the time series of the TVReturns.} 
        \label{fig:SymmTVReturns005}
\end{figure}

\clearpage
\nocite{*}
\bibliography{SymmTrendsTnRev}

\end{document}